\index{\cite{\footnote{\pageref{\ref{•}}}}}

\documentclass[iop]{emulateapj-rtx4}

\usepackage{psfig}
\usepackage{epsf}
\usepackage{graphics}
\usepackage{color}
\usepackage{lscape}
\usepackage{graphicx}
\usepackage{verbatim}

\usepackage{morefloats}

\newcommand{\HII}{H\,{\sc ii}}
\newcommand{\arcs}{\arcsec}

\newcommand{\Spitzer}{{\it Spitzer }}

\shorttitle{WISE Enhanced Resolution Galaxy Atlas}
\shortauthors{Jarrett et al. 2012}

\begin{document}

\title{
Extending the Nearby Galaxy Heritage with WISE:\\
First Results from the WISE Enhanced Resolution Galaxy Atlas
}

\author{T.H. Jarrett\altaffilmark{1,2},  F. Masci\altaffilmark{1},   C.W. Tsai\altaffilmark{1, 5},
S. Petty\altaffilmark{3}, M. Cluver\altaffilmark{4}, Roberto J. Assef\altaffilmark{5,12}, D. Benford\altaffilmark{6}, A. Blain\altaffilmark{7}, C. Bridge\altaffilmark{13},
 E. Donoso\altaffilmark{14}, P. Eisenhardt\altaffilmark{5}, B. Koribalski\altaffilmark{8},
 S. Lake\altaffilmark{3},  James D. Neill\altaffilmark{13}, M. Seibert\altaffilmark{9}, K. Sheth\altaffilmark{10},
 S. Stanford\altaffilmark{11}, E. Wright\altaffilmark{3}
 }

\altaffiltext{1}{Infrared Processing and Analysis Center, California Institute of Technology, Pasadena, CA 91125, USA}
\altaffiltext{2}{Astronomy Department, University of Cape Town, Rondebosch 7701, South Africa}
\altaffiltext{3}{Physics and Astronomy Department, University of California, Los Angeles, CA 90095, USA}
\altaffiltext{4}{Australian Astronomical Observatory, PO Box 915, North Ryde, NSW 1670, Australia}
\altaffiltext{5}{Jet Propulsion Laboratory, California Institute of Technology, 4800 Oak Grove Drive, Mail Stop 169-221, Pasadena, CA 91109, USA}
\altaffiltext{6}{NASA Goddard Space Flight Center, Code 665, Greenbelt, MD 20771, USA}
\altaffiltext{7}{Physics \& Astronomy, University of Leicester, University Road, LE1 7RH, UK}
\altaffiltext{8}{CSIRO Astronomy \& Space Science, Australia Telescope National Facility (ATNF), P.O. Box 76, Epping, NSW 1710, Australia}
\altaffiltext{9}{Observatories of the Carnegie Insititution for Science, 813 Santa Barbara Street, Pasadena, CA 91101, USA}
\altaffiltext{10}{NRAO, 520 Edgemont Road, Charlottesville, VA 22903-2475, USA }
\altaffiltext{11}{Department of Physics, University of California, One Shields Avenue, Davis, CA 95616, USA}
\altaffiltext{12}{NASA Postdoctoral Program Fellow}
\altaffiltext{13}{Department of Astronomy, California Institute of Technology, Pasadena, CA 91125, USA}
\altaffiltext{14}{Spitzer Science Center, IPAC, California Institute of Technology, Pasadena, CA 91125, USA}

\begin{abstract}

The Wide-field Infrared Survey Explorer (WISE)
mapped the entire sky at mid-infrared wavelengths
3.4 $\mu$m, 4.6 $\mu$m, 12 $\mu$m and 22 $\mu$m.
The mission was primarily designed to extract point sources,
leaving resolved and extended sources, for the most part, unexplored.
Accordingly,
we have begun a dedicated WISE Enhanced Resolution Galaxy Atlas (WERGA) project
to fully characterize large, nearby galaxies and produce a legacy image atlas
and source catalogue.
Here we demonstrate the first results
of the WERGA-project for a sample of 17 galaxies,  chosen to
be of large angular size, diverse morphology, and covering a range in color, stellar mass
and star formation.  It includes many well-studied galaxies, such as
M\,51, M\,81, M\,87, M\,83, M\,101, IC\,342.  Photometry and surface brightness decomposition is
carried out after special super-resolution processing, achieving spatial
resolutions similar to that of \Spitzer-IRAC.  The enhanced resolution method
is summarized in the first paper of this two part series.
In this second work, we present WISE, \Spitzer and GALEX photometric and characterization measurements for the sample galaxies,
combining the measurements to study the global properties.
We
derive star formation rates using the PAH-sensitive 12 $\mu$m (W3) fluxes, warm-dust sensitive
22 $\mu$m (W4) fluxes, and young massive-star sensitive UV fluxes.   Stellar masses are estimated using the
3.4 $\mu$m (W1) and 4.6 $\mu$m  (W2) measurements that trace the dominant stellar mass content.
We highlight and showcase the detailed results of M\,83, comparing the WISE/Spitzer results
with the ATCA H{\sc i} gas distribution and GALEX UV emission, tracing the evolution
from gas to stars.
In addition to the enhanced images,
WISE's all-sky coverage provides a tremendous advantage over
\Spitzer for building a
complete nearby galaxy catalog, tracing both stellar mass and star-formation histories.
We discuss the construction of a complete mid-infrared catalog of galaxies
and its complementary role to study the assembly and evolution of galaxies in the local universe.

\end{abstract}


\section{Introduction}

Galaxies are the basic building blocks of the baryonic universe, self-contained laboratories to study
the complexity and sustainability of converting gas into stars.
Comprising the local universe,
nearby galaxies represent a {\it fossil record} of galaxy evolution,
the boundary condition for cosmological models that explain past and current state of
the universe.
Although galaxies come in all shapes, sizes, mass and morphological cases,
the physics governing their structure and evolution should be elemental:  we expect that the gravitational collapse of the interstellar medium (ISM, i.e. gas) to form new stars is held in check by angular momentum and the energetic ``feedback'' from recently formed hot massive stars, supernovae explosions, and active nuclei.
However, even the basic processes in this evolution are poorly understood, leaving major questions unanswered, including: What is the relationship between the density of the ISM and the amount of stars formed?  What governs the mass spectrum of stars formed in a single event?
How does the internal mass distribution within a galaxy affect its distribution of ISM and star formation? What role does the dominant mass constituent -- the dark matter, play?
Probing galactic structure  and star formation history requires understanding the distribution of stars and gas among galaxies of all types and luminosities across a complete range of environments.

The modern study of nearby galaxies
is characterized by multi-wavelength sensing that probes the diverse physical processes that drive galaxy evolution.
Each window of the electromagnetic spectrum, from the radio to the X-ray, provides a complementary set of tools
that, in combination, reveal the internal life cycle of galaxies.  The infrared window, for example, has dual capability: sensitive to
stellar light from the evolved population of stars and relatively low-temperature processes from the interstellar medium
and star formation regions.  It is ideally suited for studying the stellar mass distribution and obscured star-formation history in galaxies.
The \Spitzer Infrared Nearby Galaxies Survey
(SINGS; Kennicutt et al. 2003) represents the
most complete study of nearby galaxies, employing every infrared instrument of \Spitzer to study in detail the properties
of 75 nearby `representative' galaxies.  With imaging only, a larger sample is found in the SINGS follow-up project, Local Volume Legacy (LVL).
In progress, the \Spitzer Survey of Stellar Structure in Galaxies (S4G; Sheth et al. 2010) expands the
sample to several thousand galaxies through the two short (near-infrared) wavelength bands of IRAC (3.6 and 4.5 $\mu$m), focusing on the internal stellar
structure of galaxies.

Following closely in succession to the AKARI all-sky survey (Murakami et al. 2006),
the Wide-field Infrared Survey Explorer (WISE;  Wright et al. 2010) is the
latest generation infrared space telescope.  As with AKARI but unlike {\it Spitzer}, it was designed and implemented
to map the entire sky.  It was thus
capable of constructing large, diverse
and complete statistical samples.   In common with \Spitzer imaging, it has both
near-infrared and mid-infrared channels, sensitive to both stellar structure (as with S4G) and
interstellar processes (as with SINGS).
The WISE Allsky Data Release of March 2012 is available
through the Infrared Science Archive (IRSA)
\footnote{http://irsa.ipac.caltech.edu/}, and includes includes imaging and source catalog.
It should be emphasized that the WISE Source Catalog
is designed, optimized and calibrated for point sources.
The complexity of detecting and measuring resolved
sources was beyond the resources of the WISE Science Data Center (WSDC) processing.  As a consequence, the WISE archive
and public release catalogs have either completely missed nearby galaxies or, even worse, their integrated fluxes are systematically
underestimated (because they are measured as point sources) and often chopped into smaller pieces.
However, the WISE public-release
imaging products do capture
resolved and complex objects.  One of the goals of this current study is to use new image products to characterize and assess the
quality of source extraction for resolved galaxies observed by WISE.

This present work demonstrates how WISE imaging can be utilized to study nearby galaxies by
focusing on a sample of large, well-studied galaxies.
Notably, we apply a technique to enhance the spatial resolution of WISE,
known as the Maximum Correlation
Method (MCM;  Masci \& Fowler 2009),
extracting information on
physical scales
comparable to those of \Spitzer imaging, thereby enabling detailed study of the internal anatomy of galaxies.
The WISE-designed MCM method is summarized in Jarrett et al. (2012a; hereafter referred to as Paper I),
and its performance is demonstrated using both simulations and real WISE imaging
of the spiral galaxy NGC\,1566.  The interested reader should refer to Paper I for more details
on WISE imaging and the MCM.
One of the goals of this current study is to use new image products to characterize and assess the
quality of source extraction for resolved galaxies observed by WISE.  We apply image resolution-enhancement and compare
the resulting measurements with those extracted using \Spitzer imaging.

The modest sample presented in
this work represents the pilot study of the more complete WISE Enhanced Resolution Galaxy Atlas (WERGA), consisting of several thousand
nearby galaxies.
The WERGA will comprise a complete mid-infrared source catalog and high-resolution image Atlas
of the largest (diameter $>$ 1$\arcmin$) angular-sized galaxies in the local universe.
The galaxies that have an
optical or near-IR  angular diameter greater than 2$\arcmin$ will undergo special MCM processing in which
super-resolution methods are used to create the highest angular (spatial) resolution WISE images,
comparable to the spatial resolution of Spitzer-IRAC and Spitzer-MIPS24, while the remaining sources will be co-added
using a `drizzle' method that significantly improves upon the nominal WISE co-added imaging
\footnote{Public release WISE co-added images are referred to as ``Atlas" images, available through
http://irsa.ipac.caltech.edu/applications/wise/}.
Source characterization and extraction will be carried
out with these enhanced images, comprising a source catalog that will be part of a public release through
NASA Extragalactic Database (NED).

We have chosen a sample of nearby galaxies all observed by \Spitzer and GALEX, to focus on the detailed WISE performance
relative to these missions.  In Section 2 we introduce the sample, and in Section 3 the observations and data.
In Section 4 we present the results and performance, starting with a case study of
early-type galaxies with their fossilized stellar populations, and then for the entire sample that consists of a diverse galaxy `zoo', comparing and
contrasting the photometric results of WISE with those of \Spitzer and IRAS (additional detail are provided in the Appendices).
Section 5 presents a detailed source characterization study of star-forming galaxy M\,83 (NGC\,5236), comparing WISE with \Spitzer,
GALEX and radio observations.
In Sections 6 and 7 we present the
star formation rates and stellar mass estimation, respectively, derived from the global UV and IR photometry, demonstrating the capabilities of
WISE to study these two important components of galaxy evolution. 
Finally in Appendix D, we discuss ongoing
work in building the WERGA.
The legacy value of these
high-resolution images will span decades. WISE is likely to be the most sensitive mid-IR all-sky survey
available for many years to come.

All reported magnitudes are in the Vega System (unless otherwise specified).
Conversion to the monochromatic AB system entails an additional
2.699, 3.339, 5.174, and 6.620 mag added to the Vega magnitudes for W1, W2, W3 and W4, respectively
(Jarrett et al. 2011).

\begin{table*}[ht!]
{\scriptsize
\caption{Observed Sources  \label{tab:observed}}
\begin{center}
\begin{tabular}{c c c c c c c c c}
\\[0.25pt]

Name  & Distance & Type & Log M$_{\rm HI}$     & Field Size & Reference &Comment \\
           &  Mpc      &          & Log M$_\odot$        & arcmin    &                  &        \\

\hline
\\[0.25pt]

NGC\,584 & 19.0 & E4 & -- & 10 & 1 & IRAC flux calibrator;  SINGS galaxy\\
NGC\,628 (M 74) & 8.2 & SA(s)c & 9.91 & 23 & 2,3 &   SINGS galaxy\\
NGC\,777 & 54.5 & E1 Sy2 & -- & 10 & 4 & IRAC flux calibrator\\
NGC\,1398 & 20.5 &  SB(r)ab & 9.46 & 12& 5,6 & S4G galaxy\\
NGC\,1566 & 9.5 &  SAB(s)bc Sy1.5 & 9.99 & 15& 4,5,6  & SINGS galaxy\\
NGC\,2403 & 3.6 & SAB(s)cd & 9.32 & 25 & 4,5,6 & SINGS galaxy\\
NGC\,3031 (M\,81) & 3.7 & SA(s)ab & 8.93 & 43 & 7,8 & Bode's Galaxy; SINGS galaxy\\
NGC\,4486 (M\,87) & 16.7 & E0 pec; Sy & -- & 30 & 20 & Virgo A\\
NGC\,5194 (M\,51a) & 8.4 & SA(s)bc pec & 10.01 & 20 & 5,9 & Whirlpool Galaxy;  SINGS galaxy\\
NGC\,5195 (M\,51b) & 8.4 & SB0 pec & -- & 20 & 5& SINGS galaxy\\
NGC\,5236 (M\,83) & 4.7 & SAB(s)c & 9.92 & 17 & 2,5,9 & Southern Pinwheel Galaxy\\
NGC\,5457 (M\,101) & 7.2 & SAB(rs)cd & 10.30 & 33 & 9,10,11 & Pinwheel Galaxy \\
NGC\,5907& 16.5 & SA(s)c & 10.39 & 30 & 4, 5,12,13  & edge-on disk \\
NGC\,6118 & 23.1 & SA(s)cd & 9.58 &10 &  5,9 & S4G galaxy\\
NGC\,6822 & 0.5 & IB(s)m  & 8.16 & 30 & 5,6, 14,15 & Barnard's Galaxy;  SINGS galaxy\\
NGC\,6946 & 6.1 & SAB(rs)cd  & 10.10 & 23 & 5, 16,17 & Fireworks Galaxy;  SINGS galaxy\\
IC\,342 &3.1 & SAB(rs)cd; Sy-2 & 9.71 & 27 & 5,18,19 & nuclear starburst\\

\hline
\end{tabular}
\end{center}
\tablecomments{
(1) Tonry et al. (2001); (2) Herrmann, K., et al. (2008);
(3) Briggs (1982); (4) Willick et al. (1997); (5) Tully (1988) \& (2009);
(6) Kilborn et al. (2004); (7) Kanbur et al (2003); (8) Chynoweth et al. (2008);
(9) Bohnensiengel \& Huchtmeier (1981);  (10) Paturel et al. (2002);
(11)Rogstad (1971); (12) Shang et al. (1998);
(13) Sancisi \& van Albada 1987;
(14) Clementini et al. (2003);
(15) Cannon et al. (2006);
(16) Poznanski et al. (2009);
(17) Boomsma et al. (2008);
(18) Crosthwaite \& Turner, J. (2000);
(19) Saha, Claver \& Hoessel (2002);
(20) Larsen et al. (2001).
}
}
\end{table*}

\section{The Sample}
The sample selection was driven by the primary science goal of this study: to demonstrate the
scientific performance of extended source characterization using WISE imaging.
The photometric performance is assessed by comparison with \Spitzer measurements, both from
the literature and from this study.   We therefore require that the sample have previous, high
quality observations from \Spitzer IRAC and MIPS-24 imaging.  Moreover, the sample should consist
of galaxies of various morphologies,  orientations and sizes, including a few large cases (e.g., M\,83) to
study their internal anatomy.

As such, we have chosen a total of 17 galaxies for this pilot study,
presented in Table \ref{tab:observed}. The bulge-dominated
elliptical galaxies in this sample: NGC\,584, NGC\,777 and NGC\,4486 (M\,87), can be simply modeled due to their fossilized population, and thus provide a test case of
photometric calibration, color and aperture corrections. The grand design spirals: NGC\,628, M\,81, M\,51, M\,83, M\,101,
NGC\,6946 and IC\,342, all possess strong ISM and star-formation emission detected in the WISE long wavelength
channels of 12 and 22 $\mu$m.  Strong nuclear starburst emitting galaxies are included in the form of NGC\,1566, NGC\,6946
and IC\,342.  The interacting system, M\,51, consisting of a late-type spiral and early-type elliptical galaxy,
represents a challenging case for deblending the photometric components.
Edge-on disk galaxy, NGC\,5907, clearly reveals the transverse bulge and halo components.
In addition, the flocculent disk galaxy, NGC\,2403 and ringed-spiral galaxy NGC\,1398 are included.  Finally, the magellanic barred galaxy,
NGC\,6822, represents the nearest galaxy in the sample (part of the Local Group), and one of the most challenging
to characterize due to the low surface brightness, amorphous shape and severe foreground stellar contamination.
The basic properties are listed in Table \ref{tab:observed}, including the distance, morphology and
neutral hydrogen content.  Molecular hydrogen masses are tabulated in Leroy et al (2008) for
several galaxies in the sample (M51a, NGC\,628, NGC\,2403, NGC\,6946).
Most of the sample have been observed by SINGS, S4G,  or were part of targeted \Spitzer
programs (e.g., M\,83, M\,101, IC\,342).
The distances, as compiled in NED, are derived from relatively accurate ($\sim$5\%) distance-ladder methods,
including Tully-Fisher (TF),
period-luminosity Cepheid variables, tip of the red giant branch (TRGB), Type II radio supernovae,
planetary nebula luminosity function (PNLF), RSV stars and surface brightness fluctuations (SBF).
We have eliminated outlier distance measurements and re-computed the average distance in Mpc; the table also
specifies the primary reference for the distance measurements.  The neutral hydrogen mass (with references
given in the table) were rescaled to the adopted distance listed in the table.  Finally, the Hubble-type morphologies
are taken from NED and T-Types (e.g., RC3 of de Vaucouleurs et al. 1991; de Vaucouleurs 1994) are assigned using these morphologies.

\begin{figure*}
\begin{center}
\includegraphics[width=18cm]{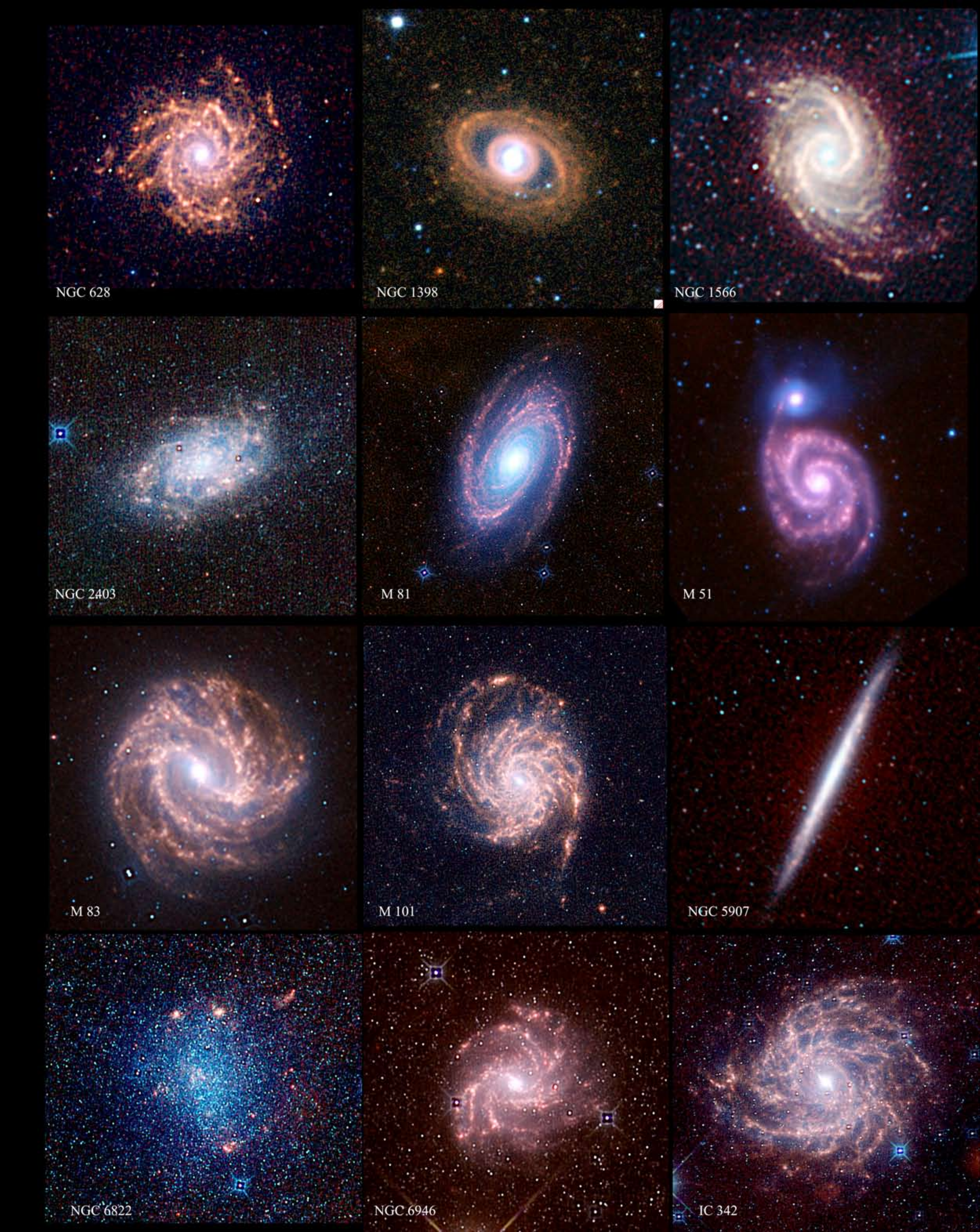}
\caption[Large Galaxies]
{\small{WISE montage of nearby galaxies,
showing resolution-enhanced images of the sample galaxies (Table \ref{tab:observed}).
The colors correspond to WISE bands:  3.4 $\mu$m (blue),
4.6 $\mu$m (cyan/green),
12.0 $\mu$m (orange),
22 $\mu$m (red).
}}
\label{Fig1}
\end{center}
\end{figure*}

\smallskip
For the galaxy sample presented in this work, we find that between 10 and 20
reconstruction iterations provides the necessary balance between
CPU demands, resolution enhancement, and artifact mitigation
(artifact-ringing associated with bright sources).
We have constructed HiRes mosaics
limiting the total number of iterations to 20 for W1,  W2 and
W3, and 10 iterations for W4 (due to lower quality sampling and S/N for this channel).
The resulting images, with pixel scale 0.6875$\arcsec$,
are presented in
Fig. \ref{Fig1}.  The angular resolution is comparable to that of \Spitzer imaging,
explained in detail in the next section.  The images are shown with four colors, where each color is assigned to
a WISE band:  blue $\leftrightarrow$ W1 (3.4 $\mu$m), cyan $\leftrightarrow$ W2 (4.6 $\mu$m),  orange $\leftrightarrow$ W3 (12 $\mu$m)
and red $\leftrightarrow$ W4 (22 $\mu$m).   Stellar light from the old, evolved population will appear blue/green,
and tends to concentrate in the nucleus and bulge regions.   The ISM, warmed and excited by star formation, will
appear yellow/orange, delineating \HII\ and photo-dissociation regions (PDRs) as well as warm dust emission (red) from
the disk.   The early-type galaxies, discussed in Section 4.1, are shown in Fig. \ref{ellipticals}.

\section{Observations, Calibration and Source Characterization Method}

\subsection{WISE Observations}

The NASA-funded Medium-Class Explorer mission, WISE, consists of a 40-cm primary-mirror space infrared telescope,
whose science instrumentation
includes 1024x1024 pixel Si:As and HgCdTe arrays, cooled with a two-stage solid hydrogen cryostat.
Dichroic beamsplitters allow simultaneous images in four mid-infrared bands, each covering a $47' \times 47'$ field of view.
The duty cycle was 11\,s,
achieved using a
 scan mirror that stabilizes the line-of-sight while the spacecraft scans the sky, achieving an angular resolution of
  $\sim$6$\arcsec$ in the short bandpasses and $\sim$12$\arcsec$  in the longest bandpass.
 Multiple, overlapping frames are combined to form deeper co-added images.
Launched in December of 2009 into a sun-synchronous polar orbit, over a time span of eight months WISE
completed its primary mission to survey the  entire sky in the  3.4, 4.6, 12 and 22 $\mu$m infrared bands
with 5\,$\sigma$ point-source sensitivities of at least 0.08, 0.11, 0.8, and 4 mJy, respectively
(Wright et al. 2010) and considerably deeper sensitivities at higher ecliptic latitudes (Jarrett et al. 2011).

\smallskip

Detailed in the WISE Explanatory Supplement (Cutri et al. 2012)\footnote{
http://wise2.ipac.caltech.edu/docs/release/allsky/},
``Atlas" Images are created from single-exposure frames that touch a pre-defined
1.56$^\circ\times$1.56$^\circ$ footprint on the sky.
For each band, a spatially registered image is produced by interpolating and co-adding multiple 7.7/8.8\,s single-exposure images
onto the image footprint.
To suppress copious cosmic rays and other transient events that populate the single-exposure frames, time-variant pixel outlier
rejection is used during the co-addition process. The resulting sky intensity ``Atlas" mosaics are 4095$\times$4095 pixels with 1.375$\arcsec$/pixel scale,
providing a 1.56$^\circ\times$1.56$^\circ$ wide-field.
In addition to the sky intensity mosaics, 1\,$\sigma$ uncertainty maps (tracking the error in intensity values) and
depth-of-coverage maps
are part of the standard products.
The number of frames that are co-added depends on the field location relative
to the ecliptic: those near the equator will have the lowest coverage (typically 12 to 14 frames), while those near the poles have the
highest coverage ($\gg$1000 frames).

For the nominal WISE survey, the co-addition process uses a resampling method based on
a matched filter derived from the WISE point spread function.  It is designed to optimize detection of point sources, the
prime objective of the WISE survey.  But this interpolation method tends to smear the images,
making them less optimal for detection and characterization of resolved sources.  Hence, for the WERGA,
we have created new mosaics using a resampling kernel that enhances angular resolution, known as
Variable-Pixel Linear Reconstruction, or `drizzling',  improving the spatial resolution performance
compared to nominal WISE Atlas Images by $\sim$30 to 40\% ,  depending on the depth of coverage.
Most importantly, we employ a deconvolution technique known as the Maximum Correlation
Method (Masci \& Fowler 2009), which can improve the resolution by factors of 3 to 4 (see Paper I).

\subsection{\Spitzer Observations}

The bulk of our sample are from the SINGS, LVL and S4G projects (see Table \ref{tab:observed}),
which have provided enhanced-quality spectroscopy and imaging mosaics\footnote
{http://data.spitzer.caltech.edu/popular/sings/20070410\_enhanced\_v1/}.
For those galaxies that are not from SINGS or S4G, we have obtained the \Spitzer
data from the \Spitzer Heritage Archive\footnote
{http://sha.ipac.caltech.edu/applications/Spitzer/SHA/}, curated by the Infrared Science Archive (IRSA).
Here we have used the \Spitzer pipeline-produced post-basic-calibrated-data (pbcds)
IRAC and MIPS mosaics.   Analysis for many
of these galaxies have been published, including NGC\,777
({\it IRAC Instrument Handbook}),
NGC\,4486 (Shi et al. 2007),
M\,83 (Dong et al. 2008),
M\,101 (Gordon et al. 2008),
and IC\,342 (IRS spectra from Brandl et al. 2006).   For IC\,342, we used a combination of both the short and long
exposures (also known as high-dynamic range, or HDR observations), because the bright nucleus saturated
in the long-exposure IRAC-1 and IRAC-2 imaging.  The resultant saturation-corrected images
have a larger flux uncertainty due to this `grafting' process.\\

For IRAC, aperture corrections are required to correct the photometry of extended sources (e.g., galaxies) whose absolute calibration is tied to point sources
with the use of finite aperture.
These corrections not only account for the ``extended" emission from the PSF outer wings (that is outside of
the finite calibration aperture), but also from the scattering of the diffuse emission across the IRAC focal plane
(see Reach et al. 2005; IRAC Instrument Handbook).  For large apertures, the corrections\footnote
{http://spider.ipac.caltech.edu/staff/jarrett/irac/calibration/ext\_apercorr.html}
are
roughly 8\%, 5\%, 23\% and 25\% for IRAC 1, 2, 3 and 4, respectively.
For MIPS-24, the photometric calibration uses an ``infinite" aperture calibration, and hence aperture corrections are generally
only needed for small apertures (relative to the size of the object being measured; see the MIPS Instrument Handbook).  We do, however, apply a small
color correction ($\sim$4\%) for the most spectrally ``red" galaxies, as recommended and tabulated in the MIPS Instrument Handbook.
Further details of the \Spitzer photometry and comparison to WISE is presented in Appendix A.

\subsection{GALEX Observations and Measurements}

GALEX FUV (0.1516 $\mu$m) and NUV (0.2267 $\mu$m) images were obtained from the GALEX Medium
Imaging Survey (MIS; Martin et al. 2005), which were
processed using the standard
GALEX pipeline (Morrissey et al. 2005, 2007).
The MIS reaches a
limiting NUV magnitude of
23 (AB mag)
through multiple eclipse exposures that are
typically 1 ks or greater in
duration, while azimuthal averaging reaches surface brightness depths of $\sim$30 to 31 mag\,arcsec$^{-2}$
(AB mag).

In order to carry out source characterization measurements, foreground stars
were identified and removed from the GALEX images.
Aperture photometry was then carried out on the
images using an elliptical annulus to determine the
median background and a nested set of elliptical apertures to perform a curve of growth analysis.
The reported integrated flux is taken from the aperture that corresponds to
the RC3 D25 diameter, whose isophote has a typical UV surface brightness of 30 mag\,arcsec$^{-2}$.
Further details are presented in Appendix B.

\subsection{Optical Imaging}

For the M\,83 analysis, we compare with optical H$\alpha$ and
B-band imaging.   The H$\alpha$ (0.657 $\mu$m) imaging was extracted
from NED and derived from Meurer et al. (2006), who used the CTIO 1.5-m
to obtain H$\alpha$(+ continuum) and R-band imaging.    The continuum subtracted
H$\alpha$ image has a seeing FWHM of 1.2$\arcsec$.
The B-band (0.44 $\mu$m) was acquired with the IMACS instrument aboard the
6.5-m Baade Telescope of the Carnegie Observatory in Las Campanas
on May 29, 2011, and was kindly provided by B. Madore (Carnegie).
It has a seeing FWHM of 0.5$\arcsec$, representing the highest angular resolution
imaging of M\,83 that is presented in this work.

\subsection{WISE Calibration and Aperture Corrections}

The WISE photometric calibration, described in detail by Jarrett et al. (2011), relies
upon stars that are located at both ecliptic poles and are common to the {\it Spitzer}, AKARI,
and Midcourse Space Experiment (MSX) infrared missions.  The stars are well characterized
K-M giants,
which have a profile that follows a Rayleigh-Jeans (R-J)
distribution, F$_\nu \sim \nu^{2}$, at mid-IR wavelengths.
WISE measures the fluxes of stars using optimal profile (PSF) fitting, and the calibration zero point
magnitude is derived from these measurements of the calibration stars.
The uncertainty in the zero point flux-to-magnitude conversion
is about 1.5\% for all bands.
Both the nature
of the calibration stars and the method used to measure their flux has important implications
toward photometric calibration of extended sources.

\smallskip

There are three different kinds of corrections that are required for aperture photometry measurements
using WISE co-added mosaics.   The first is an aperture correction that accounts for the
WISE absolute photometric calibration method using PSF profile fitting.
Similar to the \Spitzer aperture correction (Reach et al. 2005),
for large aperture measurements of WISE images the resulting integrated fluxes must be reduced by a small amount
to be consistent with the standard photometric calibration.
{\bf The All Sky Release Image Atlas
requires
the
following corrections (mag units): 0.03, 0.04, 0.03 and -0.03 mag, for W1, W2, W3 and W4, respectively, which
are to be added to the measured magnitudes.}
The uncertainty in these aperture corrections is $\sim$1\%.

The second correction is a ``color correction" that
accounts for the spectral signature of the source convolved with the WISE Relative system response (RSR).
For sources with a constant mid-IR power-law spectrum (F$_{\nu}$ $\sim \nu^0$), the standard zero magnitude flux density,
F$_{\nu 0}$,
provides the conversion from WISE magnitudes to flux density;  namely:
309.54, 171.79, 31.64 and 8.63 Jy, for W1, W2, W3 and W4 respectively.   If the source
has a spectrum that steeply rises in the mid-IR (e.g., dusty star-forming galaxies),
a color correction is required, especially for W3 due to its wide bandpass.
For example, star-forming galaxies typically have a spectral shape
index that is 1 or 2 (e.g., F$_{\nu}$ $\sim \nu^{-2}$), and thus require a color correction
that is roughly -9\% in W3.  In this case, the
zero magnitude flux density is 306.68,
170.66, 29.05, 8.28 Jy,  for W1, W2, W3 and W4 respectively.
The color correction tables and details are given in Wright et al. (2010)
and in the WISE Explanatory Supplement
\footnote{http://wise2.ipac.caltech.edu/docs/release/allsky/expsup/},
section IV.4.h.  We apply this color correction to our WISE photometry based on
the WISE color and the Hubble Type of the galaxy; note:  for bands W1, W2 and
W4, this corrections is always very small, $\sim$1\%;  but as noted in the above example,
for W3 it can be as large as 9\%.

The third correction is related to a calibration discrepancy between the WISE photometric standard ``blue" stars
and ``red" galaxies (e.g., AGN and star-forming galaxies).  For the  W4 measurements,  sources
with rising spectrum in the mid-IR, F$_\nu \sim \nu^{-2}$,  appear brighter than sources that are dominated by
R-J emission,  corresponding to the standard stars used to calibrate WISE photometry.   This discrepancy is
likely related to an error in the W4 RSR, as described in Wright et al. (2010) and Jarrett et al. (2011).
Based on work summarized in Jarrett et al. (2011), we have
adopted a factor of 0.92 correction to the W4 flux of all of our galaxies,
except those dominated by the old stellar population
(NGC\,584, NGC\,777, M\,87 and NGC\,6822) in which case no flux correction is needed.
In general, we recommend that this 8\% flux correction be applied to all spiral and disk galaxies:
translating to WISE colors:  [W2-W3] $>$ 1.3 mag.
Conversely, it should not be applied to those that with bulge-dominated populations:  [W2-W3] $<$ 1.3 mag.
The photometric uncertainty in W4 measurements due to this RSR correction
is likely to be at the 3 to 5\% level.

\subsection{Infrared Source Characterization}

Basic measurements including position, size, extent and integrated flux are carried out
for all galaxies in the sample, for both WISE and \Spitzer imaging sets.
As with the IRAC imaging, the WISE MCM-HiRes imaging allows more detailed measurements:
including the surface brightness distribution, total flux, effective or half-light radius and surface brightness,
concentration metrics and bulge-to-disk separation.  We have adapted tools and algorithms
from the 2MASS XSC pipeline (Jarrett et al. 2000)
and the WISE Photometry System (Cutri et al. 2011),
developing an interactive system that is used to
identify  foreground (contaminant) stars and assist in shape/extent characterization, surface brightness and
integrated flux measurements.  It also assists in deblending close  galaxy pairs (e.g., M\,51,  presented below).

The interactive steps are as follows: (1) display images (four bands), including an RGB-color image;
(2) identify foreground stars for removal (using colors and proximity to spiral arms);
(3) demarcate the central location,
size and shape of galaxy (used as a first guess to the 2-D fitting routines); and (4)
demarcate the location for estimation of the local background.  The annulus should
be well outside the influence of the galaxy.  These initial inputs are supplied to the
source characterization processor which carries out the measurements.

For each band, the local background is determined from the pixel value distribution within an elliptical annulus centered on the
galaxy.  Stars are excluded from the distribution through masking.
The histogram mode -- most common binned histogram value -- is a robust
metric for the ``sky", and thus is adopted as the local background value.
The local background for each band is then removed from the mosaic images.
Stars are then removed by subtracting them using the WISE point spread function (PSF)
appropriate for each band. For sources that do not subtract well (due to strong background
gradients), the regions are masked and then recovered using
local background and isophotal substitution extracted from the galaxy ellipsoid model.
With stars removed, the next
step is to determine the size and shape of the galaxy. The best fit axis
ratio and ellipticity are determined using the $3\,\sigma$
isophote.  The previous steps are then repeated using the updated galaxy shape, and the
process is iterated several times until convergence.

The next set of steps it designed to characterize the surface brightness distribution,
beginning with  the azimuthal-averaged elliptical radial surface brightness profile,
ultimately bounded by the
location of the background annulus.   The adopted 1\,$\sigma$ elliptical isophotal radius then corresponds
to the 1\,$\sigma$ (sky RMS) isophote for each band.  WISE W1 (3.4 $\mu$m) is the most sensitive for nearly all
types of galaxies and thus has the largest isophotal radius;  e.g., it is typically much larger, 3 -- 4$\times$, than
in the 2MASS Ks-band.   Depending on the total coverage (and hence, depth), the typical 1\,$\sigma$ isophotal surface brightness is
23.0, 21.8, 18.1 and 15.8 mag arcsec$^{-2}$ (Vega), respectively for W1, W2, W3 and W4 bands.
The isophotal fluxes correspond to the integral of the elliptical shape defined by the axis ratio, position angle
and isophotal radius.  Additionally,  in order to compare across bands to deriving colors, we adopt W1 as the fiducial aperture
for bands W1, W2 and W3,
integrating these other bands using this W1 aperture.  Since the sensitivity of W4 is sufficiently less than that of the
other three bands, the W4 isophotal radius is considerably smaller than that of W1 and is therefore the most appropriate
aperture to use for reporting W4 integrated fluxes.   For all four bands, the axis-ratio and position angle shape parameters
used in the fiducal aperture photometry are taken from the W1 3\,$\sigma$ isophote.  The exception, however, to this
convention applies to early-type galaxies:
since the R-J tail is falling fast in the mid-IR
for the old stellar population as traced by the W3 and W4 bands, the W1 fiducial aperture is not
appropriate to measurements of elliptical galaxies, and hence we use the
the W3 and W4 isophotal fluxes.

The azimuthal elliptical-radial surface brightness is then characterized using a double S\'ersic function, where one S\'ersic
is fit to the inner galaxy region (i.e., the bulge) and the second S\'ersic is fit to the outer region (i.e., the disk).
Separately integrating these contributions to the total light, a bulge-disk fraction is estimated.  This is a very
simplistic approach, e.g., it does not account for spiral arms or bars, but serves as a rough estimate for
the two general populations.  Nearing completion, the fit to the radial surface brightness is used to estimate the
total flux by extrapolating the fit to larger radii, corresponding to three times the disk scale length beyond
the isophotal radius.  This threshold represents a practical balance between capturing the outer disk light
and mitigating fit errors that accumulate with radius. The resultant  extrapolation or ``total" flux can be used to estimate the half-light flux
and its corresponding radius.  Finally, both half-light and concentration indices follow from integrating the radial
surface brightness profile.

\section{Results}

In this section we present the resolved-source photometric performance derived from WISE,
first focusing on the relatively simple case of early-type galaxies, comparing the extracted fluxes with the expectation based on SED models, 
followed by the full sample results.   A more detailed comparison with \Spitzer and IRAS photometry
is presented in Appendix A, notably comparing photometric results across the (approximately equal) paired bands: W1 and IRAC-1; W2 and
IRAC-2; W3, IRAC-4 and IRAS-12; and W4, MIPS-24 and IRAS-25.  Appendix B presents the GALEX photometric measurements.


\subsection{Photometric Performance: case study of Elliptical Galaxies}

Early-type ``red-sequence" galaxies have conveniently homogeneous properties, resulting in smooth, featureless spectral energy distributions that are dominated by the evolved luminous population.   Ellipticals and spheroids have low star formation, minimal dust content, and relatively high surface brightness.
The bolometric luminosity is dominated by the R-J distribution that spans the
near-infrared (1 to 5 $\mu$m) window,  sampled by the
2MASS J, H and Ks bands, the IRAC-1 and IRAC-2 bands, and the WISE W1 and W2 bands.  For nearby galaxies,
the mid-IR light is bright enough to detect the long R-J tail, sampled by the IRAC-3, IRAC-4, MIPS-24, WISE W3
and WISE W4 bands.  Because of their simple R-J profiles, ellipticals are relatively easy to model.
Single burst population synthesis models do an adequate job of describing the mid-infrared properties of spheroidal galaxies,
and therefore can be used to validate the absolute calibration of the WISE and \Spitzer photometry of resolved sources
(see also Petty et al. 2012, submitted, for detailed modeling of early-type galaxies using WISE, SDSS and GALEX measurements).

\begin{figure*}[ht!]
\begin{center}
\includegraphics[width=17cm]{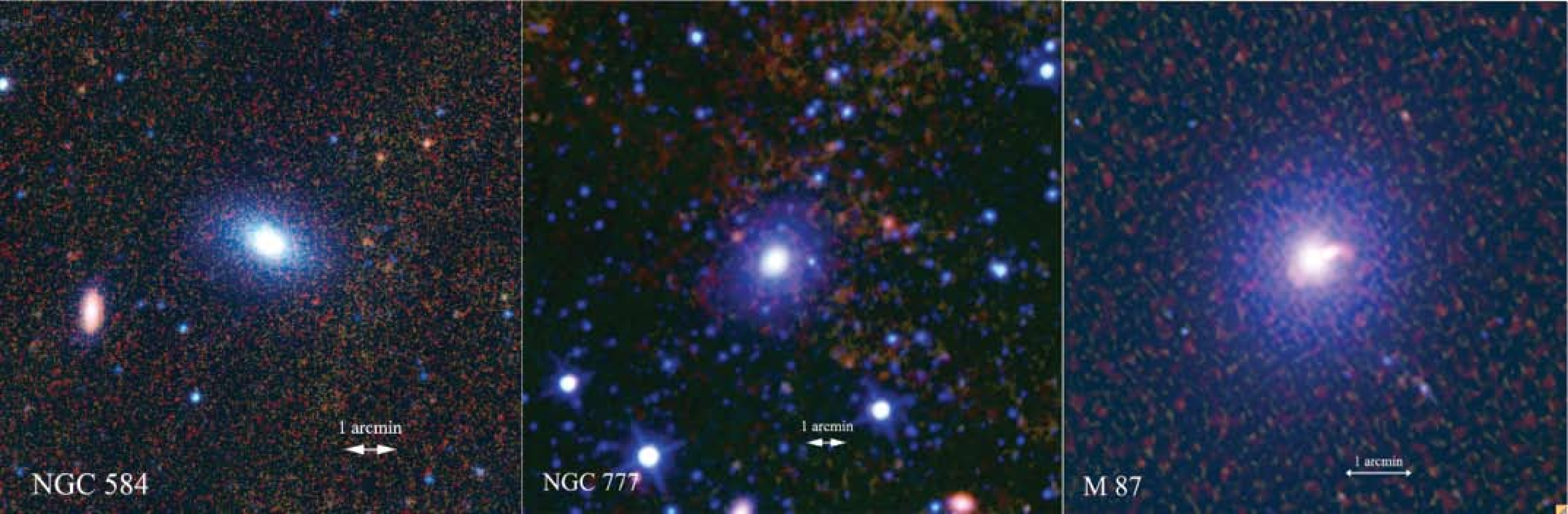}
\caption[ellipticals]
{\small{WISE  view of early-type, elliptical galaxies NGC\,584, NGC\,777 and M\,87.
The
colors correspond to WISE bands:  3.4 $\mu$m (blue),
4.6 $\mu$m (green),
12.0 $\mu$m (orange),
22 $\mu$m (red).
}}
\label{ellipticals}
\end{center}
\end{figure*}

The sample includes three nearby early-type galaxies, shown in Fig. \ref{ellipticals},
NGC\,584, NGC\,777 and M\,87.    The first is a SINGS galaxy with enhanced \Spitzer observations,
including IRS spectra of the nuclear light.  The second appears to have nominal properties for
galaxies of this type.  The third, M\,87, is the giant Virgo Cluster elliptical galaxy.  It harbours a massive blackhole
at its center that is powering an AGN, whose signature non-thermal radio lobes and jets (Virgo A) are also visible in the optical and
infrared.  All three have mid-IR light that is dominated by the old stellar population, brightest
in the W1 band, and thus appearing
as ``blue" light in the WISE color images (Fig. \ref{ellipticals}).

\begin{figure*}[ht!]
\begin{center}
\includegraphics[width=16cm]{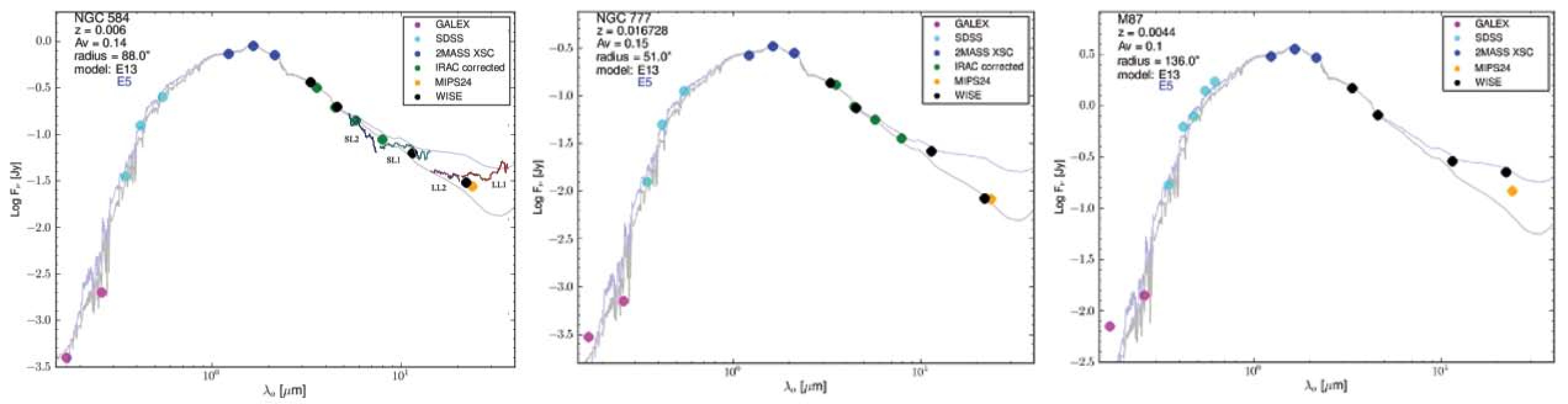}
\caption[ellitpicals]
{\small{
The mid-infrared SED for elliptical-type galaxies NGC\,584, NGC\,777 and M\,87.
The diagram includes 2MASS XSC, IRAC, MIPS and WISE photometry, where
the 2MASS isophotal radius and shape were used as the fiducial aperture.
The NGC\,584 spectra are from SINGS Spitzer-IRS (SL and LL modules) of the nucleus.
There is no IRAC data for M\,87.
The blue and grey lines are old 13 Gyr and 5 Gyr galaxy models, respectively,
adapted from the GRASIL code (Polletta et al. 2006 \& 2007; Silva et al. 1998) normalized to the near-infrared.
}}
\label{sed_ellipticals}
\end{center}
\end{figure*}

For each galaxy we construct the spectral energy distribution (SED) from GALEX (see Appendix B), SDSS,  2MASS (Jarrett et al. 2003),
and mid-infrared measurements from \Spitzer (Appendix A) and WISE; Fig. \ref{sed_ellipticals}.  A fiducial aperture, based on the 2MASS Ks-band
standard isophote, is applied to each band of WISE and \Spitzer.  A small correction is
made for the foreground Galactic extinction, estimated from the Schlegel et al. (1998) dust maps and
tabulated by NED.  Lastly,
aperture corrections have been applied to both the WISE and \Spitzer measurements.
For comparison, we show the
expected light distribution of an old galaxy at t = 5 and 13 Gyrs, adapted from the GRASIL population synthesis
models (Polletta et al. 2006 \& 2007; Silva et al. 1998).  The major difference between the two models
is the presence of AGB stars in the t=5 Gyr model, bumping up the mid-IR emission beyond 10 $\mu$m,
demonstrating that evolutionary age is an important consideration when modeling elliptical galaxies.
The model SEDs are fit to the data using the K-band (2.2 $\mu$m) measurement.
Also for comparison we show the SINGS-IRS spectral measurements of the nuclear light,
which are consistent with the broad-band photometric measurements.

For all three galaxies, the SEDs show similar behavior for the near-IR bands (1 to 5 $\mu$m):  the flux density peaks
in the H-band (1.6 $\mu$m), thereafter steeply falling toward longer wavelengths. The photometric data matches well with
the models, and in the case of NGC\,584, the \Spitzer spectroscopy of the nuclear emission.  In the mid-IR, departures from
the R-J tail arise from the presence of warm dust, either from the interstellar medium (star forming
galaxies) or from AGB stars.   NGC\,584 hints at a warm dust excursion beyond 20$\mu$m, traced
by both the IRS-LL1 spectroscopy and the MIPS-24/W4 photometry.  NGC\,777 shows no such excursions
and thus appears to be dust free, shining by the light of very old stars.
Not surprisingly,
Virgo's M\,87 exhibits an infrared excess due to the stellar light giving way to non-thermal emission from the jet at 22 and 24 $\mu$m bands.
Both the forward (approaching) and backside (receding) of the central jet are clearly seen in the
long wavelength imaging (Fig \ref{ellipticals}), notably bright in the W4 band compared
the MIPS-24 band.
Since this emission arises from a powerful, non-thermal source, the models
are not relevant to the emission beyond 10 $\mu$m, and other than variability between
the WISE and MIP-24 observations of M\,87, it is unknown why W4 is brighter
than MIPS-24.

To summarize, all three early-type galaxies in the sample show near-IR photometric properties that are consistent
among the data sets: 2MASS vs WISE vs \Spitzer, and among the population synthesis models.
Likewise, at longer wavelengths, the galaxies NGC\,584 and NGC\,777
display photometric results that are consistent between data sets and models.  Only M\,87 exhibits peculiar properties
at the longest wavelengths due to the infrared-bright jet that originates from the nucleus and
dominates the light beyond 20 $\mu$m.   We conclude that, for these three galaxies, the
WISE photometric measurements are consistent with those of \Spitzer, and that the absolute measurements
are consistent with the population synthesis models normalized to the 2MASS photometry.

\subsection{WISE Full Sample Measurements}

In this sub-section we present the photometry and characterization measurements for the entire sample of 17 galaxies,
extracted from the WISE HiRes imaging.  Appendix A presents a comparison between the matched \Spitzer IRAC and MIPS-24 imaging,
as well as between archival IRAS 12 and 25 $\mu$m measurements.  Appendix B presents the GALEX measurements.

Similar to the \Spitzer and MIPS bands, all four WISE bands provide valuable spectral information, tracing both
stellar light and that arising from the interstellar medium associated with star formation.
The WISE W1 3.4 $\mu$m and W2 4.6$\mu$m bands
are nearly ideal tracers of the stellar mass distribution in galaxies because they image the Rayleigh-Jeans
limit of the blackbody emission for stars 2000 K and hotter.  These bands are relatively extinction free, and have a W1-W2 color that
is constant and independent of the age of the stellar population and its mass function (Pahre 2004; Jarrett et al. 2011).
Thus a combination of the W1 and W2 luminosities with the corresponding mass-to-light ratio, which varies much less in the
mid-IR compared to the optical, leads to the aggregate stellar mass or `backbone' mass assembly of the galaxy
(detailed results are presented in Section 6).

At longer wavelengths, the stellar light gives way to the cooler emission of the interstellar medium.
The 22 $\mu$m  band is sensitive to warm dust emission, arising in the vicinity of hot \HII\ regions.
Calzetti et al. (2007) using similar 24 $\mu$m  data from the {\it Spitzer} Space Telescope, showed that mid-IR fluxes when combined with H$\alpha$ fluxes provide a powerful reddening free indicator of O/B star formation.  Similarly, the 12 $\mu$m data are sensitive to polycyclic aromatic hydrocarbon (PAH) emission arising from the photon-dominated regions (PDRs) located at the boundaries of \HII\ regions  and molecular clouds and thus excited by far-UV photons.  UV photons that manage to
escape the dust trap will be traced by GALEX imaging, thus completing the census of the current rate of star formation.
Star formation rates derived from the GALEX, 12 and 22 $\mu$m bands are presented in Section 6.

\subsubsection{WISE Source Characterization}

Measurements extracted from HiRes imaging is presented in three separate tables:
the fiducial isophotal photometry in Table \ref{tab:flux},
the extrapolated fluxes in Table \ref{tab:total}, and the
half-light and concentration indices in Table \ref{tab:eff}.
The quoted flux uncertainties include contributions from the Poisson errors
and background estimation errors (but do not include the calibration errors).
All reported flux densities and their formal uncertainties are in mJy units.

We adopt a fiducial aperture to report integrated fluxes
since cross-band comparisons (i.e., colors) may be directly computed
from the reported flux measurements.   As discussed in Section 3.5, for most galaxies, WISE W1 (3.4 $\mu$m) is the
most sensitive band to the faint lower surface brightness emission
in the outer disks; consequently, we adopt the W1 1\,$\sigma$ isophotal
aperture  as the fiducial aperture for aperture measurements
in the W1, W2 and W3 WISE bands, and use the W4 isophotal aperture for
the W4 integrated fluxes.  Note that the axis ratio and orientation are based on the
higher S/N isophote at 3\,$\sigma$ in W1;  see Section 3.5. The typical 1\,$\sigma$ surface brightness in
W1 is $\sim$23 mag/arcsec$^2$ (Vega).

We employ a different strategy for elliptical galaxies (and also the nearby irregular galaxy NGC\, 6822).  
The light observed in the long wavelength bands
is too faint to obtain a reliable flux in the (relatively) large W1 aperture (see Section 4.2);
the integrated flux tends to bias high when using a large aperture due to contamination
from faint, foreground stars that are not subtracted from the images.  Consequently,
for NGC\,584, NGC\,777, M\,87 and NGC\,6822, the reported W1 and W2 fluxes are derived from
the W1 isophote, and the W3 flux from the W3 isophote, and the W4 flux from the W4 isophote.
The typical 1\,$\sigma$ isophotal surface brightness for the W2, W3 and W4 bands, respectively, is
21.8, 18.1 and 15.8 mag/arcsec$^2$ (Vega).
Details of the isophotal apertures are specified in Table \ref{tab:flux} and its notes.

\begin{figure*}[ht!]
\begin{center}
\includegraphics[width=17cm]{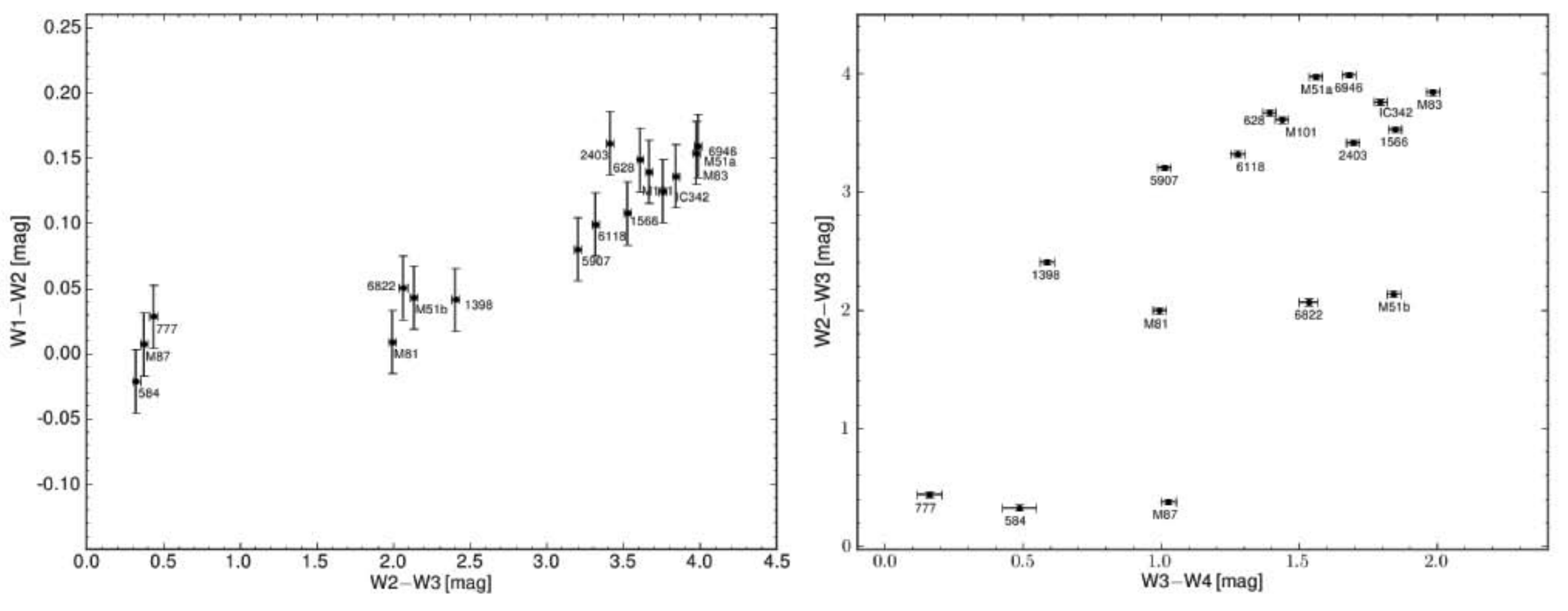}
\caption[colors]
{\small{
WISE colors for the sample, derived from matched aperture photometry.
The units are Vega magnitudes.  The error bars represent the formal uncertainties
(Table \ref{tab:flux}) and a 1.5\% photometric calibration uncertainty.
Measurements have been corrected for Galactic and internal extinction.
The galaxy name is indicated next to the measurement, where single numbers represent the NGC \#.
}}
\label{colors}
\end{center}
\end{figure*}

Derived from the integrated fluxes (Table \ref{tab:flux}), the WISE colors and their
estimated formal plus calibration uncertainties are presented in Fig \ref{colors}.
Here the magnitudes
have been corrected for the estimated foreground Galactic extinction, where we have adopted the following
coefficients based on merging the Cardelli, Clayton, \& Mathis (1990), Flaherty et al. (2007)
and Indebetouw et al. (2005) relations for the near and mid-IR windows:
A$_K$ = 0.114 A$_V$;  A$_{3.4\mu m}$ = 0.056 A$_V$;
A$_{4.6\mu m}$ = 0.049 A$_V$;
A$_{12\mu m}$ = 0.049 A$_V$;
A$_{22\mu m}$ = 0.
For the inclined galaxies (M\,81, NGC\,1566, NGC\,2403, NGC\,6118, NGC\,5907),
the expected internal extinction is estimated using the prescription from Masters,
Giovanelli \& Haynes (2003):  A$_K \simeq$ 0.26 Log(a/b), where a/b is the inverse
of the axis ratio.

The color W1$-$W2 is a metric for the steepness of the falling R-J tail and is also sensitive to
nuclear activity (Stern et al. 2012),  for the sample is relatively narrow
in range, $\sim$0.2 mag, between the elliptical galaxies and the star-forming galaxies --
for the most part sampling the same evolved stellar population.  In contrast, the color W2$-$W3 ranges
across $\sim$4 mag, from the relatively blue, R-J dominated, elliptical galaxies,
to the red, molecular PAH-dominated star-forming galaxies.  A linear normal-galaxy `sequence'  is formed
from early to late-type galaxies.
Similarly, the W3$-$W4 color
traces the ISM emission from star-forming galaxies, W3 dominated by the PAH emission
and W4 by the warm dust emission arising from the UV radiation field.  There are three galaxies
that stand-out in the W3$-$W4 color diagram:  M\,51b, M\,87 and NGC\,6822.  The mid-IR light of M\,87 is dominated
by the evolved stellar population, but in the 22 $\mu$m band the central jet begins to dominate
the emission, pushing the color to (relatively) redder levels.  The early-type galaxy M\,51b has blended
ISM dust/PAH emission from it's larger, late-type companion, M\,51a, as well as possessing a nuclear
molecular structure (cf. Kohno et al.,  2004), thereby exhibiting hybrid infrared colors that reflect stellar and ISM emission.
NGC\,6822 is the only dwarf galaxy in the sample,
possessing an old stellar population that dominates the short wavelength bands ($\sim$similar in color to early-type spirals)
and isolated star formation regions that give it a W3$-$W4 color that is comparable to late-type spirals.

\begin{table*}
{\scriptsize
\caption{WISE Isophotal-Aperture Photometry  \label{tab:flux}}
\begin{center}
\begin{tabular}{r r r r r r r r r r r}

\hline
\hline
\\[0.25pt]

Name  & R.A. & Dec. & axis$^1$& p.a.$^1$ & R1$^1_{iso}$ & R4$_{iso}$ & W1 & W2 & W3 & W4\\
  & (deg) & (deg) & ratio &  (deg) & (arcsec) & (arcsec)   &  (Jy) & (Jy) & (Jy) & (Jy)   \\

\hline
\\[0.25pt]

NGC\,584$^2$&22.83626&-6.86801&0.67&66.1&149.2&30.8&0.381$\pm$0.004&0.207$\pm$0.002&0.051$\pm$0.001&0.021$\pm$0.001\\
NGC\,628&24.17406&15.78373&1.00&0.0&354.6&198.8&0.885$\pm$0.009&0.559$\pm$0.006&3.022$\pm$0.033&2.864$\pm$0.032\\
NGC\,777$^2$&30.06247&31.42939&0.83&144.7&176.9&30.8&0.179$\pm$0.002&0.102$\pm$0.001&0.028$\pm$0.000&0.009$\pm$0.000\\
NGC\,1398&54.71692&-26.33771&0.71&93.3&326.2&140.0&0.897$\pm$0.009&0.518$\pm$0.005&0.874$\pm$0.010&0.396$\pm$0.005\\
NGC\,1566&65.00161&-54.93800&0.68&32.2&374.6&374.6&0.733$\pm$0.007&0.451$\pm$0.005&1.927$\pm$0.021&2.942$\pm$0.033\\
NGC\,2403&114.21203&65.59946&0.56&-57.8&627.2&366.5&1.701$\pm$0.017&1.099$\pm$0.011&4.699$\pm$0.052&5.966$\pm$0.066\\
NGC\,3031&148.88857&69.06503&0.56&156.4&954.6&495.0&11.317$\pm$0.115&6.366$\pm$0.064&7.359$\pm$0.075&4.966$\pm$0.055\\
NGC\,4486$^2$&187.70595&12.39088&0.73&158.1&538.5&65.3&2.206$\pm$0.022&1.233$\pm$0.012&0.321$\pm$0.004&0.219$\pm$0.003\\
NGC\,5194$^3$&202.46976&47.19496&0.67&21.3&402.7&290.3&2.444$\pm$0.025&1.564$\pm$0.016&11.196$\pm$0.113&12.324$\pm$0.125\\
NGC\,5195$^3$&202.49825&47.26591&0.95&130.4&285.3&72.0&1.153$\pm$0.012&0.666$\pm$0.007&0.878$\pm$0.009&1.255$\pm$0.013\\
NGC\,5236&204.25330&-29.86576&1.00&0.0&572.8&338.9&6.272$\pm$0.064&3.953$\pm$0.040&22.605$\pm$0.249&41.175$\pm$0.417\\
NGC\,5457&210.80225&54.34859&1.00&0.0&605.2&328.7&2.573$\pm$0.026&1.638$\pm$0.017&8.384$\pm$0.085&8.245$\pm$0.084\\
NGC\,5907&228.97302&56.32840&0.17&154.9&413.0&235.6&0.794$\pm$0.008&0.499$\pm$0.005&1.760$\pm$0.018&1.663$\pm$0.017\\
NGC\,6118&245.45239&-2.28323&0.42&52.8&177.6&121.5&0.184$\pm$0.002&0.113$\pm$0.001&0.441$\pm$0.005&0.384$\pm$0.005\\
NGC\,6822&296.23502&-14.80102&0.87&21.8&562.0&310.0&1.920$\pm$0.019&1.122$\pm$0.012&1.248$\pm$0.021&1.472$\pm$0.030\\
NGC\,6946$^4$&308.71735&60.15389&1.00&0.0&470.0&310.0&3.177$\pm$0.032&2.056$\pm$0.021&14.914$\pm$0.151&19.187$\pm$0.194\\
IC\,342&56.70229&68.09621&1.00&0.0&675.0&527.0&7.671$\pm$0.078&4.830$\pm$0.049&28.363$\pm$0.287&41.821$\pm$0.424\\

\hline
\end{tabular}
\end{center}
\tablecomments{$^1$The fiducial aperture for bands W1, W2 and W3 is based on the W1 1\,$\sigma$ isophotal radius.
The typical 1\,$\sigma$ surface brightness in
W1 is $\sim$23 mag/arcsec$^2$ (Vega) or 25.7 mag /arcsec$^2$
(AB). For W4, the isophotal radius of W4 is used to define the aperture;
The typical 1\,$\sigma$ surface brightness in
W4 is $\sim$15.8 mag/arcsec$^2$ (Vega) or 22.4 mag /arcsec$^2$
(AB).
$^2$ For the early-type elliptical galaxies, the W3 aperture radii are based on
the W3 1\,$\sigma$ isophotal radius due to the lack of strong emission at 12 and 22 $\mu$m; they
are the following:  NGC\,584:  52.9$\arcs$ ;  NGC\,777: 64.0$\arcs$ ; M\,87:  154.7$\arcs$.
$^3$Photometry of NGC\,5194/5 is uncertain due to blending.
$^4$The MIPS-24 mosaic image is too small to capture the total flux, so a slightly smaller aperture for W4 and MIPS-24 is used measure the isophotal flux.
Measurements are not corrected for Galactic or internal extinction.
}
}

\end{table*}

\begin{figure*}[ht!]
\begin{center}
\includegraphics[width=15cm]{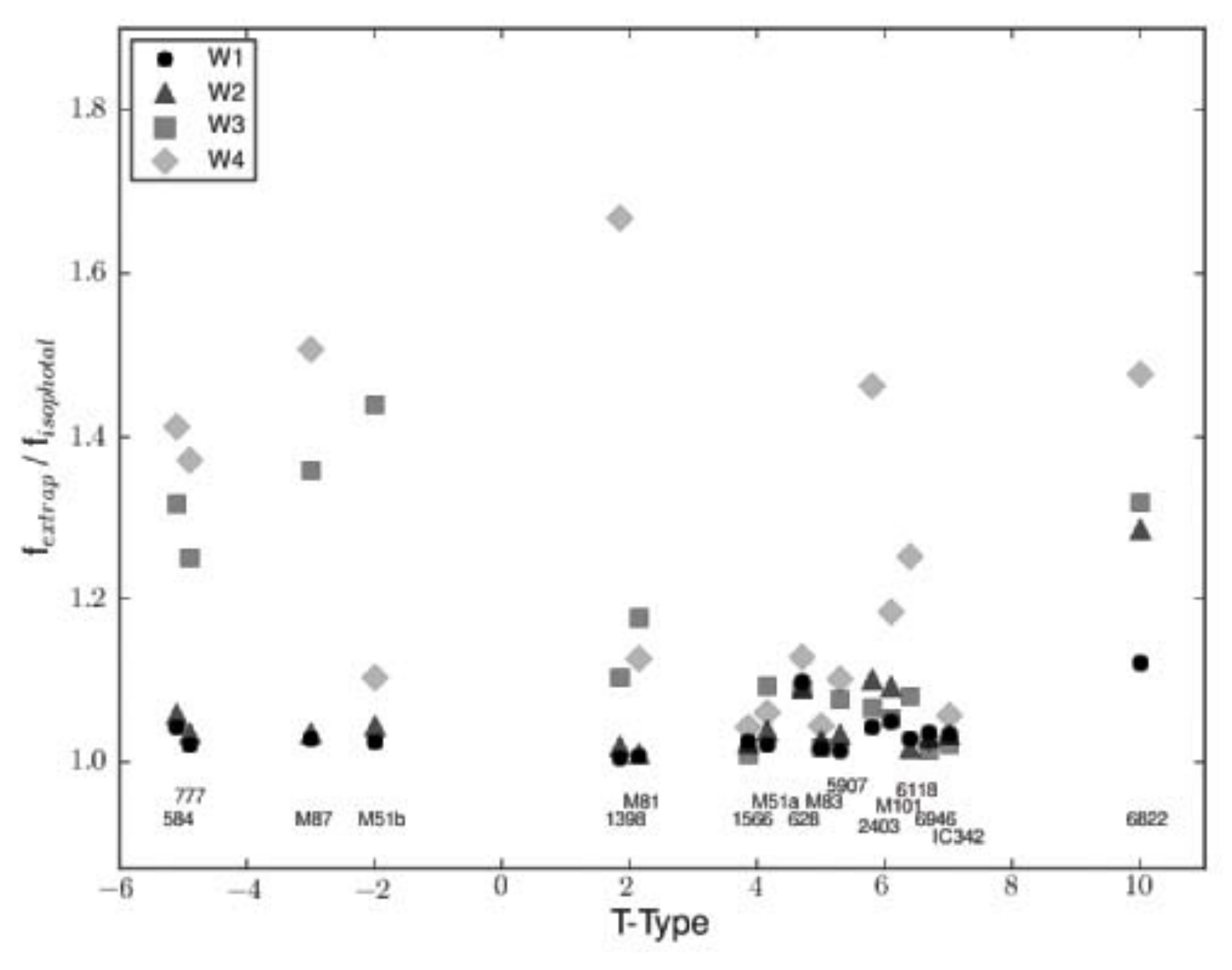}
\caption[extrapolation]
{\small{
The ratio of the extrapolation (``total") flux to the isophotal (1\,$\sigma$) flux.
The galaxy name is indicated below the measurement, where single numbers represent the NGC \#.
The Hubble T-Type is derived from the morphology (Table 1).
}}
\label{extrap}
\end{center}
\end{figure*}

To estimate the total flux, the double-S\'ersic fit to the radial surface brightness (see Appendix C) is used to extrapolate  the integrated flux
to lower levels, corresponding to a radius that is
3 disk scale lengths extended beyond
the 1\,$\sigma$ isophotal radius.  The resulting photometry is presented in
Table \ref{tab:total}. In W1, the flux densities range from
0.2 Jy (NGC\,777) to 12 Jy (M\,81).  The brightest flux,
$\sim$45 Jy, is recorded for
IC\,342 in the W4 band, and the largest sizes, $\sim$1 degree, are
derived for NGC\,2403 and M\,81.   The extrapolation adds only a small correction
to the isophotal fluxes in the W1 and W2 bands, typically only a few \%;
see Fig.  \ref{extrap}.  This result suggests that the W1 and W2 isophotal
radii basically capture the total flux of the systems, attributed to the sensitivity
and depth of these bands in WISE.  In contrast, the W3 and W4 bands require
large corrections: W3 ranges from a $\sim$few \% to 30\%, and W4 up to 70\% corrections,
notably for the early-type galaxies, NGC\,2403, the ringed-spiral NGC\,1398
and the low surface brightness `flat' NGC\,6822.

\begin{table*}
{\scriptsize
\caption{WISE Extrapolated Photometry  \label{tab:total}}
\begin{center}
\begin{tabular}{r r r r r r r r r r r}

\hline
\hline
\\[0.25pt]

Name  &  R1$_{ext}$ & W1$_{ext}$ & R2$_{ext}$ & W2$_{ext}$ & R3$_{ext}$ & W3$_{ext}$ & R4$_{ext}$ & W4$_{ext}$ \\
  &(arcsec) &  (Jy) & (arcsec) &  (Jy) & (arcsec) &  (Jy) & (arcsec) &  (Jy)   \\

\tableline
\\[0.25pt]

NGC\,584&286.3&0.398$\pm$0.004&225.0&0.220$\pm$0.002&144.2&0.068$\pm$0.002&75.2&0.030$\pm$0.001\\
NGC\,628&565.5&0.973$\pm$0.012&528.0&0.610$\pm$0.007&615.8&3.293$\pm$0.044&531.1&3.233$\pm$0.046\\
NGC\,777&244.1&0.183$\pm$0.002&231.7&0.106$\pm$0.001&147.6&0.035$\pm$0.001&68.2&0.012$\pm$0.000\\
NGC\,1398&554.0&0.902$\pm$0.010&462.0&0.528$\pm$0.006&562.6&0.966$\pm$0.013&280.0&0.661$\pm$0.023\\
NGC\,1566&567.2&0.752$\pm$0.008&548.5&0.461$\pm$0.005&506.7&1.944$\pm$0.021&589.8&3.069$\pm$0.038\\
NGC\,2403&1555.6&1.776$\pm$0.019&1070.6&1.212$\pm$0.015&979.2&5.014$\pm$0.064&1099.5&8.721$\pm$0.239\\
NGC\,3031&1213.9&11.408$\pm$0.121&1205.7&6.440$\pm$0.069&916.9&8.659$\pm$0.138&1331.0&5.599$\pm$0.080\\
NGC\,4486&707.0&2.271$\pm$0.024&722.2&1.277$\pm$0.014&271.2&0.436$\pm$0.011&195.8&0.330$\pm$0.009\\
NGC\,5194&549.9&2.498$\pm$0.027&558.2&1.625$\pm$0.018&1070.7&12.245$\pm$0.156&497.8&13.086$\pm$0.151\\
NGC\,5195&393.0&1.182$\pm$0.013&410.8&0.696$\pm$0.008&359.9&1.263$\pm$0.034&216.0&1.386$\pm$0.018\\
NGC\,5236&768.3&6.376$\pm$0.066&749.9&4.064$\pm$0.042&736.7&23.006$\pm$0.249&507.4&43.062$\pm$0.494\\
NGC\,5457&870.2&2.703$\pm$0.029&900.3&1.792$\pm$0.022&891.3&8.839$\pm$0.103&684.0&9.768$\pm$0.156\\
NGC\,5907&561.0&0.805$\pm$0.009&664.6&0.516$\pm$0.006&852.8&1.897$\pm$0.023&706.9&1.834$\pm$0.023\\
NGC\,6118&326.7&0.190$\pm$0.002&355.3&0.115$\pm$0.001&333.2&0.477$\pm$0.006&364.6&0.481$\pm$0.010\\
NGC\,6822&1568.8&2.154$\pm$0.027&1686.0&1.444$\pm$0.028&1686.0&1.645$\pm$0.038&805.5&2.174$\pm$0.062\\
NGC\,6946&657.3&3.295$\pm$0.035&660.6&2.119$\pm$0.023&636.1&15.121$\pm$0.171&583.6&19.760$\pm$0.219\\
IC\,342&1431.0&7.932$\pm$0.085&1386.2&4.993$\pm$0.054&1432.6&28.977$\pm$0.328&1299.6&44.193$\pm$0.506\\

\tableline
\end{tabular}
\end{center}
\tablecomments{The extrapolation photometry is the sum of the isophotal aperture photometry (Table \ref{tab:flux}) and the integral
of the double-S\'ersic fit to the elliptical radial surface brightness carried out to a maximum radius of
R$_{ext}$.
Measurements are not corrected for Galactic or internal extinction.}


}
\end{table*}

\begin{figure*}[ht!]
\begin{center}
\includegraphics[width=17cm]{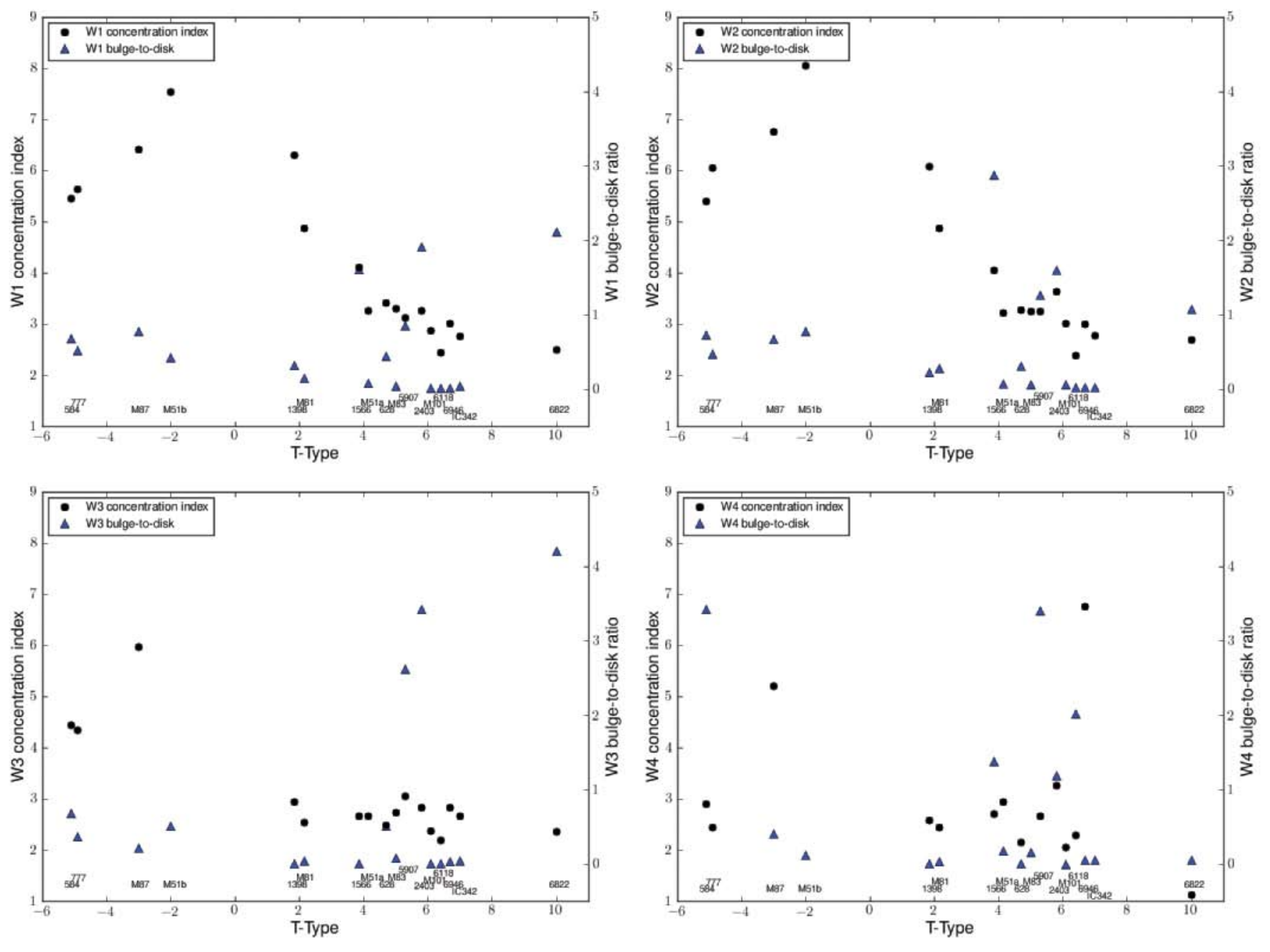}
\caption[central concentration]
{\small{
The central concentration index for the galaxy sample, corresponding to
the ratio of the 3/4 light-radius to the 1/4 light-radius.
For comparison, also shown is the bulge-to-disk
ratio based on the double-S\'ersic fit to the azimuthally-averaged elliptical-radial profile
(see Appendix C).
The galaxy name is indicated below the measurement, where single numbers represent the NGC \#.
The Hubble T-Type is derived from the morphology (Table 1).
}}
\label{concentration}
\end{center}
\end{figure*}

Derived using the extrapolation fluxes as a proxy for the total flux,
the half-light, concentration and disk fraction indices are presented
in Table \ref{tab:eff}.  As expected, the highest W1 and W2 surface brightnesses
(likewise, concentration indices) are observed for the early-type galaxies, dominated by
the bulge light that is tracing the large stellar masses (see Fig. \ref{concentration}).
The contrast seen between the central light and
the total light is much less so in the long wavelength bands, W3 and W4, the mid-IR emission
largely originates from the disk ISM.
W3 exhibits somewhat higher concentration and B/D ratios for the early-type galaxies
(due to the R-J component),
while
for W4, with a few exceptions (e.g., M\,87 due to the central AGN), the concentration index is basically the same for all Hubble types,
presumably because W4 is insensitive to the R-J emission from stars.
As viewed in these ISM-sensitive bands, the mid-IR light is more evenly distributed throughout the systems.

\begin{table*}\setlength{\tabcolsep}{2pt}
{\scriptsize
\caption{WISE Half Light Surface Brightness, Concentration \& Disk Fraction \label{tab:eff}}

\begin{center}
\begin{tabular}{l r r r r r r r r r r r r r r r r }

\tableline
\tableline
\\[0.25pt]

Name  &  R1$_e$ & W1$_e$ & C1 & f1$_{disk}$ & R2$_e$ & W2$_e$ & C2 & f2$_{disk}$ &
R3$_e$ & W3$_e$ & C3 & f3$_{disk}$ & R4$_e$ & W4$_e$ & C4 & f4$_{disk}$  \\
 & (arcsec) &  (mag/as$^{2}$) &  &  & (arcsec) &  (mag/as$^{2}$) &  &   & (arcsec) &  (mag/as$^{2}$) &  &  & (arcsec) &  (mag/as$^{2}$) &  &    \\

\tableline
\\[0.25pt]

NGC\,584&25.2&15.796&5.46&0.595&27.0&15.947&5.40&0.579&24.0&15.132&4.44&0.594&20.3&14.200&2.90&0.226\\
NGC\,628&114.7&18.550&3.41&0.692&117.5&18.470&3.28&0.761&124.7&14.934&2.48&0.661&114.1&13.300&2.15&0.993\\
NGC\,777&24.3&16.791&5.64&0.658&27.9&17.043&6.06&0.679&26.8&16.317&4.35&0.731&21.3&15.555&2.45&0.053\\
NGC\,1398&55.2&16.679&6.30&0.758&63.4&16.922&6.09&0.815&140.1&16.153&2.95&0.990&124.8&14.854&2.59&0.998\\
NGC\,1566&66.5&17.227&4.11&0.383&65.0&17.068&4.05&0.258&76.0&13.919&2.66&0.997&78.4&12.123&2.71&0.420\\
NGC\,2403&176.9&18.198&3.27&0.343&194.1&18.175&3.64&0.384&181.3&14.650&2.83&0.226&217.7&12.986&3.26&0.457\\
NGC\,3031&168.7&16.082&4.88&0.869&179.5&16.197&4.88&0.778&324.3&15.325&2.54&0.962&305.1&14.207&2.45&0.969\\
NGC\,4486&87.5&16.694&6.41&0.563&94.1&16.838&6.77&0.596&79.4&15.799&5.97&0.824&35.8&12.928&5.21&0.709\\
NGC\,5194&117.6&17.145&3.27&0.928&123.2&17.073&3.22&0.930&139.0&13.308&2.66&0.074&128.5&11.607&2.94&0.851\\
NGC\,5195&43.6&16.184&7.54&0.702&44.6&16.166&8.05&0.561&39.1&13.397&0.00&0.660&0.0&0.090&0.00&0.889\\
NGC\,5236&146.5&17.040&3.30&0.960&145.4&16.873&3.25&0.940&130.4&12.826&2.73&0.928&98.1&10.162&12.28&0.861\\
NGC\,5457&201.7&18.667&2.87&0.984&218.2&18.645&3.02&0.937&207.2&14.963&2.37&0.998&220.3&13.529&2.05&0.999\\
NGC\,5907&79.8&16.045&3.13&0.540&82.6&15.962&3.25&0.441&93.4&12.982&3.06&0.276&81.5&11.261&2.67&0.227\\
NGC\,6118&65.1&18.152&2.44&0.981&65.5&18.075&2.39&0.980&72.6&14.916&2.19&0.990&75.8&13.541&2.29&0.331\\
NGC\,6822&298.8&19.615&2.50&0.321&335.8&19.663&2.69&0.481&316.4&17.464&2.36&0.192&302.6&15.713&1.12&0.951\\
NGC\,6946&142.1&17.692&3.01&0.981&139.4&17.490&3.00&0.980&131.7&13.396&2.84&0.966&110.4&11.262&6.77&0.949\\
IC\,342&279.5&18.206&2.76&0.959&277.4&18.053&2.78&0.974&270.0&14.249&2.66&0.954&179.6&11.446&35.99&0.946\\

\tableline
\end{tabular}
\end{center}
\tablecomments{The half light is relative to the extrapolated integrated flux; see Table \ref{tab:total}.  The concentration
index is ratio of the 3/4 light-radius to the 1/4 light-radius.
The disk fraction, f$_{disk}$, is the fraction of the total integrated light
that is attributed to the disk component characterized by the S\'ersic function fit to the extended light.
Measurements are not corrected for Galactic or internal extinction.}


}
\end{table*}

\subsubsection{Comparing WISE with \Spitzer and IRAS}

Building on the analysis that was carried for NGC\,1566 in Paper I, Appendix A presents a detailed comparison between WISE and ancillary matched photometry
for the entire sample presented in this work.  It is
augmented with SED analysis using population synthesis models and the bandpass information,
with results that validate the WISE photometric calibration and WERGA imaging at the 5$-$10\% relative to \Spitzer and IRAS.

\section{Case study of M\,83}

In this section we showcase the science potential of the WISE and \Spitzer imaging for studying the galaxy NGC\,5236 (M\,83),
comparing the stellar, gas and dust distribution that the infrared observations provide,
to complementary tracers of star formation and gas distribution,
including UV, H$\alpha$, molecular CO and neutral H\,{\sc i} mm/radio observations.

\subsection{M\,83: The Infrared and UV Properties}

M\,83 is one of the most spectacular nearby galaxies, possessing a face-on, grand-design spiral arm system
that is anchored by a large-scale bar.   Its beautiful symmetry and prominent spiral arms has earned the moniker, the {\it Southern Pinwheel Galaxy}.
It is also notable for its extended H\,{\sc i} disk, spanning more than a degree along its
major axis (Tilanus \& Allen 1993; Koribalski et al. 2004; Koribalski 2008), and its central nuclear region that is almost
exclusively molecular gas (Crosthwaite et al. 2002; Lundgren et al. 2008).  In global terms, the molecular gas
contributes about 25\% to the total gas content.

\begin{figure*}[ht!]
\begin{center}
\includegraphics[width=17.5cm]{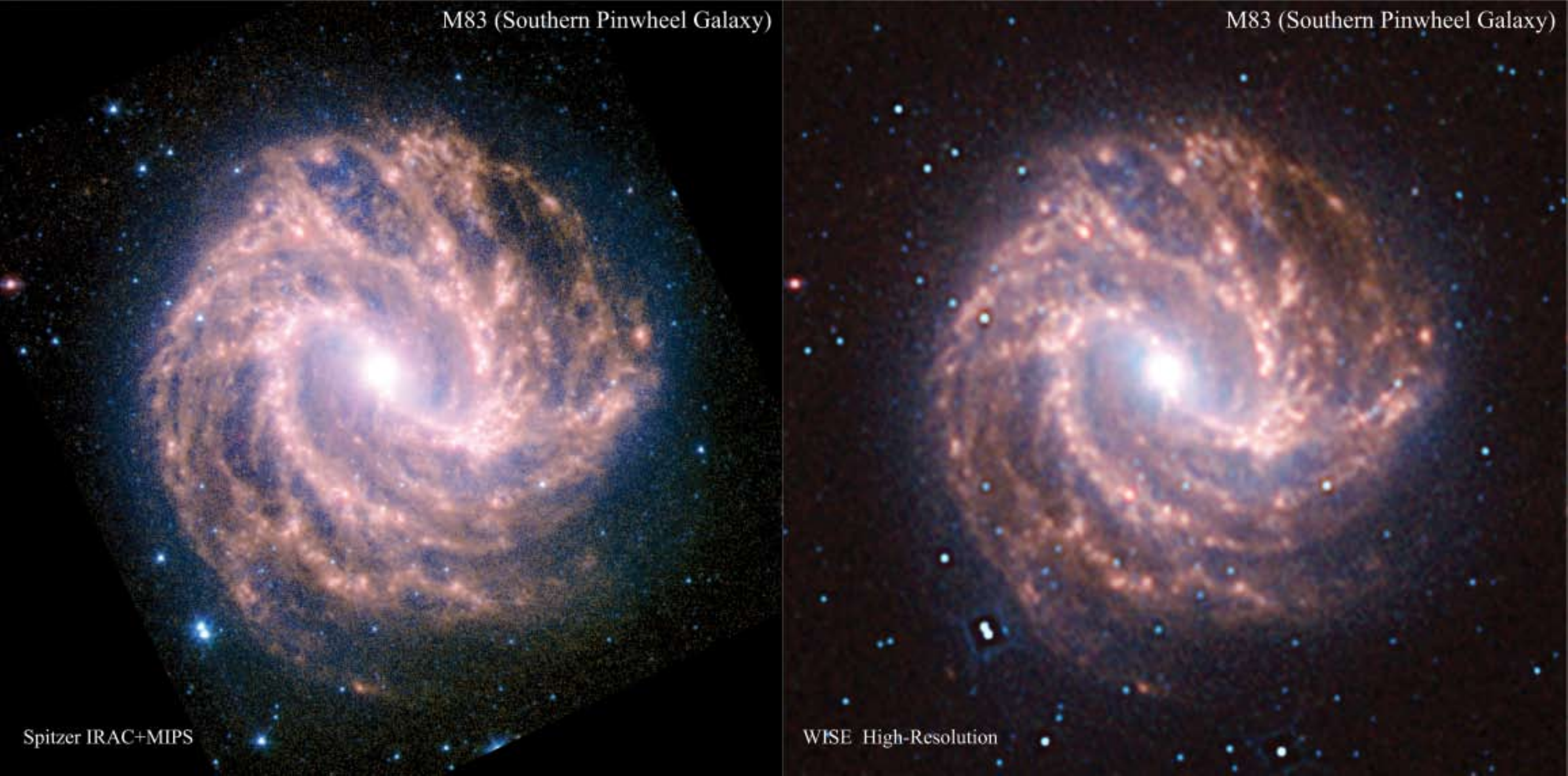}
\caption[M\,83]
{\small{{\it Spitzer} and WISE view of M\,83 (NGC\,5236).
The left panel shows IRAC+MIPS mosaic, where
the colors correspond to:  3.6 $\mu$m (blue),
4.5 $\mu$m (green),
5.8 $\mu$m (yellow),
8.0 $\mu$m (orange),
24 $\mu$m (red).  The right panel
shows the WISE HiRes (MCM) mosaic, where the
colors correspond to WISE bands:  3.4 $\mu$m (blue),
4.6 $\mu$m (green),
12.0 $\mu$m (orange),
22 $\mu$m (red).
The field of view for both images is
13.5 arcmin.
}}
\label{m83all}
\end{center}
\end{figure*}

The mid-IR disk spans about 18$\arcmin$ down to the 1\,$\sigma$ background level in W1.
Fig \ref{m83all} compares the WISE HiRes
imaging and the IRAC + MIPS-24 imaging for the central 14$\arcmin$ disk.
As with the NGC\,1566 comparison between IRAC and WISE HiRes (Paper I),
the M\,83 comparison
impressively shows how well the WISE HiRes reconstructions improve the spatial resolution,
nearly matching the $\sim$2$\arcsec$-resolved imaging of IRAC.  For both
color representations, the faint blue light is tracing the evolved stellar population,
smoothly distributed across arms and inner-arms, while the orange/red light is
is arising from star formation, localized to the spiral arms, bar and nuclear regions,
as traced by the molecular (PAH) and warm dust emission.  The WISE and \Spitzer angular resolution and
optimal sensitivity to the gas/ISM reveals sub-structure between the spiral arms, likely related to the
shear-formed ``spurs" that arise from differential compression of  gas as it flows through the arms
(Shetty \& Ostriker 2006).  Other grand design spirals in the sample also
show transverse spurs, including M\,101, M\,51a, NGC\,6946, and most notably, IC\,342, who's disk/spiral pattern may be described
as a lace-work of nodes (massive star formation sites) and filaments (spiral and transverse arms;   see Fig. 1).
Finally, the bar itself appears bounded, to the northeast and to the southwest, by the highest concentration of star formation in the disk
-- these are the bar transition zones, or bar ``cusps".    These zones arise from the convergence
of gas stream lines due to the interaction of the gravitational density wave with the bar (see Kenney \& Lord 1991).
Orbit crowding compresses the gas and star formation ensues in these relatively small regions of the disk.
Later we investigate the southwest cusp with our multi-wavelength data set that probes down to $\sim$100 pc scales.

\begin{figure*}[ht!]
\begin{center}
\includegraphics[width=17.5cm]{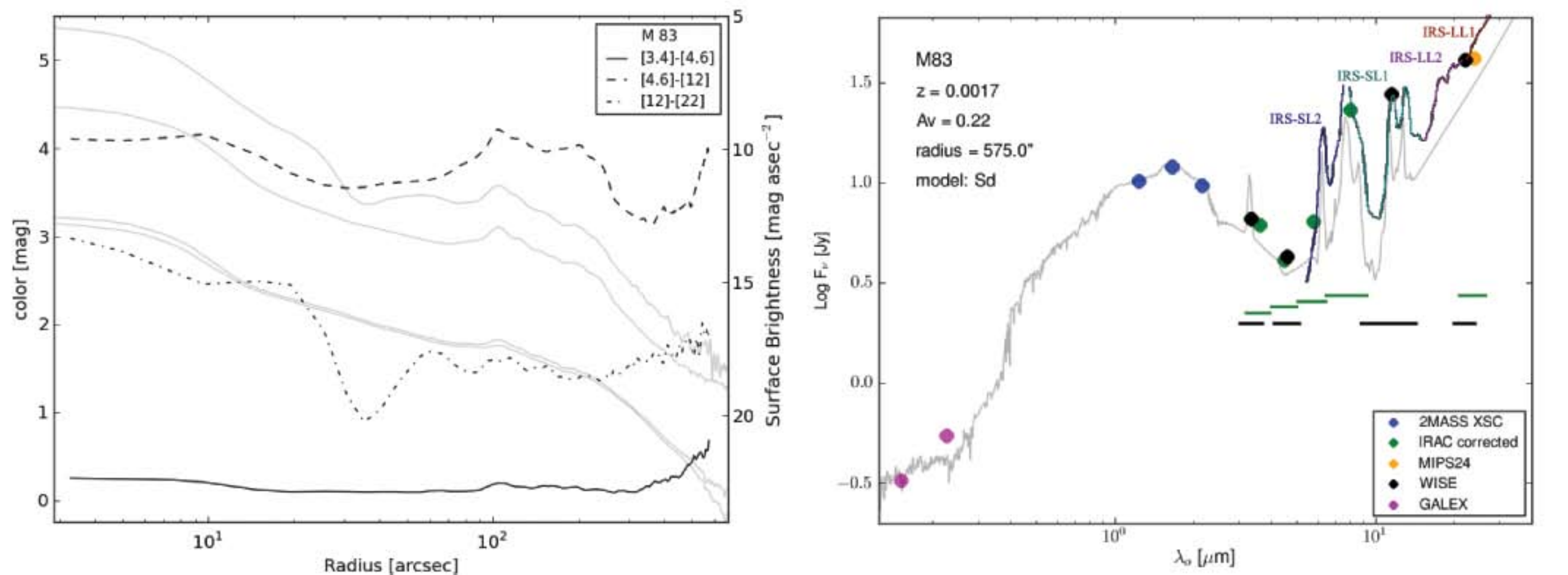}
\caption[M83color]
{\small{Radial color distribution and global SED for M\,83 (NGC\,5236).
(left) The faint grey lines correspond to the azimuthal elliptical-radial surface brightness.
The solid, dashed and dash-dot lines correspond to the difference in surface brightness between
W1 vs W2, W2 vs W3 and W3 vs W4, respectively.
(right) The UV-NIR-MIR SED for M\,83, including GALEX, 2MASS XSC, IRAC, MIPS and WISE photometry,
corrected for the foreground Galactic extinction (A$_v$ = 0.22 mag).
The spectra are from SINGS Spitzer-IRS (SL and LL modules) of the nucleus. The grey line is an Sd galaxy model,
adapted from the GRASIL code (Polletta et al. 2006 \& 2007; Silva et al. 1998) normalized to the near-infrared.
The IRAC, MIPS-24 and WISE bandpass widths are indicated with solid green and black lines, respectively.
}}
\label{m83color}
\end{center}
\end{figure*}

The infrared observations of WISE and \Spitzer provide a direct probe of the physical conditions
of the interstellar medium  responding to the present day star formation.
Confined primarily to the spiral arms, the stars form in giant molecular clouds
and complexes that are dotted along the arms and in the bar.  Fig. \ref{m83color} shows that with radial-averaging, the WISE colors
reveal the axi-symmetric arms and complexes (e.g., bar end cusps at a radius of 110$\arcs$), notably with the
W2$-$W3 color (dashed line), whereas the evolved stellar population is smoothly distributed
throughout the disk and bulge (solid line).  In a global integrated-light view (right panel of Fig. \ref{m83color}),
the SED is characterized by a significant (aggregate massive) population of old stars, forming the near-IR peak
and R-J tail in the mid-infrared,  very strong PAH emission at 6.2, 7.7 and 11.2 $\mu$m,
and a rising warm (dust-emitted) continuum that is powered by young, hot stars, as evident from
the relatively bright UV flux observed throughout the disk of M83.

For disk/spiral galaxies, the stars that provide the most feedback to the ISM are the hottest and most massive.
The O and B stars emit strong ultra-violet (UV) radiation that ionizes and heats the ISM, providing the energy
needed to warm dust, excite and break apart molecules thereby creating HII and photo-dissociation regions
(e.g. Tielens \& Hollenbach 1985; Wolfire et al. 2003).
This young population of stars is ideally probed by the GALEX
UV survey (Martin et al. 2005) using the 0.1516 $\mu$m (FUV) and
0.2267 $\mu$m (NUV) bands, obtaining a spatial resolution comparable to that of WISE.
The integrated flux of M\,83 in these two bands is
9.56 [AB mag] and 10.12 [AB mag], respectively for NUV and FUV, where the magnitudes have been corrected
for the
foreground extinction toward M\,83 (A$_V$ = 0.22 mag);  see Gil de Paz et al. (2007) for the extinction
coefficients.   Fig. \ref{m83color} presents the full integrated SED of M\,83,
plotting the UV, near-IR and mid-IR photometric results.
The figure includes a model that is generic
to Sd-type galaxies (generated using the GRASIL code of Silva et al. 1998),
simply illustrating that M\,83 possesses integrated characteristics of
late-type galaxies.

\begin{figure*}[ht!]
\begin{center}
\includegraphics[width=17.5cm]{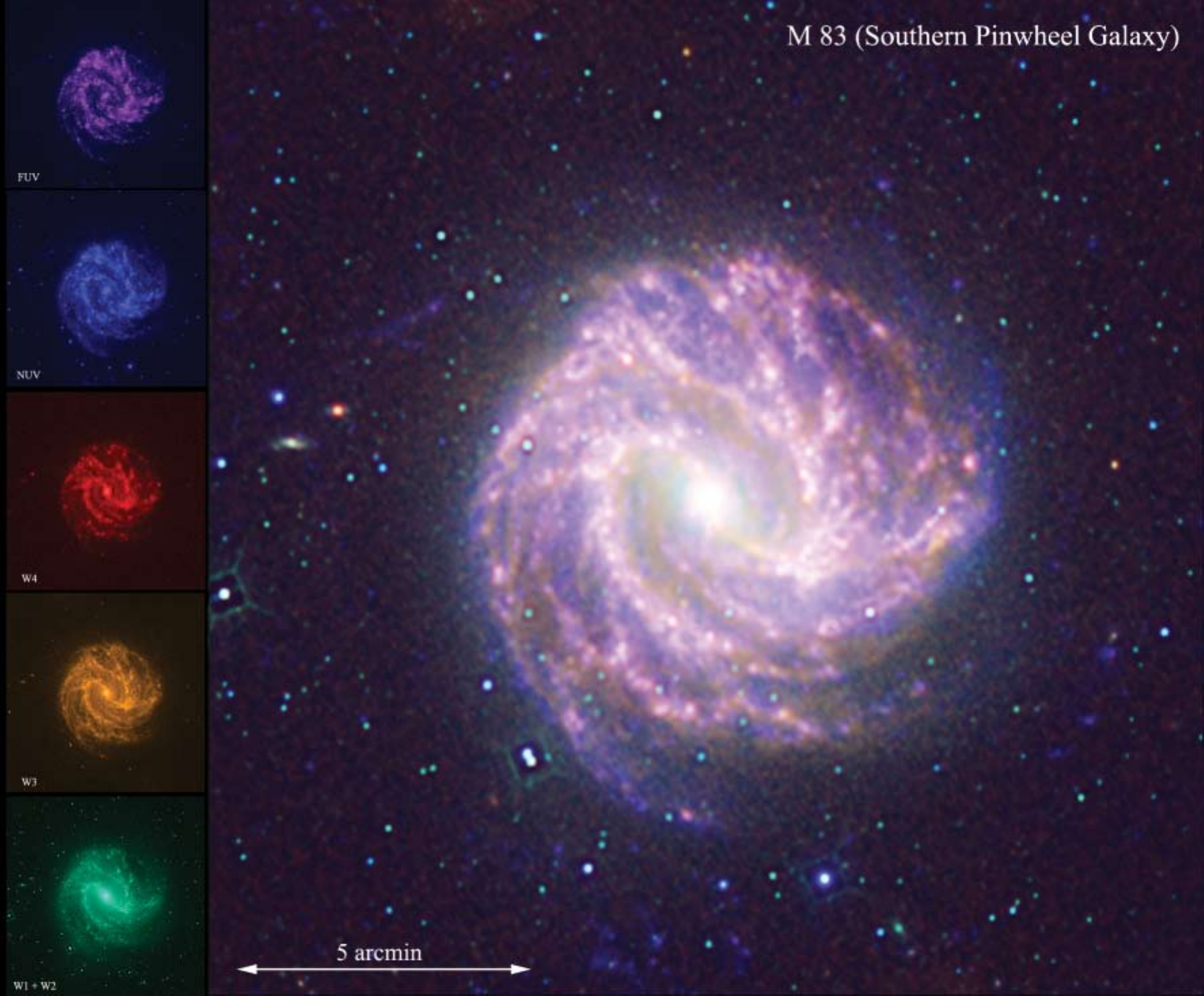}
\caption[M83galex]
{\small{GALEX and WISE color-composite view of M\,83 (NGC\,5236).
The color assignment for the WISE high-resolution
imaging (3.4 $\mu$m +
4.6 $\mu$m,
12.0 $\mu$m and
22 $\mu$m) and GALEX imaging
(0.227 $\mu$m and 0.152 $\mu$m)
 is shown with the small panels to the left.
}}
\label{m83galex}
\end{center}
\end{figure*}

Combining the GALEX UV and WISE infrared observations
provides a more complete understanding of the star formation that drives the evolution of disk galaxies.
In Fig. \ref{m83galex} we present a qualitative perspective of how GALEX and WISE observations of M\,83
form a snapshot gallery of the galaxy anatomy.
The color scheme is such that the ISM emission appears orange/red,
the evolved stellar ``backbone"
appears green,  and the photospheric light from young, hot stars
appear magenta (FUV) and blue (NUV).  The largest star formation complexes have both
UV and infrared light strongly correlated, appearing as white blobs, notably viewed
near the bar ends and the connecting eastern spiral arm.  The UV-Infrared color-composite also reveals
striking differences
between the distribution of young stars and the warm ISM: (1) The GALEX emission is clearly extended
beyond the mid-IR limits, highlighting localized star formation that is well outside of the primary (optical R$_{25}$) disk
and spiral arm system,
a property that is also seen in many other spiral galaxies studied with GALEX (e.g., Thilker et al. 2005;
Goddard, Kennicutt \& Ryan-Weber 2010).
(2) The nuclear region and much of the central
bar have very little UV emission, while the mid-IR continuum is bright in all WISE bands.
(3) There is diffuse UV emission filling most of the disk, likely arising from B stars that have
dispersed from their birth clouds.
(4) The trailing edge of
the spirals arms have strong W3 infrared emission (note the orange/brown color inside of the spiral arms),
arising from excited 11.3 $\mu$m PAH emission integrated along the galaxy's line-of-sight.
(5) To the north of the nucleus, near the edge of the mid-IR disk,
there is a large mid-IR `void' or fork in the spiral arms, in contrast the entire area is filled with UV emission
arising from massive star clusters.

\begin{figure*}[ht!]
\begin{center}
\includegraphics[width=17.5cm]{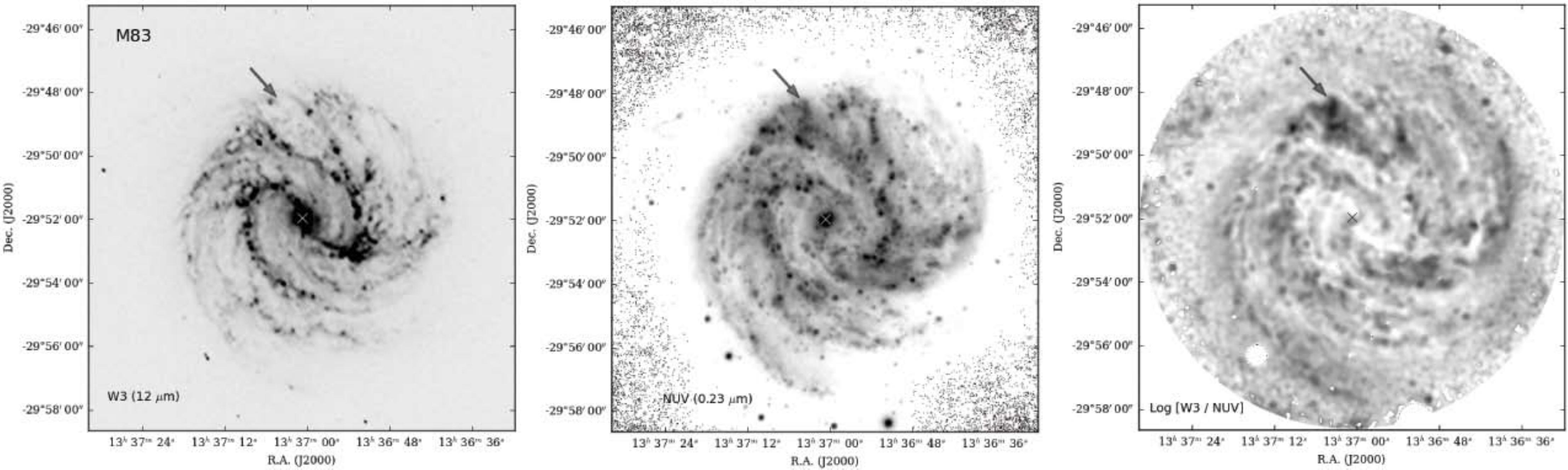}
\caption[W3vsNuv]
{\small{WISE W3 comparison with GALEX NUV of M\,83 (NGC\,5236).
(left) The first panel shows a log-stretch of the WISE W3 image, the (middle) panel
a log-stretch of the GALEX NUV image, and the (right) panel
is the flux ratio between the two bands.  Dark grey-scale values indicate
strong NUV, while light values indicate relatively strong W3; e.g.,
the arrow points to high NUV and low W3.
The nucleus is denoted with an $\times$ symbol.
}}
\label{W3vsNUV}
\end{center}
\end{figure*}

The distinct separation of the ultraviolet from the infrared emission
is further illuminated in
Fig. \ref{W3vsNUV}, comparing the WISE W3 12 $\mu$m image and the GALEX NUV 0.23 $\mu$m
image of M\,83.   The primary bar+spiral arms are clearly seen in the 12 $\mu$m image,
wrapping around the nucleus and extending to the north where it forks into two
separate arms.   Between these two arms there is hardly any mid-IR emission.
Comparing to the NUV image:  the bar+spiral arms are not prominent (the
south-western end of the bar has the brightest emission seen in the NUV),
but the inter-arm `void' to the north is easily apparent at these wavelengths.  The contrast between
the W3 and the NUV is demonstrated in the flux ratio, F$_\nu$(NUV) / F$_\nu$(W3),
shown in the third panel of Fig. \ref{W3vsNUV}:  the greyscale is such that strong UV
(relative to W3) appears grey or black, and inversely, strong W3 appears white.
The `white' structures show where the UV radiation has been fully absorbed by the dust
and re-emitted at longer infrared wavelengths.  The `grey' complexes have less dust extinction and
some large fraction of the UV light escapes.  The `black' traces where the UV light is fully escaping from
the veil of dust, due to either the absence of gas/dust (perhaps blown out by SNe winds) or the
dust geometry; e.g., super star cluster located foreground to the molecular cloud, or located
outside of a spiral arm, photoionizing gas and destroying the mid-IR PAHs in the local vicinity.

The northern inter-arm fork/void region is enormous, over 2$\arcmin$ ($>$ 3 kpc) in length.
It has two clumps of prominent UV emission (i.e., likely OB associations)
and relatively strong diffuse emission filling the void.  How does such a large area
come to be evacuated of dust and gas (see below) while also filled by light from
young, hot stars?  The size of such a region seems to preclude a simple wind-blown
event coming from a super star cluster; for example, the prominent (and resolved by HST) stellar cluster, NGC 206, seen
in the disk of M\,31 has cleared out a much smaller physical area ($sim$500 pc diameter hole;  see Hunter et al. 1996) by contrast.   Could the M83 void be a remnant of some tidal interaction
or similar dynamical disturbance due to accretion of a low-dust, gas-rich satellite galaxy?
We note that M\,83, the largest member of its galaxy group, has a large optically-detected tidal stream to the north of this region
(Malin \& Hadley 1997; Pohlen et al. 2003), tracing the disruption
of a dwarf galaxy in the strong gravitational field of M\,83.  These IR-UV properties are not
unique to M\,83.
Other massive  galaxies,
including M\,101 (see Fig. 1) and, as noted above, M\,31,
also exhibit tidal streams and large unveiled (dust-free) UV star clusters and associations
(e.g., Thilker et al. 2005).

There is another UV-filled void to the south-west
of the nucleus, just beyond the bar cusp, that is somewhat symmetric in location with the northern void
relative to the nucleus.   Alternative to UV transparency or a simple dust geometric effect, the diffuse UV radiation may
arise from older (less massive) star complexes that have either dispersed or blown away their birth cradle environment,
revealing open clusters that are dominated by A and B stars (not unlike the Pleiades star cluster).
Such regions would be much smaller in area than the northern fork/void discussed above.
This mechanism was proposed by Thilker et al. (2007) as one of the explanations for the striking differences
observed
between the infrared and UV emitted by the nearby star forming galaxy NGC\,7331.

\begin{figure*}[ht!]
\begin{center}
\includegraphics[width=17.5cm]{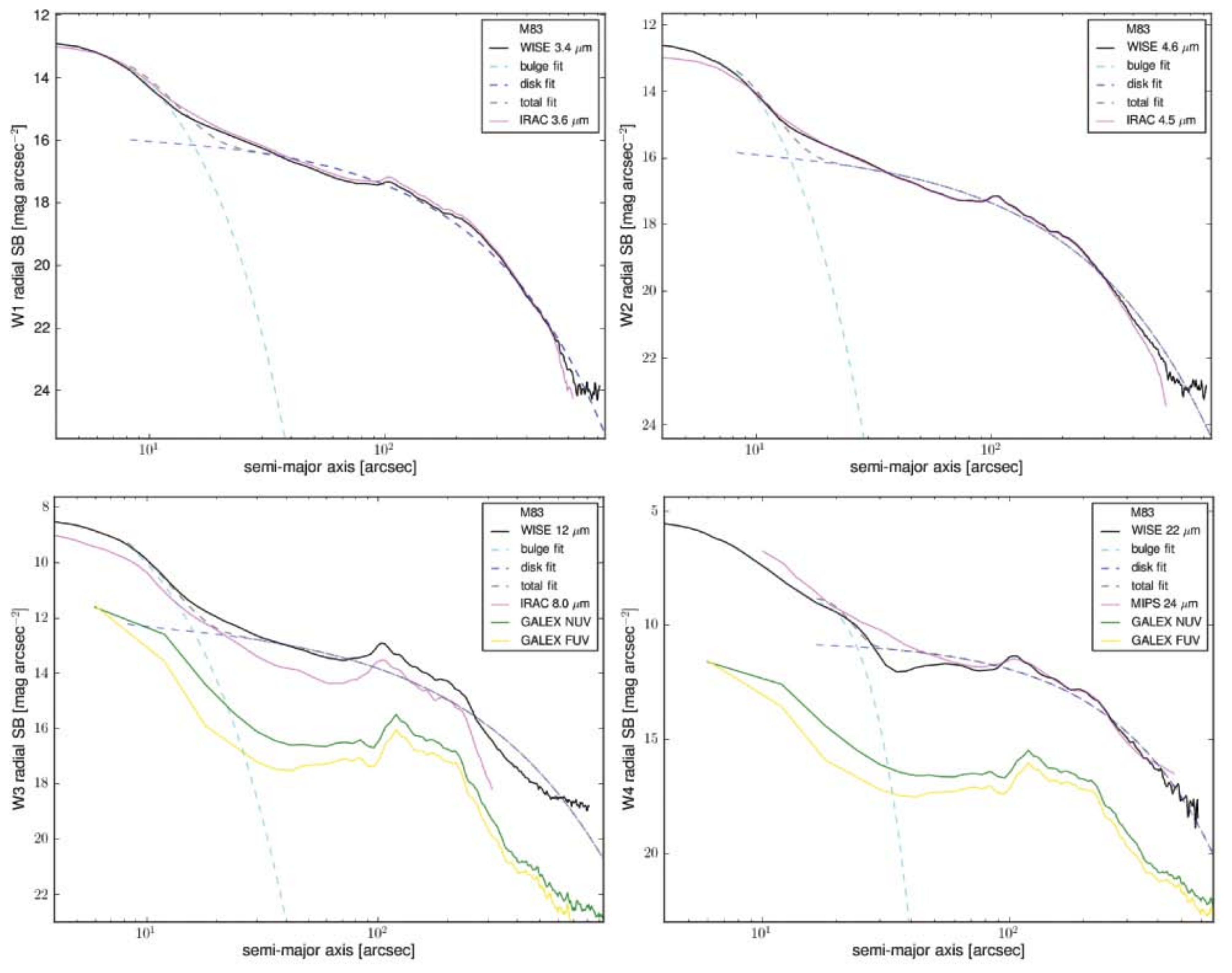}
\caption[M83radial]
{\small{M\,83 (NGC\,5236) azimuthal elliptical-radial surface brightness profiles
[Vega mag arcsec$^{-2}$],
comparing WISE (black line) with
IRAC and MIPS-24 (magenta line).
 A double S\'ersic function (grey dashed line) is fit to the WISE radial profile,
where the blue dashed line is the 'bulge' component and
the magenta dashed line the 'disk' component.
Additionally, the GALEX NUV and FUV
radial profiles are shown in the W3 and W4 panels
(note:  the GALEX AB magnitudes have been offset by $\sim$7 magnitudes to
fit within the Y-axis dynamic range).

}}
\label{m83radial}
\end{center}
\end{figure*}

Averaging with elliptical-radial axial symmetry, the resulting azimuthal surface brightness profiles of
M\,83 also reveal the spatial offsets between the UV and infrared emission.
Fig. \ref{m83radial} presents the surface brightness for WISE (W1 and W2) and \Spitzer
(IRAC-1 and IRAC-2), and the addition of GALEX (NUV) to the W3+IRAC-4 profiles
and GALEX (FUV) to the W4+MIPS-24 profiles.    The WISE and \Spitzer magnitudes
are in Vega units, and the GALEX magnitudes are in AB units plus an offset of
$-$7 magnitudes to fit within the plot.  The figure panels also show the
double-S\'ersic fit to the radial profile (bulge in cyan, disk in blue).

The first panel shows that W1 3.4 $\mu$m
and IRAC-1 3.6$\mu$m have similar profiles, both reaching depths of
24 mag arcsec$^{-2}$ (corresponding to 26.7 mag arcsec$^{-2}$ in AB).
The sharp `bump' at a radius of $\sim$110$\arcs$ corresponds to the bright
continuum emission arising from the bar end cusps where the gas has piled up from
shock focusing, thereby fueling
the observed active star formation (Lord \& Kenney  1991;  see also below).
The second panel shows the W2 4.6 $\mu$m and IRAC-2 4.5$\mu$m
profiles, which are also nicely co-aligned (as expected).  The bar end cusps are even
more prominent in this view of the mid-IR R-J tail.  The W2 depth reaches a limit
of 23 mag arcsec$^{-2}$ (corresponding to 26.3 mag arcsec$^{-2}$ in AB),
while the IRAC-2 is about 1 mag deeper in sensitivity.
The third panel shows the W3 12 $\mu$m, IRAC-4 8 $\mu$m,
and GALEX NUV + FUV.  Both W3 and IRAC-4 have the same shape, but are offset slightly
due to band-to-band differences between WISE and \Spitzer:  we would expect
a flux ratio of 1.1 for M\,83-type galaxies (e.g., see Fig \ref{WISEvSpitzer}, third panel).
The W3 depth reaches a limit
of 19.4 mag arcsec$^{-2}$ (corresponding to 24.6 mag arcsec$^{-2}$ in AB).
In the UV window, the profiles have considerably different shape;  notably, the absence of
UV emission in the central core (note the shallow profiles between 15 and
80$\arcs$ radius) and the bar cusps have shifted outward by 15 to 20$\arcs$
($\sim$0.5 Kpc), exhibiting a more pronounced, localized or compact, signal relative to the infrared.
The UV light then falls off steeper than the infrared light between 200 and 800$\arcs$;
but thereafter it remains constant and extends well beyond the infrared disk, to radii
exceeding  ($>$ 1000$\arcs$), punctuated by dense knots that are unmistakable markers for
star formation.  The last panel shows
the W4 22 $\mu$m, MIPS 24 $\mu$m,
and GALEX NUV + FUV.   W4 exhibits sharper profiles than MIPS-24 (note the dip
at 30$\arcs$) and extends to lower levels because the MIPS-24 mosaic is not large enough
to capture all of the light from M\,83.  The W4 depth reaches a limit
of 18.5 mag arcsec$^{-2}$ (corresponding to 25.1 mag arcsec$^{-2}$ in AB).
Likewise with the W3 comparison to the UV, the W4 band appears shifted relative to
the GALEX bands, although it is not as distinct as 12 $\mu$m comparison.

\smallskip

To summarize our IR-UV imaging results thus far:  although there is a clear physical connection between
the star formation activity and the warm dust emission arising from the interstellar medium,
the M\,83 IR-to-UV flux ratio diagram and surface brightness profiles emphasize that the distribution of young stars and UV diffuse
emission is only partically correlated with mid-infrared radiation, emphasizing the importance of tracking the
obscured (infrared) and unobscured (UV)
star formation.   We next investigate the IR to gas connection.

\smallskip

\subsection{M\,83:  Radio H\,{\sc i}  and Infrared Connection}

\begin{figure*}[ht!]
\begin{center}
\includegraphics[width=17.75cm]{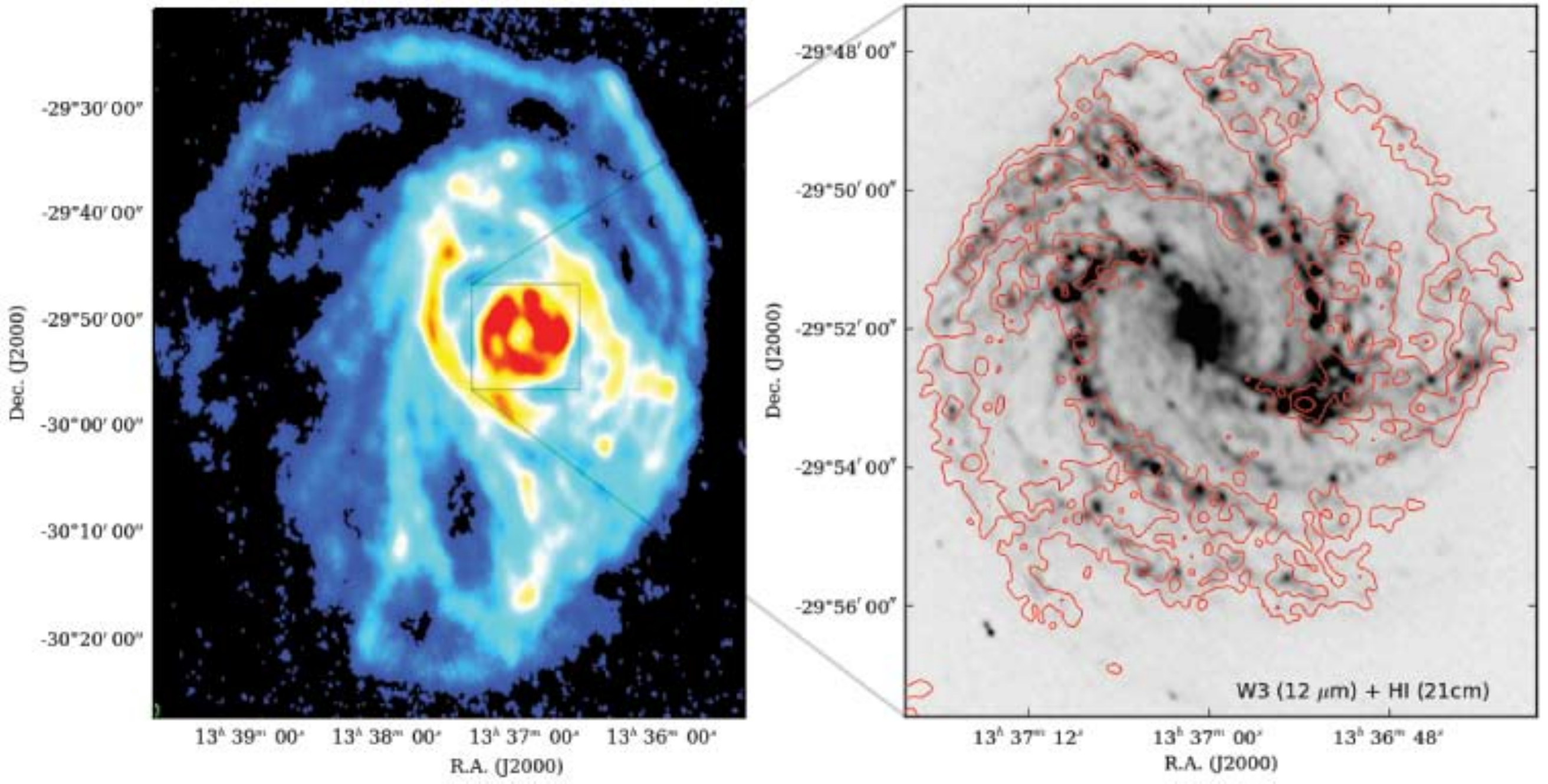}
\caption[W3vsHI]
{\small{The neutral hydrogen distribution of the southern spiral galaxy M\,83.
(Left) Large-scale moment-0 map of the extended H\,{\sc i}  gas distribution,
 constructed from mosaic data obtained with both the 21-cm multibeam system
 on the 64-m Parkes dish and the Australia Telescope Compact Array (ATCA).
(Right) ATCA high-resolution H\,{\sc i}  moment-0 map (red contours) overlaid
 onto the WISE 12$\mu$m map (grey-scale, log-stretch) of the central disk
 (denoted with a grey box in the left panel). The H\,{\sc i}  contour levels are
 0.1, 0.27, 0.43 and 0.6 Jy/beam km/s. The ATCA beam is approx. 10",
 comparable to the WISE 22 $\mu$m band, but note that the diffuse gas (see left panel) is
 resolved out when emphasizing the longer ATCA baseline measurements.
}}
\label{W3vsHI}
\end{center}
\end{figure*}

\smallskip

The neutral hydrogen gas reservoir of M\,83 extends $\sim$1 degree (80 kpc) in length,
containing a HIPASS-measured total of 8.3$\times$10$^9$ M$_\odot$ of H\,{\sc i} (Koribalski et al. 2004).
Another 4$\times$10$^9$ M$_\odot$ of molecular hydrogen resides in the central interior and bar of M\,83
(Lundgren et al. 2004 and 2008, for a distance of 4.66 Mpc).  The total dynamical mass is estimated
to be 6$\times$10$^{10}$~M$_\odot$ (Lundgren et al. 2008) and 8$\times$10$^{10}$~M$_\odot$  (Crosthwaite et al. 2002).

We have obtained new high spatial resolution 21-cm imaging of the M\,83 region using the Australia Telescope Compact Array (ATCA)
as part of the Local Volume HI Survey (LVHIS; Koribalski, B.S. 2008),
and in combination with single-beam observations using the Parkes 64-m dish
\footnote{Applying uniform weighting of the $uv$-data gives more
more weight to the longest baselines in the compact array ($\la$6~km)}, constructed a map of
the H\,{\sc i} distribution that is comparable in resolution to that of WISE and GALEX (10$\arcsec$ vs 5$\arcsec$).
Fig. \ref{W3vsHI} presents the neutral gas distribution of M\,83, viewed to its fullest extent (left panel)
and focusing on the inner disk in which the infrared is detected (right panel).
At the largest scales, the gas is well extended beyond the optical/infrared disk,
forming a warped structure that indicates tidal disturbance.
To the north, a long H\,{\sc i} filament corresponds spatially and kinematically to the tidal stream of stars
tracing a dwarf galaxy that is shredding and spiraling into the M\,83 gravitational well.
The highly asymmetric extended gas distribution suggests that other small galaxies may also be
accreting onto the disk.

Closer to the central region, more ordered spiral arms are discerned.
The right panel of Fig. \ref{W3vsHI} shows the H\,{\sc i} overlayed (red contours) on the
W3 12 $\mu$m image.  The spatial correspondence between the neutral gas and
the integrated (line-of-sight) PAH-ISM emission is striking: except for the nuclear region, where the H\,{\sc i} is absent
(having been converted to molecular gas for consumption),
the atomic hydrogen traces the current star formation on large angular scales.
Since it is the molecular hydrogen gas that serves as the primary fuel for star formation, we would
not necessarily expect the neutral hydrogen to be so closely correlated with the PAH emission arising from
PDRs and \HII\ regions (e.g., see below).  However, the Kennicutt-Schmidt Law (K-S;  cf. Kennicutt 1989), which relates the SFR surface density to the
gas surface density, tells us that the total gas (neutral plus molecular) controls the star formation --
here traced by the 12 $\mu$m maps which highlight the sites of enhanced star formation density.

This gas-to-infrared dependence is further emphasized in the extended Schmidt Law detailed in Shi et al. (2011)
High resolution maps of molecular CO emission (tracing the cold H$_2$ distribution)
show concentrated, clumpy associations that trace the embedded star formation, particularly along
the inner eastern arm (see Lord \& Kenney 1991; Rand 1999).  What leads to the strong correlations between
 neutral H\,{\sc i}
and the infrared emission?   Beyond replenishment through accretion and bar-directed feeding from the massive H\,{\sc i} reservoir,
Tilanus \& Allen (1993;  see also Lundgren et al. 2008) propose
that the atomic hydrogen in the central disk of M\,83 is in fact the by-product of molecular hydrogen
that has been dissociated by the strong radiation fields arising from the massive star formation in
the spiral arms.  The gas is likely to pass though many neutral to molecular to neutral cycles during
the lifetime of star formation in the spiral arms.  In any event, 
it is clear that the interplay between molecular gas and neutral gas can produce
a complex physical association with the star formation rate and its efficiency.

\begin{figure*}[ht!]
\begin{center}
\includegraphics[width=17.1cm]{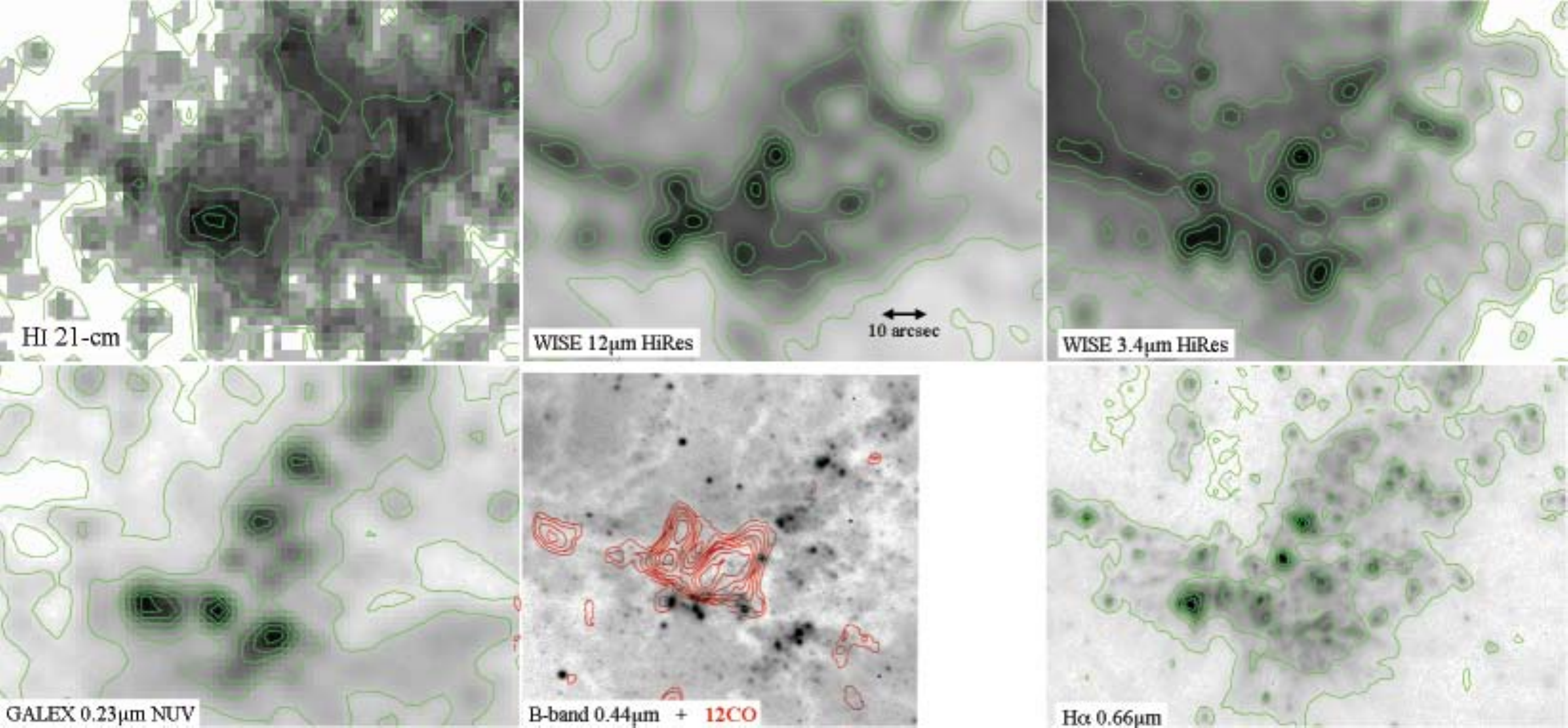}
\caption[M83zoom]
{\small{Close-up view of the southwestern bar cusp of M83,
centered on J2000 coordinate 13h36m52.86s, -29d52m55.3s.
Clockwise starting from left:  high spatial resolution 21-cm image
(ranging from 0.01 to 0.60 Jy km s$^{-1}$,
WISE HiRes 12 $\mu$m,  WISE HiRes 3.4 $\mu$m,
H$\alpha$, B-band (with 12CO contours) and GALEX NUV.  The green contours emphasize the
grey-scale contrast.  Overlaying the B-band image, the red contours are 12CO(1-0) with levels:  2, 3, 4, 5, 6,
7, 8, 9 $\times$ 2.3 Jy km s$^{-1}$ (Kenney \& Lord 1991).
At the distance of M83, the physical scale of 10$\arcsec$ is 213 pc.
}}
\label{M83zoom}
\end{center}
\end{figure*}

To investigate the gas to mid-IR emitting dust relationship at the finest scales that our
multi-wavelength data enable, we zoom into the southwest bar transition zone, located roughly
at 13h36m52.86s, -29d52m55.3s (J2000).  The region contains several massive star formation
complexes, as revealed in Fig. \ref{M83zoom},  generating strong radiation fields that 
heat the gas and dust that has funnelled into this region, recently mapped with far-infrared imaging from Herschel (cf. Foyle et al, 2012).
The neutral gas content is represented by
the H\,{\sc i}  21-cm map (upper left), and the molecular gas by the $^{12}$CO contours (red).
The obscured star formation is traced by the PAH emission as seen at 12 $\mu$m, and
the unobscured star formation by the UV and H$\alpha$ imaging.  Finally, the evolved population
emission peaks in the near-infrared, here represented by W1 3.4 $\mu$m.  The ``cusp" is clearly
evident with these probes, the bar ends at  center and then promptly turns to the northwest as it transitions to
the massive spiral arm that gives M\,83 its distinctive morphology.  It is clear that the molecular gas
(red contours) more closely aligned with the star formation (both W3 and the UV) than the neutral gas;
nevertheless, the total gas is concentrated in the same area as the bright star formation knots.
The density wave and bar kinematics have focused the fuel that is now driving the star formation.
The B-band image, whose angular resolution discerns features as small as 10 pc, exhibits dark
dust lanes that represent the thickest concentrations of gas, roughly outlining the UV (transparent) knots
and coincident with the W3 infrared (obscured) emission.   The recombination H$\alpha$
shows that the star formation complex breaks into several small clumps bathed in diffuse emission
that W3 clearly detects.   Kenney \& Lord (1991) make the case that the southwest bar cusp is composed
of distinct
kinematic gas components that originated from different regions of the disk but have streamed into
the transition zone forming dense molecular clouds.  The red contours also show that they are spatially distinction,
 resolved by both WISE and GALEX.    A wealth of information is provided by the multi-wavelength data sets
 presented here,  highlighting the importance of extracting both spatial
 and spectral elements to study and decode the physics in star formation.\\

\begin{figure*}[ht!]
\begin{center}
\includegraphics[width=17.1cm]{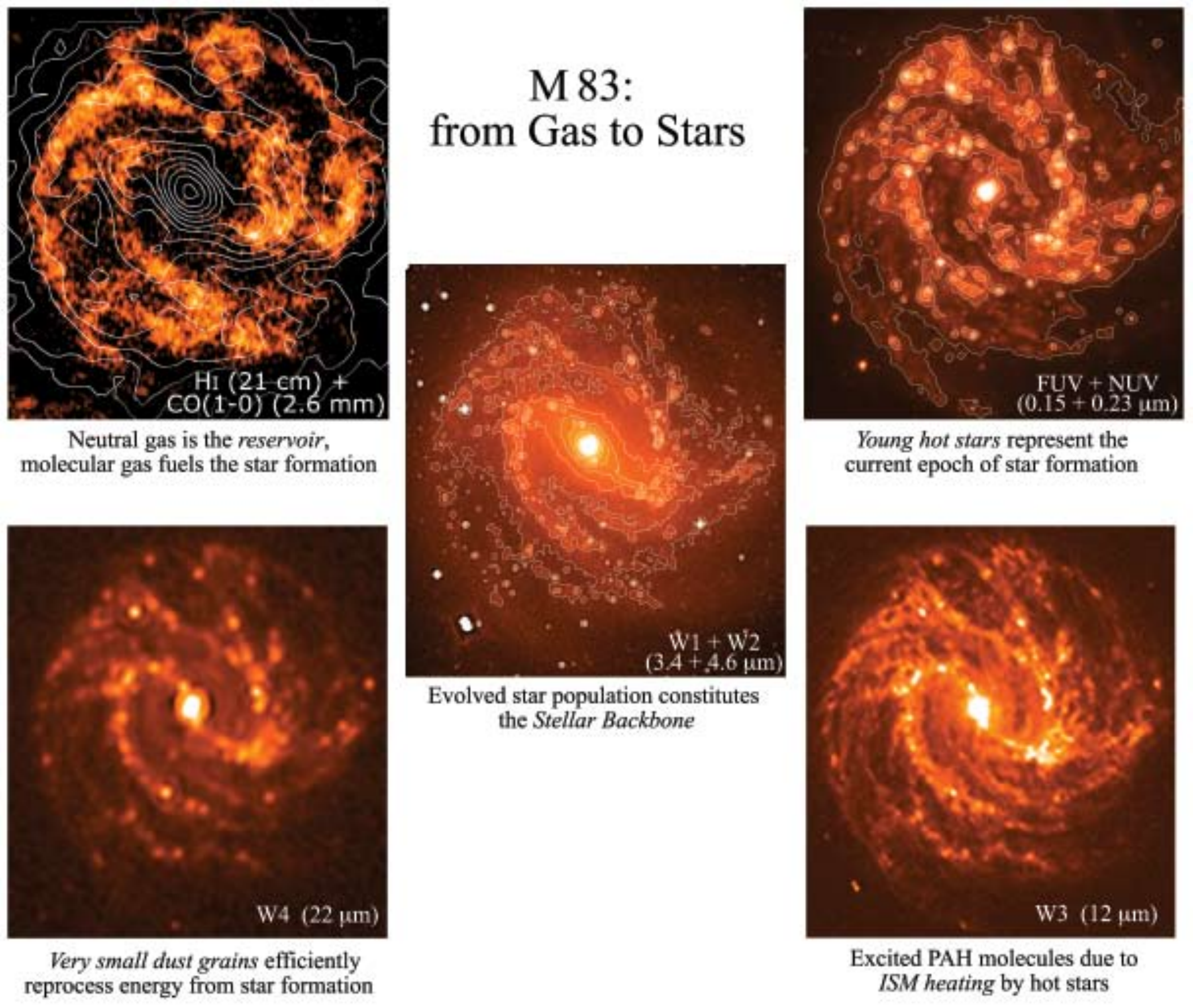}
\caption[M83all]
{\small{The many faces of M\,83, highlighting the evolution from
gas to stars.  The 10 acmin panels show:  the neutral (H\,{\sc i} grayscale) and molecular hydrogen (CO contours) gas content,
massive star formation as viewed by GALEX NUV (grayscale) and FUV (white contours), WISE view of 11.3 $\mu$m PAH emission (W3 band)
and reprocessed
starlight (W4 band) both associated with star formation, and the center panel shows
the stellar distribution of the previous generations of star formation as viewed
with the W1 (3.4 $\mu$m; grayscale) and W2 (4.6 $\mu$m;  white contours) bands.   The CO(1-0) is from
Crosthwaite et al. (2002).
}}
\label{M83views}
\end{center}
\end{figure*}

\newpage

\subsection{M\,83:  From Gas to Stars}

Understanding the process that fuels the star formation, requires both the neutral and
molecular hydrogen content be mapped and studied with respect to the young, hot population
of stars (GALEX) and the interstellar medium that responds to the energy input (WISE and {\it Spitzer}).
Fig. \ref{M83views} summarizes the evolution from gas to stars using the maps presented
in this work, starting in the upper left and moving clockwise, completing the cycle in the center panel.
We start with the fuel: the primordial H\,{\sc i} combines to form molecular H$_2$,
which gravitationally collapses (e.g., via spiral density waves) to form new stars.
Massive O and B stars radiate copious amounts of UV photons which are detected by
GALEX in the FUV and NUV bands.  Most of the radiation, however, is scattered and absorbed
by dust grains that are distributed throughout the spiral arms and birth clouds.  The radiation
excites the complex PAH molecules that form on dust grains located outside of \HII\ regions.
These photodissociation regions are detected by WISE in the W3 12 $\mu$m band and by
\Spitzer in the IRAC 8 $\mu$m band.  The radiation warms small dust grains to temperatures
of $\sim$100$-$150 K, radiating (modified blackbody) photons in the mid-infrared window
that is detected by WISE with the W4 22 $\mu$m band and by
\Spitzer with the MIPS 24 $\mu$m band.   Over time the stars disperse from their birth clouds, becoming
visible to detection at optical wavelengths.  Massive stars burn hot and bright, living for only a few to tens of
millions years before exploding and sending their gas back into the ISM for recycling.  Less massive
stars live much longer: solar-type stars burn for billions of years, evolving into luminous red giants.
These stars, far more numerous that the massive O and B stars, form the aggregate massive-backbone of the galaxy.
These cool giants emit most of their photospheric radiation in the near-infrared window, detected and studied by
WISE with the W1 3.4 $\mu$m and W2 4.6 $\mu$m bands and by
\Spitzer with the IRAC 3.6 $\mu$m and 4.5 $\mu$m bands.  While some galaxies undergo dynamical interactions
(e.g., within group and cluster environments), the basic ingredients and processes that drive M\,83, sketched
in Fig. \ref{M83views}, are
how galaxies form and evolve over vast stretches of time.


\section{Star Formation Rate}

Both the present star formation rate and total stellar mass (M$_{*}$) are fundamental parameters for the study of galaxy evolution.
Direct comparison of the global SFR (from the WISE long-wavelength bands)
with stellar mass (short-wavelength bands) enumerate the
present-to-past star formation history,
or effectively,
how fast the galaxy is building.
This specific star formation rate (sSFR) is
a critical metric of
morphological evolution.   In this penultimate section, we explore the SFR and stellar mass properties
of our large galaxy sample, providing a preliminary prescription for estimating these quantities from
the spectral luminosities.   We focus on the global properties in this work, with an extension to
much larger samples and to 
more detailed work on
the internal star formation, gas and stellar density properties left to future work.
Table  \ref{tab:lum} presents the $\nu$L$_\nu$ luminosity densities for
the near-infrared (2MASS Ks-band), mid-infrared (WISE and MIPS-24) and GALEX FUV/NUV,
corrected for foreground Galactic extinction and the expected internal extinction
(see Section 4.3.1).   The expected uncertainty in the luminosity is $\sim$10\%, assuming
5\% uncertainty in the distance estimate.  Here the $\nu$L$_\nu$ values have been normalized
by the total solar luminosity (L$_\odot$): 3.839$\times$10$^{33}$ ergs s$^{-1}$.

To convert the luminosities to the ``inband" equivalent
\footnote{The ``inband" luminosity, L$_{\lambda}$ should not be confused with the `spectral'
luminosity ($\nu$L$_{\nu}$).  The M/L ratio is calibrated using
the inband luminosity:  the computed luminosity (as measured in the band) normalized by the absolute Solar luminosity as
measured in the band.  We derive the absolute in-band magnitude of the Sun: 3.32, 3.24, 3.27, 3.23 and 3.25 magnitude 
for K$_s$, W1, W2, W3 and W4 respectively;  we will refer to these luminosites as L$_K$, L$_{W1}$, etc.},
scale the $\nu$L$_\nu$ by a factor of 7.37, 22.883, 58.204, 38.858 and 282.50 for
the K-band, W1, W2, W3 and W4 bands, respectively, where the scaling factor takes into account the
difference between the total Solar luminosity and the inband value (as measured by the 2MASS
and WISE bands; see section below on Stellar Masses).

\begin{table*}[ht!]
{\scriptsize
\caption{Broad-band Spectral Luminosities\label{tab:lum}}
\begin{center}
\begin{tabular}{r r r r r r r r r r r}

\hline
\hline
\\[0.25pt]

Name   & K$_s$ 2.2 $\mu$m & W1 3.4 $\mu$m  & W2 4.6 $\mu$m  & W3 12$\mu$m & W4 22 $\mu$m & MIPS 24 $\mu$m & FUV 0.16 $\mu$m & NUV 0.23 $\mu$m  \\
           & Log [$\nu$L$_\nu$/L$_\odot$] &   Log [$\nu$L$_\nu$/L$_\odot$] & Log [$\nu$L$_\nu$/L$_\odot$] & Log [$\nu$L$_\nu$/L$_\odot$] &  Log [$\nu$L$_\nu$/L$_\odot$] & Log [$\nu$L$_\nu$/L$_\odot$]  & Log [$\nu$L$_\nu$/L$_\odot$] & Log [$\nu$L$_\nu$/L$_\odot$] \\

\hline
\\[0.25pt]

NGC\,584     &  10.101&   9.586&   9.183&   8.178&   7.513&   7.440&   7.910&   8.387\\
NGC\,628     &   9.559&   9.225&   8.887&   9.220&   8.911&   8.883&   9.494&   9.463\\
NGC\,777     &  10.594&  10.176&   9.794&   8.834&   8.039&   7.988&   8.739&   8.892\\
NGC\,1398    &  10.498&  10.027&   9.650&   9.477&   8.846&    --&   9.384&   9.401\\
NGC\,1566    &   9.677&   9.273&   8.922&   9.161&   9.048&   9.004&   9.455&   9.388\\
NGC\,2403    &   9.110&   8.791&   8.462&   8.693&   8.505&   8.442&   9.293&   9.226\\
NGC\,3031    &  10.099&   9.652&   9.262&   8.925&   8.457&   8.454&   9.172&   9.145\\
NGC\,4486    &  10.585&  10.236&   9.845&   8.861&   8.412&   8.309&   9.047&   9.200\\
NGC\,5194    &  10.116&   9.685&   9.353&   9.808&   9.566&   9.550&   9.759&   9.796\\
NGC\,5195    &   9.814&   9.359&   8.983&   8.702&   8.574&   8.601&   8.295&   8.464\\
NGC\,5236    &   9.963&   9.587&   9.248&   9.613&   9.578&   9.558&   9.639&   9.688\\
NGC\,5457    &   9.969&   9.567&   9.233&   9.542&   9.251&   9.226&  10.023&   9.944\\
NGC\,5907    &  10.264&   9.815&   9.472&   9.619&   9.281&   9.264&   9.132&   9.145\\
NGC\,6118    &   --&   9.455&   9.101&   9.294&   8.939&    --&    --&    --\\
NGC\,6822    &   7.188&   7.127&   6.754&   6.408&   6.175&   6.137&   7.727&   7.728\\
NGC\,6946    &   9.934&   9.543&   9.213&   9.674&   9.480&   9.472&   9.821&   9.850\\
IC\,342      &   9.710&   9.362&   9.018&   9.387&   9.239&   9.184&   9.797&   9.654\\

\hline
\end{tabular}
\end{center}
\tablecomments{The $\nu$L$_\nu$ luminosity (normalized by the total Solar luminosity)
is derived from the integrated flux density and the distance listed in
Table \ref{tab:observed}, where  the luminosity uncertainty is completely dominated by the distance uncertainty;
assuming a distance uncertainty of 5\%, it follows that the luminosity uncertainty is 10\%.
The fluxes have been corrected for the Galactic extinction and
the expected internal extinction.
The K$_s$ flux densities are from the 2MASS Large Galaxy Atlas (Jarrett et al. 2003).
}

}

\end{table*}

For nearby galaxies, the present star formation is traditionally studied using tracers of massive stars,
notably Balmer emission arising from \HII\ regions (e.g., Kennicutt 1998), which requires significant extinction
corrections to account for absorbed or scatter light.
With the advent of space-based UV observations of galaxies in the local universe, notably from GALEX,
these studies are now carried out on large, diverse and statistically significant samples.
Since the UV photons originating from or associated with
hot, young stars are predominately absorbed by dust grains and re-radiated at longer wavelengths,
the infrared window may be used to effectively
trace the underlying star formation, although ideally a combination of the infrared and GALEX UV provides the
most complete estimate.
Studies using IRAS, ISO and \Spitzer observations have correlated the mid and far-infrared emission
with the present young stellar population that is embedded within molecular clouds that embody spiral arms and disks
of galaxies.  Most relevant to WISE, global star formation rates may be directly estimated using the
warm dust grain and integrated PAH emission (alternatively, gas column density-normalized PAH emission)
or combinations of UV or H$\alpha$ and infrared tracers to capture both the unobscured and
obscured star formation (Calzetti et al. 2007; Leroy et al. 2008; Calzetti 2011;  Kennicutt et al. 2009;
Rieke et al. 2009; Treyer et al. 2010).    Below we present both infrared and combined
IR + UV star formation rates.

\begin{figure*}[ht!]
\begin{center}
\vspace{-5pt}
\includegraphics[width=17.5cm]{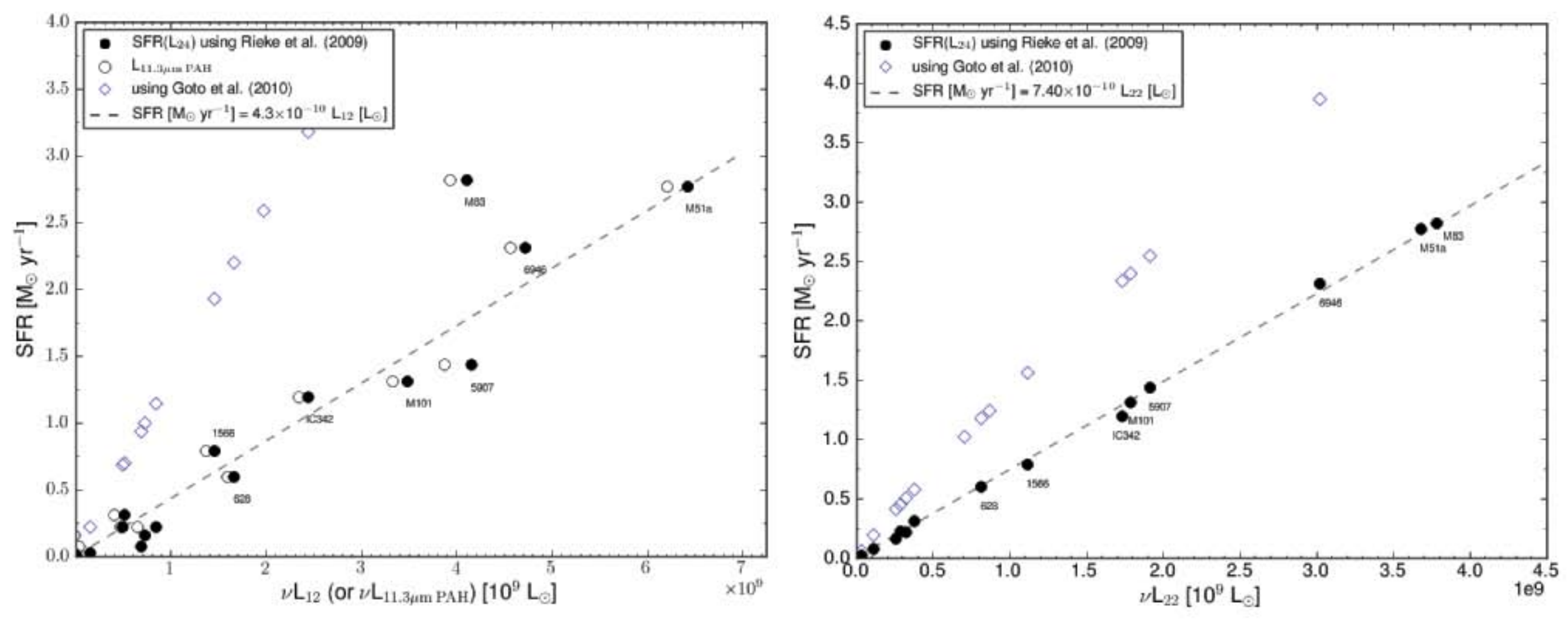}
\caption[WISE SFR calibration]
{\small{Correlating the Star Formation Rate with the WISE mid-IR luminosity.
The global SFR$_{IR}$ is derived from the MIPS 24 $\mu$m luminosity and the
Rieke et al. (2009) calibration.
(left) The global SFR$_{IR}$ as a function of the WISE 12 $\mu$m luminosity
(filled circles) and the integrated 11.3 $\mu$m PAH luminosity (open circles).
(right) The global SFR$_{IR}$ as a function of the WISE 22 $\mu$m luminosity.
A simple linear fit to the data is shown (dashed line).
The galaxy name is indicated below the measurement, where single numbers represent the NGC \#.
}}
\label{SFR0}
\end{center}
\end{figure*}

We establish the global WISE SFR$_{IR}$ relation by bootstrapping from the well-studied MIPS-24 relation
(e.g., Calzetti et al. 2007; Rieke et al. 2009; Rujopakarn et al. 2011).
Although using only the 24 $\mu$m luminosity
to estimate the SFR$_{IR}$ is not as accurate as using the L$_{\rm TIR}$, or some combination of infrared, UV and optical, it still
provides a tight correlation for galaxies with moderate to high metallicities
(but see Relaño et al 2007 for SFR analysis of individual H II regions).  Using the Rieke et al. (2009) relation to derive the SFR$_{IR}$
from MIPS-24 luminosity, we plot the WISE $\nu$L$_{12}$ and $\nu$L$_{22}$ against the MIPS-estimated SFR$_{IR}$; Fig. \ref{SFR0}.
For the SFR range between zero and 3, we find the best fit $\nu$L$_{12}$ and $\nu$L$_{22}$ relations to be:

{\scriptsize
\begin{eqnarray}
{\rm W3: \,\, SFR_{IR}} \, (\pm 0.28) \, [M_\odot \, yr^{-1}]  &=& 4.91 ( \pm 0.39) \times 10^{-10} \, \nu L_{12} [L_\odot],\\
{\rm W4: \,\,  SFR_{IR}} \, (\pm 0.04) \, [M_\odot \, yr^{-1}]  &=& 7.50 (\pm 0.07) \times 10^{-10} \, \nu L_{22} [L_\odot].
\end{eqnarray}
}

The RMS scatter between the MIPS-estimated SFR$_{IR}$ and the WISE SFRs (Eq. 1 \& 2)
is 0.28 and 0.04 M$_\odot$ yr$^{-1}$, for SFR (12 $\mu$m) and
SFR (22 $\mu$m), respectively; the actual uncertainty in any given WISE-derived SFR is at least 10\%  given
the uncertainty in the luminosity.
As expected, there is a very tight relation between the WISE $\nu$L$_{22}$ and the Rieke SFR$_{IR}$, since both WISE W4 and MIPS-24 have
very similar bandpasses, tracing the warm ISM dust continuum.   
Conversely and similar to the larger scatter observed in the IRAC
8 $\mu$m-to-SFR relation (Calzetti et al. 2007),  the
WISE $\nu$L$_{12}$-to-SFR$_{IR}$ relation exhibits a trend that reflects the complex relationship of
combined thermal dust $+$ silicate absorption $+$ PAH emission with
star formation activity.
The strength of the PAH emission bands depends on the
metallicity, column density of gas and ionizing radiation field from the young stellar population (Engelbracht et al 2008; Draine 2011),
but the 11.3 $\mu$m PAH can also be excited by spatially-diffuse, evolved ($\sim$Gyr) stellar population (Calzetti 2011).
Nevertheless, the $\nu$L$_{12}$ relation is very useful to have because the W3 (12 $\mu$m) band is so much more
sensitive than the W4 (22 $\mu$m) band, and often, for galaxies beyond the local universe, is the only detection from WISE.

We have attempted to isolate the line-of-sight integrated PAH emission in the W3 band from the underlying continuum by
subtracting from the W3 a scaled version of W1, representing the R-J stellar light.  Note that we are
ignoring the warm-dust emission in W3 that arises from the dust shells of AGB stars and from AGN;
these potentially important components are further discussed below.   Using the three elliptical
galaxies in the sample, which trace only the photospheric continuum contribution, we determine that the W3 to W1 scale factor
is $\sim$15\% (which may be compared with the IRAC-4 to IRAC-1 ratio of 23\%; Helou et al. 2004).
The  resulting $\nu$L$_{\rm PAH}$ is shown in Fig. \ref{SFR0} as open circles.  The scatter in the SFR ($\nu$L$_{11.3 \mu m PAH}$)
relative to the Rieke SFR$_{IR}$ is slightly smaller, 0.26 M$_\odot$ yr$^{-1}$ but still much larger than the SFR (22 $\mu$m) distribution.
For comparison we also show the SFR estimated using the Goto et al. (2010) relation, who attempted to
adapt the SINGS 8 and 24 $\mu$m luminosity-to-L$_{\rm TIR}$ relation to the expected WISE W3 and W4 bandpasses.
The Goto et al. (2010) SFRs are very likely overestimated,  $\sim$2$\times$ larger than those predicted by Rieke et al. (2009) for 22 $\mu$m
and $\sim$3$\times$ larger for 12 $\mu$m.

We stress that the relations presented here (Eqs 1 and 2) are only
preliminary due to the small size of our sample.  A more thorough investigation should involve a larger sample,
that includes a wide range in metallicity and $\nu$L$_{IR}$ (e.g.,
SINGS sample), calibrating the WISE luminosities to the SFR derived through UV, infrared and hydrogen recombination line analysis
as with Calzetti et al. (2007) and Kennicutt et al. (2009).  A final important caveat:  even with a more complete sample
to establish the WISE SFR relations,
the mid-IR is sensitive to only the warm tracers
of star formation (PAHS, small dust grains), and thus represents a lower limit to the total infrared SFR that also includes
the heavily-obscured star formation in dense molecular cores, best traced by the far-IR emission (e.g., observations
from IRAS, Herschel and AKARI).

\begin{table*}
{\scriptsize
\caption{Infrared and Ultraviolet Global Star Formation Rates\label{tab:SFRs}}
\begin{center}
\begin{tabular}{r r r r r r r r r r r}

\hline
\hline
\\[0.25pt]

Name   &  SFR$_{IR}$ (12 $\mu$m) & SFR$_{IR}$ (22 $\mu$m) & SFR$_{IR}$ (24 $\mu$m) & SFR$_{FUV}$ (0.15 $\mu$m) & SFR$_{NUV}$ (0.23 $\mu$m) & SFR$_{tot}$ (IR+UV)  \\
           &   [M$_\odot$ yr$^{-1}$] & [M$_\odot$ yr$^{-1}$] & [M$_\odot$ yr$^{-1}$] & [M$_\odot$ yr$^{-1}$] & [M$_\odot$ yr$^{-1}$] & [M$_\odot$ yr$^{-1}$] \\

\hline
\\[0.25pt]

NGC\,584     &     0.1&     $<$10$^{-2}$&     $<$10$^{-2}$&     $<$10$^{-2}$&     $<$10$^{-2}$&     $<$10$^{-2}$\\
NGC\,628     &     0.8&     0.6&     0.6&     0.6&     0.6&     1.1\\
NGC\,777     &     0.3&     0.1&     0.1&     0.1&     0.2&     0.2\\
NGC\,1398    &     1.5&     0.5&    --&     0.5&     0.5&     0.9\\
NGC\,1566    &     0.7&     0.8&     0.8&     0.6&     0.5&     1.3\\
NGC\,2403    &     0.2&     0.2&     0.2&     0.4&     0.3&     0.6\\
NGC\,3031    &     0.4&     0.2&     0.2&     0.3&     0.3&     0.5\\
NGC\,4486    &     0.4&     0.2&     0.2&     0.2&     0.3&     0.4\\
NGC\,5194    &     3.1&     2.8&     2.8&     1.2&     1.2&     3.5\\
NGC\,5195    &     0.2&     0.3&     0.3&     0.0&     0.1&     0.3\\
NGC\,5236    &     2.0&     2.8&     2.8&     0.9&     1.0&     3.2\\
NGC\,5457    &     1.7&     1.3&     1.3&     2.2&     1.7&     3.3\\
NGC\,5907    &     2.0&     1.4&     1.4&     0.3&     0.3&     1.5\\
NGC\,6118    &     1.0&     0.7&    --&     --&     --&     $<$1\\
NGC\,6822    &     $<$10$^{-2}$&     $<$10$^{-2}$&     $<$10$^{-2}$ &     $<$10$^{-2}$&     $<$10$^{-2}$ &     $<$10$^{-2}$ \\
NGC\,6946    &     2.3&     2.3&     2.3&     1.4&     1.4&     3.2\\
IC\,342      &     1.2&     1.3&     1.2&     1.3&     0.9&     2.4\\

\hline
\end{tabular}
\end{center}
\tablecomments{
SFR$_{IR}$ [12 $\mu$m] = 4.9 $\times$ 10$^{−10}$ L$_{12}$ [L$_\odot$]
and SFR$_{IR}$ [22 $\mu$m] = 7.5 $\times$ 10$^{−10}$ L$_{22}$ [L$_\odot$] ;  see Fig. \ref{SFR0}.
SFR$_{IR}$ [24 $\mu$m] is
derived from L$_{24 \mu m}$ and  Eqs. 10 and 11 of Reike et al. (2009).
The UV rates are estimated from the FUV luminosity (using the relation from
Buat et al. 2008; 2011) and the NUV luminosity density (using the relation from Schiminovich et al. 2007).
The total UV + IR SFR,
characterized as ($1 - \eta$)SFR$_{IR}$ + $\gamma$SFR$_{FUV}$, where $\eta$ is the fraction of mid-IR light
that originates from the dust shells of AGB stars and $\gamma$ scales the UV transparency;
here $\eta$ is assumed to be 0.17 and $\gamma$ is unity; see  Elbaz et al. (2007) and Buat et al. (2011).
}
}

\end{table*}

For the galaxy sample, the infrared SFRs are estimated using the empirical relations, Eqs. 1 and 2,
and are listed in Table \ref{tab:SFRs}.   They trace the dust-obscured star formation activity,
which is dependent on the dust geometry and total gas/dust column density.
For the UV photons -- associated with young massive stars -- that manage to escape the galaxy,
GALEX may be used to estimate the unobscured star formation;  Table \ref{tab:SFRs}
lists the SFR estimated from the FUV and NUV extinction-corrected measurements as follows.
Buat et al. (2008; 2011) calibrate
the FUV star formation rate, characterizing it in terms of the GALEX luminosity as follows:
log SFR$_{FUV}$ = log ($\nu$L$_{FUV}$/L$_\odot$) - 9.69.
The SFR based on the NUV luminosity density is provided by
Eq.6  of Schiminovich et al. (2007):
SFR$_{NUV}$ = 10$^{-28.165}\,$L$_{NUV}$[erg s$^{-1}$ hz$^{-1}$].
For the sample, about half of the UV SFRs are larger or comparable to the estimates using the infrared tracers,
which verifies the importance of accounting for the unobscured star formation.
Combining both UV and IR estimates, the ``total" SFR
may be
characterized (Elbaz et al. 2007; Buat et al. 2011) as ($1 - \eta$)SFR$_{IR}$ + $\gamma$SFR$_{FUV}$, where $\eta$ is the fraction of mid-IR light
that originates from the dust shells of AGB stars and $\gamma$ scales the UV transparency.
In effect, the value of
$\eta$ will depend on the fractional number of AGB and intermediate, post-starburst stars  (ages 1 to 3 Gyrs) that contribute to the
mid-infrared light;  Buat et al. (2011) compute an ensemble average value of 0.17 for $\eta$ using a large sample of
field galaxies.
Adopting the Buat et al. (2011) values for $\eta$ (0.17) and $\gamma$ (unity),
the UV+IR SFR is listed in the last column of Table \ref{tab:SFRs}, and is generally larger ($\sim$10-20\%) of
the infrared component.
So for example, consider the SFR of NGC\,5236 (M\,83):    reported in Section 5,
the extinction-corrected integrated FUV flux is 10.11 mag (325 mJy),
translating to a luminosity of 4.4$\times$10$^9$ L$_\odot$.    Using the relation
of Buat et al. (2011), the M\,83 SFR$_{FUV}$ is then 0.9 M$_\odot$ yr$^{-1}$, or roughly
1/3 of the estimate SFR$_{IR}$ (Table \ref{tab:SFRs}), assuming $\gamma$ is unity, and using the average
$\eta$ value, the total SFR would then
be $\sim$3.3, which is 14\%  greater than the reported SFR$_{IR}$.
We note that the resulting total SFRs have values that are generally 5 to 50\% larger than the tabulated
values in Leroy et al. (2008), but are systematically much smaller than the SFRs predicted using the combined
GALEX FUV + MIPS(24) relation in the same work (equation D10 of Leroy et al 2008).

\begin{figure*}[ht!]
\begin{center}
\vspace{-5pt}
\includegraphics[width=12cm]{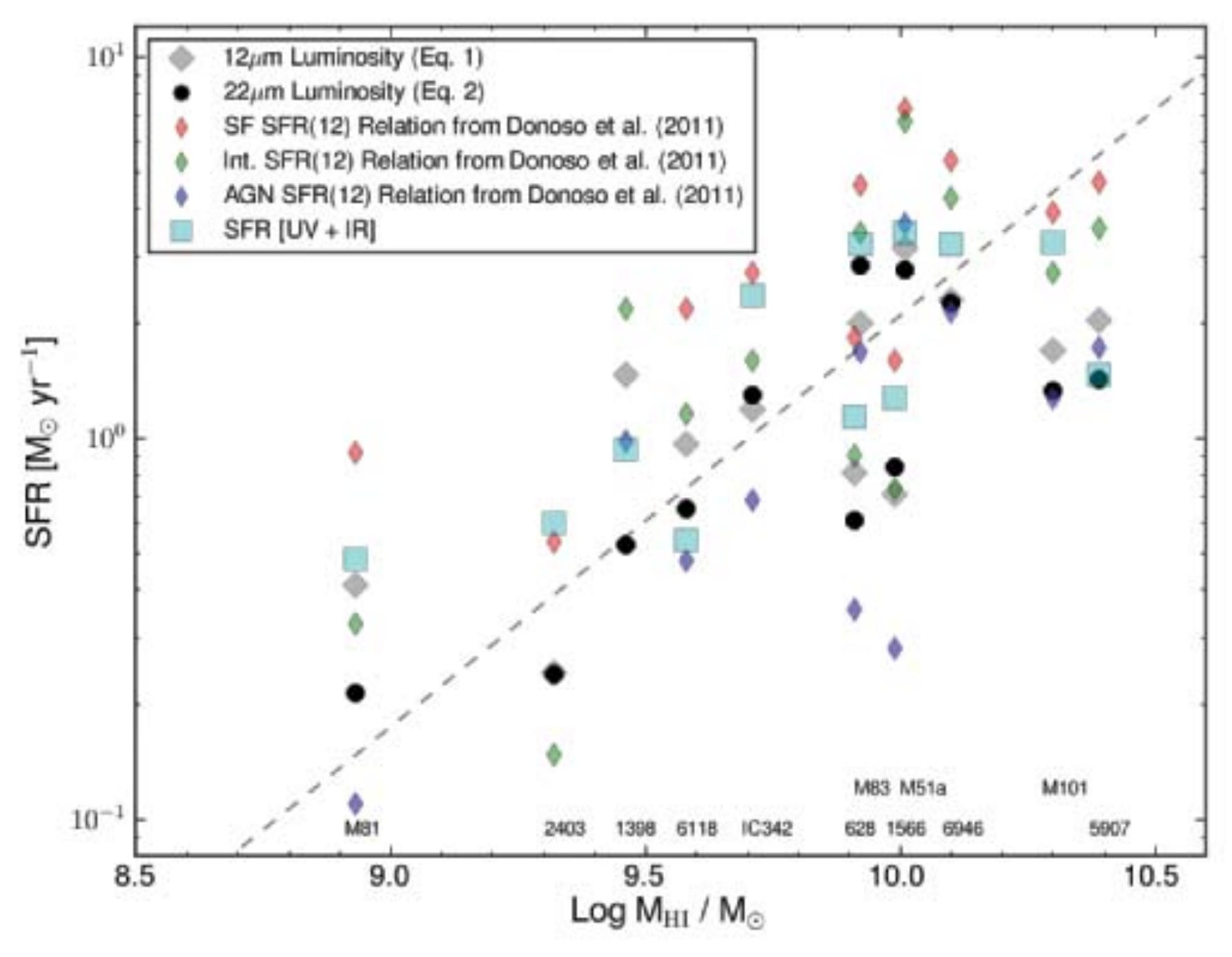}
\caption[WISE SFR]
{\small{Global star formation rates derived from the WISE
12, 22 $\mu$m and GALEX FUV and NUV luminosities, compared to
the neutral hydrogren content;  see Table \ref{tab:SFRs}
for details.  For comparison, also shown are the SFRs that follow from the
Donoso et al. (2011) 12 $\mu$m relations of starburst, intermediate and AGN
galaxy types.  The dashed line corresponds to the trend observed in the
SINGS sample (Cluver et al. 2010).
}}
\label{SFRs}
\end{center}
\end{figure*}

The global SFRs are graphically
presented in Fig. \ref{SFRs}, where we compare with the neutral hydrogen gas content to demonstrate
the essential star formation efficiency trend of the K-S scaling relation, connecting the star formation density to the
total gas density.  Although the H\,{\sc i} is usually associated with the diffuse
infrared emission, while the molecular gas (as traced by CO and HCN emission) is more closely associated with
star formation, we demonstrated
in the last section that the obscured star formation in M\,83 is associated with
the neutral gas distribution.  And indeed, we observed a
clear trend (RMS scatter $\sim$ 1 M$_\odot$ yr$^{-1}$) in both SFR$_{IR}$ and SFR$_{tot}$ and the H\,{\sc i} mass.
Also shown in the figure
is the trend quantified using the SINGS sample and a nearby high-H\,{\sc i} mass galaxy (Cluver et al. 2010) that
covered a much larger range in
neutral gas mass, 10$^{7}$ $-$10$^{11}$ M$_\odot$.   The slope in the SFR-to-gas trend will increase if the
contribution from the molecular gas is included:  gas rich galaxies have higher fractions of molecular gas
(e.g., 25 to 50\%  of the total gas) while gas-poor galaxies have much lower fractions of  H\,{\sc ii}
(e.g., Leroy et al 2008;  see also the COLDGASS scaling relations in Saintonge et al. 2012).
The highest global SFRs belong to M\,51a, M\,83, M\,101 and NGC\,6946,  while the lowest rates belong to the elliptical
galaxies and the dwarf NGC\,6822.
The lower the H\,{\sc i} content, the smaller the infrared luminosity (and hence, SFR as predicted by the global K-S relation),
while the most luminous disk galaxies have the highest gas content.   The H\,{\sc i} reservoir feeds and builds the
molecular hydrogen reservoir, which is what ultimately fuels the star formation and slowly builds the stellar mass of the galaxy (see below for discussion of the scaling relation between SFR and M$\_*$).

For comparison purposes, Fig. \ref{SFRs} includes the infrared SFRs derived using the relations given by
Donoso et al. (2011), who cross-matched WISE 12 $\mu$m fluxes with SDSS photometry and
spectroscopy of a large, statistically significant sample of nearby galaxies ($\bar{z} \sim$ 0.08).
The 12 $\mu$m emission is sensitive to both recent SF ($\sim$few hundred 10$^6$ yr)  and to
star formation averaged over Gyr timescales.  Moreover, the emission may arise or be associated
with AGN as well as the host itself.  Consequently,
they separated their matched sample into
star-forming, intermediate, and type-2 AGN, calibrating the SFR based on optical emission lines
with the WISE 12 $\mu$m luminosity.  Using these three relations applied to the WERGA galaxy sample W3 luminosities,
the Donoso et al. relation SFRs are consistent with the calibrated SFRs from this work (Eqs. 1 and 2).   Specifically, the star-formers (red diamonds) tend to the have
the highest SFRs and are 20 - 30\% larger than those derived in this work (Eq. 1), intermediates (green diamonds) are most consistent
with the WERGA sample, and the AGN types (blue diamonds) have the lowest SFRs, but still in line with the overall scatter in W3.
They point out that most of the 12 $\mu$m emission arises from the young stellar populations, $<$0.6 Gyr in age,
consistent with the M\,83 analysis comparing in detail the W3 distribution relative to the UV and radio emission (see Section 5).
For the earlier-type galaxies, however,  the AGN types have 12 $\mu$m emission
that is dominated by the older, evolved stars (including the AGB population), with ages between 1 and 3 Gyr.   As we show below, these
populations are an important component in the SSP (simple stellar population) models that are used to derive the underlying stellar mass that is traced
by the near-IR 1$-$5 $\mu$m emission.   The larger scatter seen in the SFR$_{IR}$ based on the WISE 12 $\mu$m
(or the IRAC 8 $\mu$m) emission relative
to the SFR based on the WISE 22 $\mu$m (or MIPS 24 $\mu$m) emission is a consequence of the different emission
mechanisms (AGN, PDR and stellar) and metallicity that augment the mid-IR PAH emission strength.
Nevertheless, in the absence
of reliable SF tracers (e.g., extinction-corrected H$\alpha$ luminosity, 22 or 24 $\mu$m luminosity,
L$_{TIR}$, or radio 20\,cm continuum), the SFR$_{IR}$ based on WISE 12 $\mu$m provides an adequate proxy for the
star formation activity.

\section{Stellar Mass Estimation}

Most of the baryons that comprise a galaxy are locked up in the evolved stellar population,  low mass ($\sim$solar) stars
that emit the bulk of their light in the 1 - 3 $\mu$m near-IR window.    A photometric census of this population of cool K and M giants, in combination with
the mass-to-light ratio (M/L), renders an estimate of the total baryonic stellar mass for a galaxy.   This simplistic prescription for deriving
the stellar mass is complicated by individual variations in the IMF, star formation history and population ages, metallicity, dust extinction,
AGB (notably the luminous thermal pulsating stars) contributions and nuclear activity.  
Nevertheless, there are many studies that have successfully employed both optical and near-IR (2 $\mu$m) observations (e.g., 2MASS)  to conduct statistical studies of the extragalactic stellar mass-to-light relation (see Bell et al. 2003; Zibetti, Charlot \& Rix 2009).  
With the advent of Spitzer-IRAC imaging of nearby galaxies,
the emphasis is now turning to the
mid-infrared 3 and 5 $\mu$m window of the R-J distribution, with the major advantages being the superior sensitivity to lower surface brightness emission
(thus capturing more of the total flux)
while less sensitive to the foreground dust extinction.   On the other hand, the disadvantage of the mid-IR relative to the near-IR
is the added sensitivity to light arising from
dust emission associated with star formation and with the evolved population of TP-AGB stars, both of which will
boost the mid-IR luminosity and thereby render an over-estimate of the stellar mass (e.g., Meidt et al. 2012).  As we shall see below,  there is a strong
dependence of the M/L ratio on the mid-IR galaxy color, but we caution the results are preliminary and further
analysis with a larger sample is paramount.

The WISE ``inband" luminosity, L$_{\lambda}$ is derived using
 using the equation.:  
 
\begin{eqnarray}
 L(band) \, / \, L_\odot \,&=& \, 10^{-0.4[M(band) \, - \, M_{\odot}(band)]} \,,
\end{eqnarray}
where
M(band) is the absolute magnitude of the source and M$_\odot$(band) is the absolute magnitude of the Sun as measured
in the band.  Employing the 2MASS and WISE band RSRs (Jarrett et al. 2010), the
spectral-energy distributions of the Sun and Vega (Fukugita et al. 1995), we derive (cf. Oh et al. 2008)
the absolute in-band magnitude of the Sun: 3.32, 3.24, 3.27, 3.23 and 3.25 magnitude for K$_s$, W1, W2, W3 and W4 respectively,
and are used to 
estimate the stellar mass in conjunction with the M/L ratio.


Recent work that has attempted to derive the M/L calibration for IRAC 3.6 $\mu$m and 4.5 $\mu$m (usually by bootstrapping from the near-infrared)
includes Li et al. (2007), Oh et al. (2008), Westmeier et al. (2011) and Zhu et al. (2010). The latter used SWIRE imaging and archival `reference' (stellar) masses that were estimated by combining SDSS photometric data with models from Bruzual \& Charlot 2003.  The mid-IR M/L that Zhu et al. (2010) present depend on the IRAC luminosities and, to correct for galaxy-to-galaxy variation in the star formation history, the optical (g-r) color (see Eq. 6 and 7 in their work).
Yet other studies (e.g., Zibetti, Charlot \& Rix, 2009), employ the latest SSP models of Charlot \& Bruzual (see Bruzual 2007) that have an updated treatment of the AGB contribution, although there remains significant disagreement  between observations and the new SSP models (e.g., Kriek et al. 2010; see more discussion on this topic below).   Another method that was applied to \Spitzer observations of the
Circinus Galaxy  (For, Koribalski \& Jarrett 2012), combines near-IR derived M/L with conversion to Spitzer wavelengths through color conversions and SSP models; as follows:  using the  M/L (K$_s$) relation derived from the analysis of
Bell \& de Jong (2001) and stated in Westmeier et al.  (2011; see their Eq. 8) and 
M/L near-IR to mid-IR transformations from Oh et al. (2008), the IRAC M/L relation in
terms of the (J-K) color follows:
{\scriptsize
\begin{eqnarray}
{\rm IRAC \,\, 3.6 \mu m: \,\,} && (M/L) \,[M_\odot / L_\odot] \, = \, \nonumber \\
&&[0.92\, \times \, 10^{(1.434\,(J-K_s)-1.380)}]-0.05\,.
\end{eqnarray}
}
The WISE equivalent of this relation follows by applying a small scaling factor (near unity)
to convert the WISE W1 flux to equivalent IRAC-1 flux based on the Hubble Type (see Fig. \ref{WISEvSpitzer}).
The WISE M/L becomes:
{\scriptsize
\begin{eqnarray}
{\rm WISE \,\, 3.4 \mu m: \,\,} && (M/L) \,[M_\odot / L_\odot] \, =  \, \nonumber \\
&&[(0.92/\zeta_1)\, \times\, 10^{(1.434\,(J-K_s)-1.380)}]-(0.05/\zeta_1)\,.
\end{eqnarray}
}


\smallskip
One of the most extensive M/L investigations comes from the
S4G study of nearby galaxies (e.g., Meidt et al. 2012;  see also
Eskew, Zaritsky \& Meidt, 2012),  utilizing sensitive IRAC 3.6 $\mu$m and 4.5 $\mu$m imaging from the `warm' \Spitzer Mission.  Employing a Chabrier-type IMF (e.g., Chabrier  2003) that is optimal for early type galaxies, the SSP models of Bruzual \& Charlot (2003),
and cross-calibrating 2MASS near-IR colors of K/M giants to those measured with Spitzer-GLIMPSE,
they derive a relation between IRAC-1 3.6 $\mu$m M/L  and the [3.6]-[4.5] color:
{\scriptsize
\begin{eqnarray}
{\rm IRAC \,\, 3.6 \mu m: \,\,} && Log \, (M/L) \,[M_\odot / L_\odot] \, = \, \nonumber \\
&&-0.22 \, + \, 3.42 ([3.6]-[4.5])\,.
\end{eqnarray}
}
Although these studies are in preliminary stages, the S4G relation appears to render results that are
consistent with other metrics (see below), notably for the early-type galaxies (i.e., dominant R-J emission).
The WISE equivalent of this relation follows by applying a small scaling factor
to convert the WISE magnitudes to equivalent IRAC magnitudes.  This entails using the zero point flux to magnitude conversion, and
Hubble Type and expected WISE-to-IRAC flux
ratio based on model templates are used to derive the band-to-band scaling factors.
The M/L would become:
{\scriptsize
\begin{eqnarray}
{\rm WISE \,\, 3.4 \mu m: \,\,} && Log \, (M/L) \,[M_\odot / L_\odot] \, = \, \nonumber \\
&&-0.75 \, + \, 3.42 [ (W1-W2) - 2.5\,log(\zeta_2 / \zeta_1)]\,,
\end{eqnarray}
}
where $\zeta_1$ accounts for the W1-to-IRAC-1 band-to-band differences, and
$\zeta_2$ accounts for the W2-to-IRAC-2 band-to-band differences per Hubble Type.
Using elliptical galaxies as an example, the
W1/IRAC-1 flux ratio is 1.06, and W2/IRAC-2 flux ratio is 0.94,
the WISE M/L relation is then:
{\scriptsize
\begin{eqnarray}
{\rm WISE \,\, 3.4 \mu m\, (early \, types): \,\,} && Log \, (M/L) \,[M_\odot / L_\odot] \, = \, \nonumber \\
&&-0.31 \, + \, 3.42 (W1-W2)\,.
\end{eqnarray}
}

\smallskip

\begin{figure*}[ht!]
\begin{center}
\vspace{-5pt}
\includegraphics[width=17.5cm]{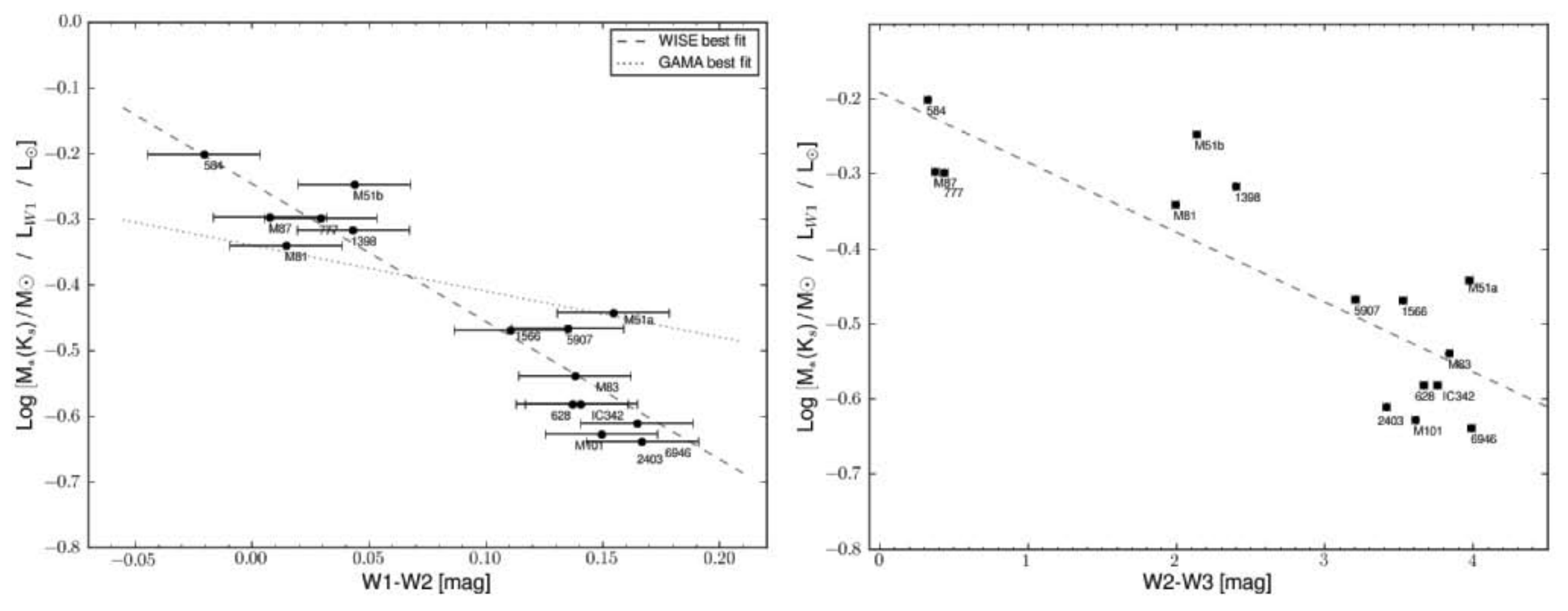}
\caption[WISE ML]
{\small{Empirical mass-to-light relation, [M$_\odot$ / L$_\odot$], derived from
WISE 3.4 $\mu$m in-band luminosity, W1-W2 color (left panel), W2-W3 color (right panel) and K$_s$-derived stellar
masses using the Zhu et al. (2010) relation (see also Fig. \ref{massdex}).  The color error bars represent the formal uncertainties
(Table \ref{tab:flux}) and a 1.5\% photometric calibration uncertainty.
The linear trend (dashed line) is
described by Eqs. 9 and 10.  The dotted line (left panel) represents the [M$_\odot$ / L$_\odot$]
derived for nearby galaxies in the GAMA survey (see text for details).
}}
\label{ML}
\end{center}
\end{figure*}

\begin{figure*}[ht!]
\begin{center}
\vspace{-5pt}
\includegraphics[width=15cm]{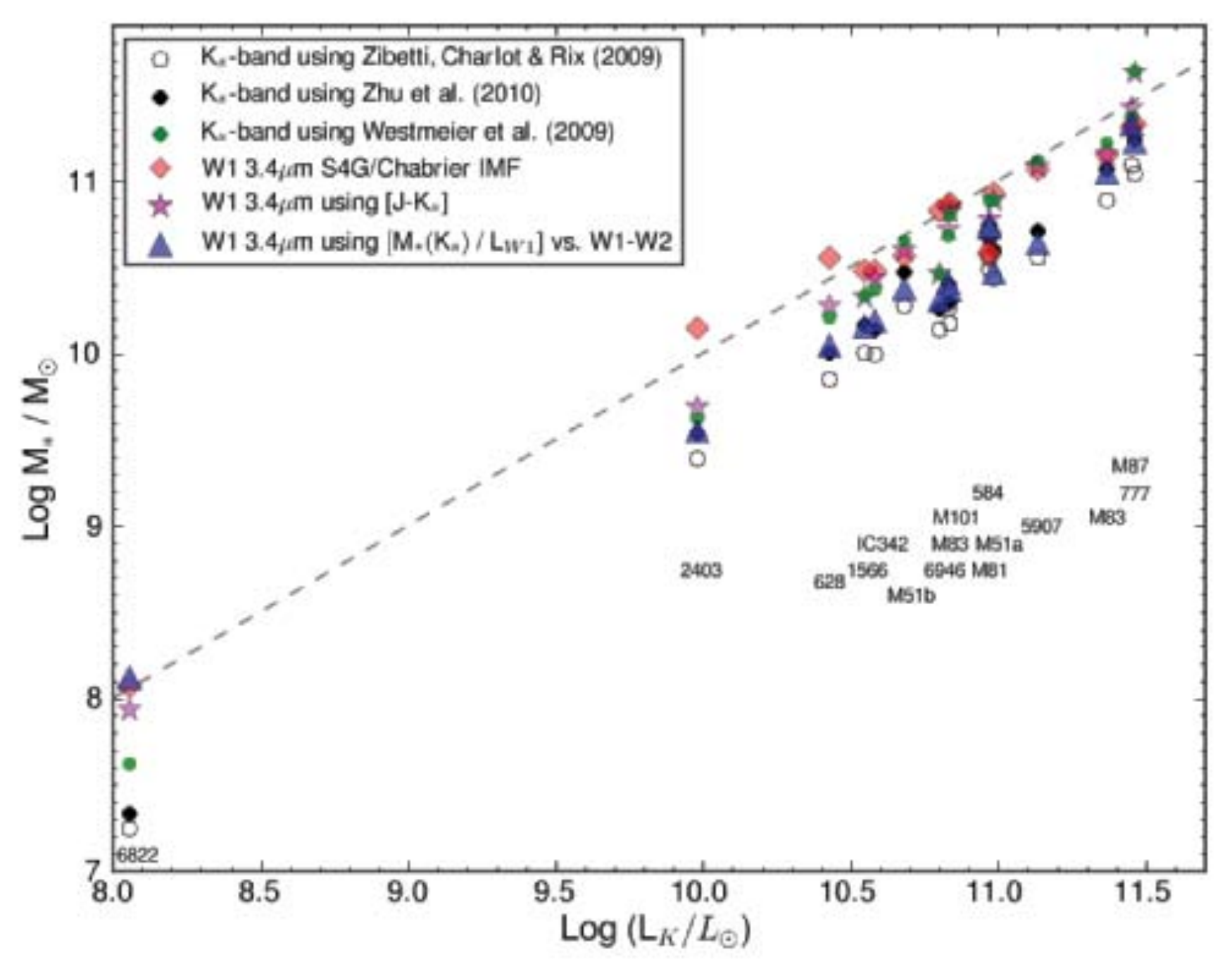}
\caption[WISE stellar mass]
{\small{Stellar mass compared to the K$_s$-band luminosity.
The masses are estimated from the K$_s$,
W1 and W2 luminosities and the M/L relations of
Zibetti, Charlot \& Rix (2009), Zhu et al. (2010),
 the S4G/Chabrier IMF relation (Eq. 7) and
 the M/L relations from this work (Eq. 4 and 9; Fig \ref{ML}).
For comparison, the dashed line represents a M/L fraction of unity.
}}
\label{massdex}
\end{center}
\end{figure*}

A more direct, but largely empirical method for using the WISE colors to adjust the M/L relation for star formation history effects, is
to adopt the stellar `mass' derived using the Ks-band in-band luminosity  and compare with the `light' traced by the
W1 (3.4 $\mu$m) luminosity.  This hybrid method then correlates the mass with the light in the sense that
the K and W1 bands are sampling the same stellar population. 
Accordingly, Fig. \ref{ML} presents the hybrid M$_*$(K$_s$)-to-L$_{W1}$ ratio
as a function of the W1-W2 color and the W2-W3 color, where the stellar mass is estimated using the
2MASS K$_s$ in-band luminosity (Jarrett et al. 2003) and the Zhu et al. (2010) relation (their Eq. 5)
that includes the (g-r) color correction\footnote{Estimated ($\sim$10 to 20\%) using SED model templates; see also the
Hubble Type vs. (g-r) color distribution
in James et al. (2008).}.
It should be stressed that since we have used the Zhu et al. (2010) for the K-band stellar masses, the resultant masses are smaller 
by $\sim$0.3 to 0.4 dex compared to those masses derived using,  for example (see Fig. \ref{massdex}a below), 
Bell et al. (2003), Leroy et al. (2008) and de Blok et al. (2008),
due to the Zhu et al. (2010) formulation which uses ``reference" stellar masses (see also Kannappan \&
Gawiser 2007).
 We see from Fig. \ref{ML} that the M/L has a clear linear trend with WISE color: 
the M/L is higher for the early types (ie., bulge-dominted blue galaxies) relative to the late-types (star forming, red galaxies).
This trend likely arises from extinction (e.g., dust geometry), metallicity and population age differences.
The linear equation
that best fits the sample distribution  (Fig. \ref{ML}, dashed lines)  is:
{\scriptsize
\begin{eqnarray}
{\rm WISE \,\, 3.4 \mu m: \,\,} && Log \, (M_*(K_s)/L_{W1}) \,[M_\odot / L_\odot] \, = \, \nonumber \\
&&-0.246 (\pm 0.027) \, - \, 2.100(\pm0.238) \, (W1-W2); \\
{\rm WISE \,\, 3.4 \mu m: \,\,} && Log \, (M_*(K_s)/L_{W1}) \,[M_\odot / L_\odot] \, = \, \nonumber \\
&&-0.192 (\pm 0.049) \, - \, 0.093(\pm0.016) \, (W2-W3).
\end{eqnarray}
}
The steepness in the M/L relation with WISE color over a relatively short range ( -0.02 $<$ W1$-$W2 $<$ 0.17 mag) 
may be influenced by the small sample of `normal' galaxies presented in this work.   Comparing to a much larger sample
of nearby ($z <$ 0.1) galaxies
extracted from the Galaxy and Mass Assembly (GAMA;  Driver et al. 2011) project from their field G15, the M/L
relation as a function of the WISE color is much flatter (see dotted line in Fig. \ref{ML}a; Cluver et al., in preparation).
The GAMA stellar masses are derived using optical (g-i) colors,  stellar population synthesis models
and careful Bayesian parameter estimation (Taylor et al. 2011),
all of which contribute to the observed WISE vs GAMA M/L slope difference; but moreover,
the GAMA G15 field has a greater variety of galaxy types,
covering a wide range in stellar population (notably giant and AGB  relative contributions), metallicity, IMF dust geometry,
nuclear activity
and star formation history.  We therefore strongly caution that the WISE M/L hybrid K-band relation presented in
Fig. \ref{ML} and Eqs. 9-10 should be viewed as preliminary and, at best, incomplete.


\medskip

Bringing it all together and comparing the different methods introduced here,
we estimate the stellar mass for our galaxy sample using the 2MASS K$_s$-band fluxes (Jarrett et al. 2003)
and WISE W1 and W2 fluxes in conjunction with the
Bell et al. (2003), Zibetti, Charlot \& Rix (2009),
Zhu et al. (2010), 2MASS color (Eq. 4),
S4G (Eq. 7) and hybrid near-IR-to-mid-IR (Fig. \ref{ML} ) M/L ratios that depend on the WISE colors.
As noted previously, for the mid-IR relations derived using IRAC photometry, we apply small scaling
factors to convert the WISE fluxes to equivalent IRAC fluxes.
The results are presented in Fig. \ref{massdex}, where the derived stellar mass is
plotted against the K$_s$ luminosity, currently the most reliable tracer of the evolved stellar population.

The observed scatter between the near-IR methods and those using the hybrid methods
is about 0.1 to 0.2 dex ($\sim$40\% RMS), with a systematic difference between the two
methods of about 0.3 to 0.5 dex.  The masses that are derived from the J-K colors tend to be larger
than those derived from the Zibetti, Charlot \& Rix (2009) and Zhu et al. (2010) (reference mass) treatments.
For example, the stellar mass of NGC 6946 is estimated to be $\sim$3.2$\times$10$^{10}$ M$_\odot$
using the near-infrared Bell et al. (2003) formulations (e.g., Leroy et al. 2008; de Blok et al. 2008), and
$\sim$1.5$\times$10$^{10}$ M$_\odot$ using Eq. 9.
Note that the WISE W1-W2 adjusted masses (Eq. 9) track closely to the near-IR stellar mass estimates of Zhu et al. (2010), which is to be expected
since the WISE+2MASS hybrid-relation employs the K$_s$ estimated mass.
The S4G/Chabrier values are closest to the Bell et al. (2003) result (M/L $\sim$ unity)
for massive galaxies.

Overall differences arise from the assumed IMF and the stellar population synthesis models; the latest generation include more sophisticated treatment of the AGB contribution (e.g., Maraston et al. 2006; Bruzual 2007; Charlot \& Bruzual 2007), although, for example, they
do not fit very well to the observed SEDs of post-starburst galaxies (cf., Kriek et al. 2010).  As of this writing, there remains large uncertainty in
modeling the mid-IR light contribution from star-forming and post-starburst galaxies, rendering equally large uncertainties in the IRAC M/L relations.
Consequently -- although there is close binding between the near-IR and
mid-IR luminosities -- the M/L relation that accounts for the galaxy color (Eq. 9) should have
the best correspondence
with the 2 $\mu$m estimates, and we thus adopt them as the stellar mass for our sample.

\section{Star Formation History}

\begin{figure*}[ht!]
\begin{center}
\vspace{-5pt}
\includegraphics[width=17.5cm]{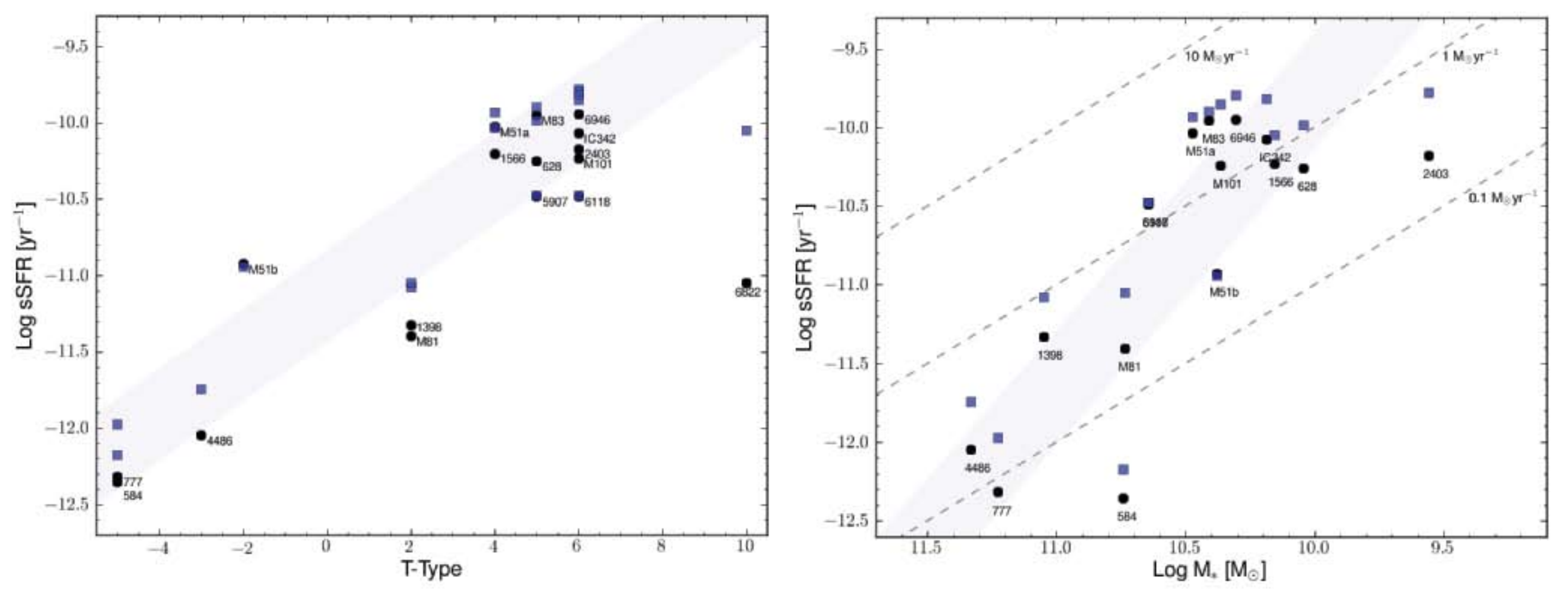}
\caption[WISE specific star formation]
{\small{Specific star formation rate
compared to the Hubble Type and stellar mass.
The stellar mass is derived using
the M/L relation from this work (Eq. 9).
Two global SFRs are shown:  black points correspond to the
SFR (22 $\mu$m; Eq. 2) and blue squares
correspond to the combined IR+UV SFR.
There is a similar linear trend
(illustrated by the blue shaded regions) for both Hubble Type and 
the host stellar mass.  A range in SFRs (from 0.1 to 10 M$_\odot$yr$^{-1}$)t
are represented by the dashed lines.
}}
\label{ssfr}
\end{center}
\end{figure*}

In this final section, we construct the scaling relation between the star formation and
the stellar mass of the galaxy sample.
Accordingly, using this mass (Eq. 9 and the W1 luminosity) in conjunction with the
SFR$_{IR}$ (22 $\mu$m; Eq. 2) and the combined IR+UV SFR, we derive the global specific star formation rate (sSFR),  gauging the
present-to-past star formation history.  Fig. \ref{ssfr} presents the sSFR compared to the Hubble type
and to the stellar mass content.  The sSFR derived using the total SFR compared to the infrared SFR
is slightly higher ($\sim$0.3 dex), yet both exhibit a similar trend with morphology, gas mass and stellar content.

\begin{figure*}[ht!]
\begin{center}
\includegraphics[width=17.5cm]{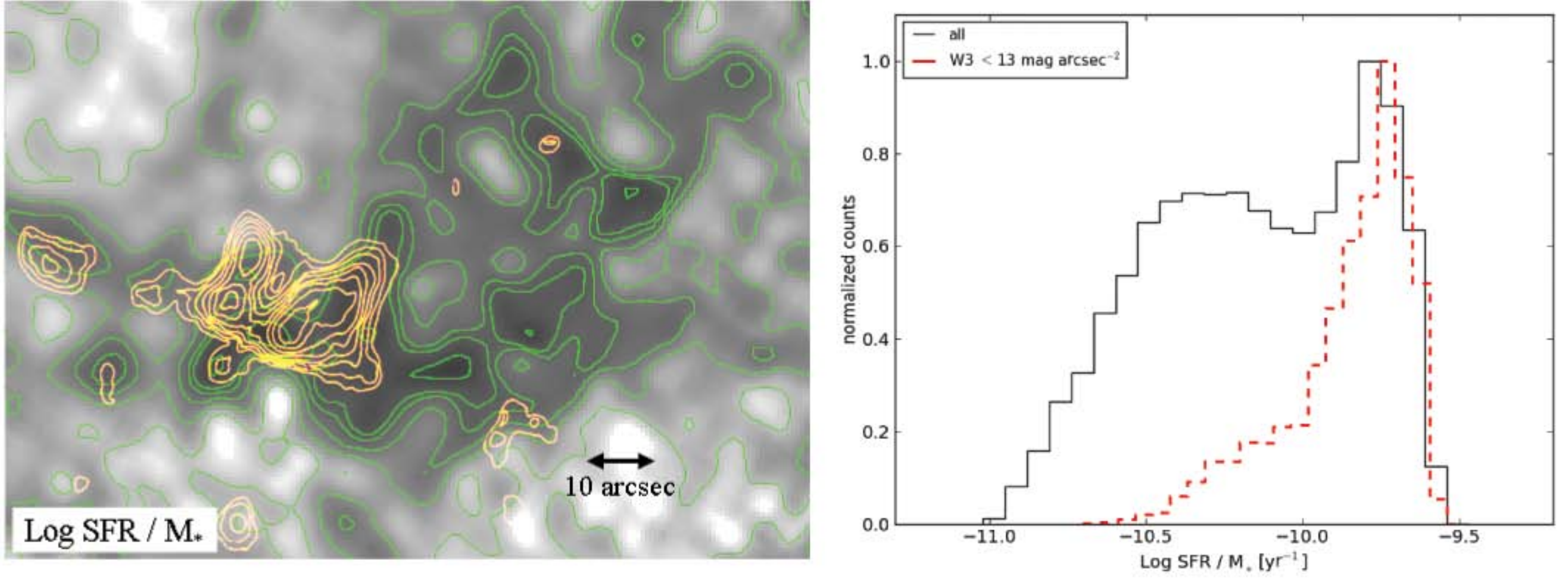}
\caption[sSFRzoom]
{\small{Specific star formation rate in the
the southwestern bar cusp of M83.
(Left)  The sSFR is derived from the
W1 (stellar mass) and W3 (SFR) imaging.  The green contours range from
-10.5 to -9.75 for Log SFR/M$_*$ [yr$^{-1}$].
The orange contours correspond to the molecular gas (see Fig \ref{M83zoom});
(Right) Histogram showing the sSFR distribution for the region.
The red dashed line denotes the 12 $\mu$m high surface brightness knots:
$<$ 13 mag (0.2 mJy) arcsec$^{-2}$.
}}
\label{sSFRzoom}
\end{center}
\end{figure*}

Reminiscent of the segregation observed in the WISE color-color plot (Fig. \ref{colors}), there are three distinct groupings:
early types ellipticals have very low sSFRs, early-type spirals have moderate sSFR,
and late-type spirals have high sSFRs.   Elliptical galaxies have exhausted their fuel supply, little if any star formation is
happening and thus there is no growth in the total stellar mass.  At the other extreme, late type spirals (e.g., M\,83) are
actively forming stars from molecular hydrogen (with the neutral hydrogen tracking the gas content),
building their stellar backbone and bulge populations (the cycle is graphically depicted in Fig. 14).  
On smaller scales, indeed this is observed in Fig. \ref{sSFRzoom},
a revisit of the southwest bar transition of M\,83 (see Fig. \ref{M83zoom} for details), but now converting
the W3 image to SFR and the W1 image to stellar mass to derive the sSFR at parsec-physical scales.
Overlayed (in yellow/orange) are the 12CO contours, tracking the molecular gas.  Both the gas and
the highest sSFR regions are coincident, with values that peak
at Log SFR/M$_* = -9.75$, as shown in the red-dashed histogram.    As would be expected, the
molecular gas is fueling the massive star formation, indirectly traced by W3 through PAH emission arising
from PDRs.  For the local universe, the existence of a tight scaling relation between the stellar mass (star formation 
history)
and the star formation rate (current activity) is a vital clue that 
the formation of stars (and hence galaxy evolution) is regulated by secular (physical) processes that
that are active over long (Gyr) periods of time.
Morever, 
this behavior where high mass galaxies form fewer stars per unit stellar
mass compared to less massive systems is typically found to exist for $z<$2,
with an overall decline in SSFR with decreasing redshift (for example,
Daddi et al. 2007; Noeske et al. 2007; Pannella et al. 2009, Karim et al. 2011).  The shift of
star formation efficiency from higher mass systems in the past to lower
mass systems observed locally is often referred to as "cosmic downsizing"
(Cowie et al. 1996).   

This observed
difference in star formation efficiency between Hubble types, gas reservoir and stellar content
is consistent with the large sample results of Donoso et al. (2011),
who compared the sSFR of star-forming, intermediate and bulge-dominated AGN-type galaxies,
as well as the GASS/COLDGASS analysis (Saintonge et al., 2011; 2012) that explored the relationship
between the atomic and molecular gas components and the star formation histories using SDSS and GALEX.
Finally, we note that the simple WISE color diagram (Fig \ref{colors}a) has a similar
behavior as the derived sSFR diagram because the colors, in fact, are tracing the same evolutionary states.
That is, the W1-W2 vs W2-W3 colors in (Fig \ref{colors}a) essentially capture the specific star formation rate.

\section{Summary}

In this paper we have presented the results of a mid-infrared and ultraviolet photometric study of 17
large, nearby galaxies,
using the WISE, \Spitzer and GALEX space telescopes.
The primary science goals were to (1) characterize and assess the
quality of source extraction for resolved galaxies observed by WISE,
(2) to validate the WISE Enhanced Resolution Galaxy Atlas measurements by comparison with those using \Spitzer imaging,
(3) to derive the star formation activity and (3) the stellar mass content for the sample.  For the galaxy M\,83 (NGC\,5236),
we examined in detail
the distribution of young and old stars, star formation activity, and gas content.
We highlight here our main results from the M\,83 analysis and from the larger sample:

\begin{itemize}

\item Employing an MCM-HiRes deconvolution technique, we have reconstructed WISE images that have
an improvement in angular resolution that is approximately a factor of three to four relative to the nominal (Atlas) WISE imaging,
and comparable (to within 30\%) in resolution to that of \Spitzer-IRAC and \Spitzer-MIPS24.  A more complete description and demonstration of MCM
is given in Paper I.

\item The typical 1\,$\sigma$ isophotal surface brightness for the WISE W1, W2, W3 and W4 bands, respectively,
is 21.8, 18.1 and 15.8 mag arcsec$^2$ (Vega).  Azimuthal averaging achieves surface brightnesses that
reach depths of 24.0, 23.0, 19.4 and 18.5 mag arcsec$^2$ for bands W1, W2, W3, and W4 respectively,
equivalent to AB mags of 26.7, 26.3, 24.6 and 25.1 mag arcsec$^2$, respectively.

\item The photometric performance using the 
MCM-HiRes reconstructed WISE imaging appears to be of high quality as 
gauged by detailed comparison with
2MASS, \Spitzer and IRAS photometric properties, augmented with SED analysis using population synthesis models.
We caution that there remain cosmetic (negative depression) artifacts enveloping high surface brightness sources (e.g., bright stars; bright
nuclei), particularly with W4 (22 $\mu$); see Paper I for further discussion of these 'ringing' artifacts.

\item For the barred spiral galaxy M\,83, the direct comparison between WISE and GALEX of the two-dimensional distribution of
the warmed ISM and the
massive star population reveals the obscured (WISE) and unobscured (GALEX) sites for star formation.
Intriguingly,  the neutral H\,{\sc i} gas correlates strongly with the 11.3 $\mu$m PAH emission,
traced by the WISE 12 $\mu$m band, which is usually
associated with the molecular gas that fuels star formation --  the H\,{\sc i} may be by-product of molecular hydrogen that has
been dissociated by the strong radiation fields arising from the massive star formation in the spiral arms of M\,83.

\item Employing the 2MASS and WISE band RSRs, and the
spectral-energy distributions of the Sun and Vega, we derive
the absolute in-band magnitude of the Sun: 3.32, 3.24, 3.27, 3.23 and 3.25 magnitude for K$_s$, W1, W2, W3 and W4 respectively.

\item We have calibrated the infrared global SFR relation that is appropriate to the WISE 12 and 22 $\mu$m bands using
the \Spitzer 24 $\mu$m relation as a bootstrap.    In combination with the GALEX luminosities, computed SFRs
that encompass both the obscured (traced by WISE) and unobscured (GALEX) present-day star formation activity.
We find that the SFR relative to the neutral gas content follows a simple linear trend in which the larger the gas reservoir
the higher the SFR, consistent with the K-S scaling relation.

\item The aggregate stellar mass is estimated using the 3.4 and 4.6 $\mu$m bands of WISE.  We have derived several
variations of the IRAC and WISE M/L relations, bootstrapped from the well-studied K$_s$-band relations and from
population synthesis models that relate the mid-IR light that arises from the evolved stellar population to
the total stellar content.

\item Combining the global SFR and the stellar content, we investigate the Specific Star Formation Rate,
gauging the present-to-past star formation history for the sample of galaxies.
We find a clear scaling relation between the sSFR and the with Hubble Type and the gas content:
early type galaxies have exhausted their fuel supply, there is either slow or no growth in the total stellar mass. Late type spirals and galaxies with
large gas reservoirs, are actively forming stars from molecular hydrogen and building their stellar disk and bulge populations.

\item We discuss the construction of WERGA that comprises a complete, volume-limited sample of the local universe.
The strengths of WISE to trace the stellar mass and the star formation activity mean the WERGA will
play a crucial and complementary role in the multi-wavelength effort to understand galaxy assembly and evolution.

\end{itemize}

\bigskip

\noindent Acknowledgements

We thank G. Meurer, S. Lord, J. Mazzarella and B. Madore for tapping their vast knowledge base of nearby galaxies.
Discussions with S. Meidt and N. Taylor were very helpful in understanding
the (on-going) difficulties with M/L modeling.
This work is based [in part] on observations made with the {\it Spitzer} and research using the  NASA/IPAC Extragalactic Database (NED) and
IPAC Infrared Science Archive,
all are operated by JPL, Caltech under a contract with the National Aeronautics and Space Administration.
Support for this work was provided by NASA through an award issued by JPL/Caltech.
R.J.A. was supported by an appointment to the NASA Postdoctoral Program at the Jet Propulsion Laboratory, administered
by Oak Ridge Associated Universities through a contract with NASA.
MEC acknowledges support from the Australian Research Council (FS110200023).
This publication makes use of data products from the Wide-field Infrared Survey Explorer, which is a joint project of the University of California, Los Angeles, and the Jet Propulsion Laboratory/California Institute of Technology, funded by the National Aeronautics and Space Administration.

\section{References}


\noindent Aumann, H. H., Fowler, J. W., \& Melnyk, M. 1990, AJ, 99, 1674\\
\noindent Bell, E. F. \& de Jong, R. S. 2001, ApJ, 550, 212\\

\noindent Bell, E., McIntosh, D., Katz, N., Weinberg, M.  2003, ApJS, 149, 289\\
\noindent Berta, S. et al. 2003, A\&A, 403, 119\\
\noindent Bohnensiengel, H. \& Huchtmeier, W. 1981, AA, 100. 72\\
\noindent Brandl, B.R., et al., 2006, ApJ, 653, 1129\\
\noindent Briggs, F.H. 1982, AJ 259, 544\\
\noindent Bruzual, G., 2007, astro-ph/0702091\\
\noindent Buat, V., et al., 2011, A\&A, 529, A22\\

\noindent Buat, V., et al., 2008, A\&A, 483, 107\\
\noindent Calzetti, D., et al., 2007, ApJ, 666.\\
\noindent Calzetti, D., 2011, EAS Publications Series, 46, 133\\
\noindent Cardelli, J., Clayton, G., \& Mathis, J., 1989, ApJ, 345, 245\\
\noindent Chabrier, G. 2003, PASP, 115, 763\\
\noindent Chynoweth, K., Langston, G., Holley-Bockelmann, K, Lockman, F., 2009, AJ 138, 287\\
\noindent Cluver, M., et al., 2010, ApJ, 725, 1550\\
\noindent Cowie, L.L., Songaila, A., Hu, E.M., \& Cohen, J.G.  1996, AJ, 112, 839\\
\noindent Crosthwaite, L., et al., 2002, AJ, 123, 1892\\
\noindent Cutri, R., et al., 2011, WISE Explanatory Supplement\\
\noindent Daddi, E., Dickinson, M., Morrison, G., et al., 2007, ApJ, 670, 156\\
\noindent Dale, D., et al. 2007, ApJ 655 863\\
\noindent de Blok, W.J.G., Walter, F., Brinks, E., Trachternach, C., Oh, S. \& Kennicutt, R., 2008, ApJ, 136, 2648.\\
\noindent Donoso, E., et al., 2011, ApJ, submitted.\\
\noindent de Vaucouleurs, G., et al., 1991, Third Reference Catalogue of Bright Galaxies (RC3), Springer-Verlag: New York\\
\noindent de Vaucouleurs, G., 1994, {\it Quantifying Galaxy Morphology at High Redshift}, STSCI workshop.\\
\noindent Dong, H. et al., 2008, AJ, 136, 479\\
\noindent Draine, B.T.  2011, EAS Publications Series, 46, 29\\
\noindent Driver, S.P., et al., 2011, MNRAS, 413, 971.\\
\noindent Elbaz, D., Daddi, E., Le Borgne, D., et al. 2007, A\&A, 468, 33\\
\noindent Eskew, M., Zaritsky, D. \& Meidt, S.  2012, AJ, submitted.\\
\noindent Flaherty, K.M., et al., 2007, ApJ, 663, 1069\\
\noindent For, B.Q., Koribalski, B.S., \& Jarrett, T.H. 2012, MNRAS, in press (astro-ph/1206.4102)
\noindent Fowler, J. W., \& Aumann, H. H. 1994, in Science with High-Resolution Far-Infrared Data, ed. S. Terebey \& J. Mazzarella (JPL Publication 94-5), 1.\\
\noindent Gil de Paz, A., et al. 2007, ApJS, 173, 185\\
\noindent Gordon, K., et al., 2008, ApJ, 682, 336\\
\noindent Goto, T., et al., 2010, arXiv:1008.0859v1\\
\noindent Helou, G., et al., 2004, ApJS, 154, 253\\
\noindent Herrmann, K., et al., 2008, 683, 630\\
\noindent Hunter, D.A, Baum, W., O'Neil, E. \& Lynds, R.  1996, 468, 633\\
\noindent Indebetouw, R., et al., ApJ, 619, 931\\
\noindent James, P.A., Knapen, J.H., Shane, N.S., Baldry, I., de Jong, R.  2008, A\&A, 482, 507  
\noindent Jarrett, T.H, Chester, T., Cutri, R., Schneider, S. \& Huchra, J. 2003, AJ, 125, 525\\
\noindent Jarrett, T.J., et al., 2011, ApJ, 735, 112\\
\noindent Jarrett, T.J., et al., 2012, AJ (Paper I;  submitted)\\
\noindent Kanbur, S.M., et al., 2003, AA, 411, 361\\
\noindent Kannappan, S. J., \& Gawiser, E. 2007, ApJ, 657, L5\\
\noindent Karim, A., Schinnerer,E., Mart{\'{\i}}nez-Sansigre, A., et al., 2011, ApJ, 730, 61\\
\noindent Kenney, J. D. P., \ Lord, S. D. 1991, ApJ, 381, 118\\
\noindent Kennicutt, R. C. Jr. 1989, ApJ, 344, 68\\
\noindent Kennicutt, R. 1998, ARAA, 36, 189\\
\noindent Kennicutt, R. et al. 2003 PASP, 115, 928\\
\noindent Kennicutt, R. et al. 2009, ApJ, 703, 1672\\
\noindent Kilborn, V., Koribalski, B., Forbes, D, Barnes, D., Musgrave, R.  2004, MNRAS, 356, 77\\
\noindent Kohno, TBD. et al., 2004, PASJ, 54.\\
\noindent Kriek, M., et al., 2010, ApJ, 722, 64\\
\noindent Laine, S. et al., 2006, AJ, 131, 701\\
\noindent Larsen, S. et al. 2001, AJ, 121, 2974\\
\noindent Leroy, A., et al, 2008, AJ, 136, 2782.\\
\noindent Li, H.-N., Wu, H., Cao, C., \& Zhu, Y.-N. 2007, AJ, 134, 1315   
\noindent Lord, S. D. \& Kenney, J. D. P. 1991, ApJ, 381, 130\\
\noindent Lundgren, A. A., Wiklind, T., Olofsson, H., \& Rydbeck, G. 2004, A\&A, 413, 505\\
\noindent Malin, D. \& Hadley, B. 1997, PASA, 14, 52\\
\noindent Maraston, S. T., et al. 2006, ApJ, 652, 85\\
\noindent Martin, D. C., et al. 2005, ApJ, 619, L1\\
\noindent Masci, F.J., \& Fowler, J.W.,  in Proceedings of Astronomical Data Analysis Software and Systems XVIII, Québec City, ASP Conference Series, Edited by D. Bohlender, P. Dowler, and D. Durand, Vol. 411, 2009, p.67\\
\noindent Masters, K., Giovanelli, R., \& Haynes, M. 2003, ApJ, 126, 158\\
\noindent Meidt, S.E., et al., 2012, arXiv:1203.0467v1.
\noindent Morrissey, P., et al. 2005, ApJ, 619, L7\\
\noindent Morrissey, P., et al. 2007, ApJS, 173, 682\\
\noindent Murakami, H., et al., 2007, PASJ, 59, S369.\\
\noindent Noeske, K. G., Weiner, B. J., Faber, S. M., et al., 2007, ApJ, 660, L43\\
\noindent Norris, R., et al., 2011, A\&A, submitted\\
\noindent Oh, S.-H., et al., 2008, AJ, 136, 2761\\  
\noindent Oey, M., 2011, ApJ, 739, L46\\
\noindent Pannella, M., Gabasch, A., Goranova, Y., et al., 2009, ApJ, 701, 787\\
\noindent Paturel, G., et al., 2002, A\&A, 389, 19\\
\noindent Pohlen, M., et al., 2003, ASP Conf. Ser 327 Satellites and Tidal Streams (astro-ph/0308142)\\
\noindent Polletta, M. et al. 2006, ApJ, 642, 673\\
\noindent Polletta, M. et al. 2007, ApJ, 663, 81\\
\noindent Rand, R. J., Lord, S. D., \& Higdon, J. L. 1999, ApJ, 513, 720\\
\noindent Reach et al. (2005)]{Reach05} Reach et al. 2005, PASP, 117, 978\\
\noindent RRelano, M., Lisenfeld, U., P\'erez-Gonz\'alez, P., Vílchez, J.,  \& Battaner, E., 2007, ApJ, 667, L141\\
\noindent Rieke, G. H., et al., 2009, ApJ, 692, 556\\
\noindent Rogstad, D., 1971, AA, 13, 108\\
\noindent Rujopakarn, W., Rieke, G., Weiner, B., Rex, M., Walth, G., Kartaltepe, J.  2011, ApJ, submitted (astro-ph/2921v1)\\
\noindent Saha, A., Claver, J. \& Hoessel, J. 2002, AJ, 124, 839\\
\noindent Saintonge, A., Kauffmann, G., Wang, J., et al. 2011, MNRAS, 415, 61\\
\noindent Saintonge, A., Tacconi, L., Fabello, S., et al. 2012 (astro-ph 1209.0476v1)\\
\noindent Shang Z., Zheng, Z., Brinks, E.1988, AJ, 504, L23\\
\noindent Sheth, K., et al, 2010, PASP, 122, 1397\\
\noindent Shetty, R. \& Ostriker, E., 2006, ApJ, 647, 997\\
\noindent Shi, Y., Rieke, G., Hines, D., Gordon, K., \& Egami, E., 2007, ApJ, 655, 781\\
\noindent Schiminovich, D., et al., 2007, ApJS, 173, 315\\
\noindent Silva, L. et al. 1998, ApJ, 509, 103\\
\noindent Taylor, E.N., et al. 2011, MNRAS, 418, 1587\\
\noindent Thilker, D., et al. 2005    ApJ, 619, 67\\
\noindent Thilker, D., et al. 2007, ApJS, 173, 572\\
\noindent Tielens, A. \& Hollenbach, D. 1985, ApJ, 291, 722\\
\noindent Tielens, A., 2008, ARAA, 46, 289\\
\noindent Tilanus, R. P. J. \& Allen, R. J.,  1993, A\&A, 274, 707\\
\noindent Tonry, J.L., et al. 2001, ApJ, 546, 681\\
\noindent Treyer, M., et al., 2007, ApJS, 173, 256\\
\noindent Treyer, M., Johnson, B., Schiminovich, D., \& O'Dowd, M., 2010 (astro-ph/1005.13164)\\
\noindent Tully, R.B. 1988, NEARBY GALAXY CATALOG\\
\noindent Tully, R.B., et al., 2009, AJ, 138, 323\\
\noindent Westmeier, T., Braun, R., \& Koribalski, B., 2011, MNRAS, 410, 2217\\
\noindent Willick, J.A., et al., 1997, ApJS, 109, 333\\
\noindent Wolfire, M. et al. 2003, ApJ, 587, 278\\
\noindent Wright, E. et al., 2010, AJ, 140, 1868\\
\noindent Zhu, Y.-N, Wu, H., Li, H., Cao, C., 2010, arXiv:1001.2627v1  
\noindent Zibetti, S., Charlot, S, Rix, H.  2009, MNRAS, 400, 1181\\

\newpage
\pagebreak

\section{Appendix A:  \Spitzer and IRAS Photometry}

This section presents both the ancillary photometry measurements and
the direct comparisons with those of WISE.  Extending the analysis presented
in Paper I, the objective is to validate the
spatially de-convolved WISE imaging global and surface brightness measurements using
ancillary infrared observations.

\subsection{A.1  \Spitzer Source Characterization}

In order to directly compare the WISE isophotal aperture photometry with that of {\it Spitzer},
the same aperture size and shape used for the WISE measurements (Table \ref{tab:flux})
are applied to the IRAC and MIPS-24 measurements.  The resulting photometry is presented
in Table \ref{tab:spitflux}.  These tabular results
do not include any Galactic or internal extinction corrections.

Measurements are carried out for those galaxies with post-BCD mosaics
available from either the SINGS archive or the \Spitzer Heritage Archive.   In a few cases the resulting
mosaic images are too small to extract the integrated flux (using the WISE fiducial aperture) or to measure a reliable local background.
NGC\,1398 and NGC\,6118 were only observed during the \Spitzer Warm Mission, and thus only
the IRAC-1 and IRAC-2 data were available.  In the case of NGC\,4486 (M\,87), only the MIPS-24 mosaics
were adequate in size to extract a complete flux.
For the case of IC\,342, the IRAC imaging does adequately cover the
field, however the nucleus of IC\,342 is so bright that it saturated the 12s exposures.   Consequently we used the
short-exposure (HDR) images to measure the IRAC-1 and IRAC-2 mosaics.  For IRAC-3 and IRAC-4, since they were
too faint to measure with the short HDR imaging, we used saturation recovery methods (developed by the \Spitzer
Science Center) to rectify the saturated nucleus of IC\,342 which is so bright at these wavelengths that
the light is dominated by the unresolved nucleus.

\begin{table*}
{\scriptsize
\caption{{\it Spitzer} Fixed-Aperture$^1$ Photometry  \label{tab:spitflux}}
\begin{center}
\begin{tabular}{r r r r r r r r r r r}

\hline
\hline
\\[0.25pt]

Name   & IRAC-1 & IRAC-2 & IRAC-3 & IRAC-4 & MIPS-24\\
           &   (Jy) & (Jy) & (Jy) & (Jy) & (Jy)   \\

\hline
\\[0.25pt]

NGC\,584&0.347$\pm$0.007&0.216$\pm$0.004&0.142$\pm$0.004&0.070$\pm$0.002&0.019$\pm$0.001\\
NGC\,628&0.822$\pm$0.017&0.589$\pm$0.012&1.025$\pm$0.028&3.344$\pm$0.083&2.882$\pm$0.051\\
NGC\,777&0.167$\pm$0.004&0.112$\pm$0.003&0.095$\pm$0.008&0.037$\pm$0.005&0.008$\pm$0.004\\
NGC\,1398&0.794$\pm$0.016&0.514$\pm$0.010&--&--&--\\
NGC\,1566&0.705$\pm$0.014&0.485$\pm$0.010&0.861$\pm$0.022&2.047$\pm$0.051&2.851$\pm$0.050\\
NGC\,2403&1.610$\pm$0.033&1.083$\pm$0.021&1.854$\pm$0.049&3.996$\pm$0.099&5.536$\pm$0.098\\
NGC\,3031&10.535$\pm$0.213&6.644$\pm$0.130&6.111$\pm$0.164&6.216$\pm$0.162&5.296$\pm$0.094\\
NGC\,4486&--&--&--&--&0.185$\pm$0.004\\
NGC\,5194$^2$&2.611$\pm$0.053&1.748$\pm$0.034&4.082$\pm$0.105&10.912$\pm$0.271&12.757$\pm$0.226\\
NGC\,5195$^2$&1.120$\pm$0.023&0.712$\pm$0.014&0.667$\pm$0.017&0.970$\pm$0.024&1.434$\pm$0.025\\
NGC\,5236&6.083$\pm$0.123&4.048$\pm$0.079&6.350$\pm$0.168&22.986$\pm$0.572&42.155$\pm$0.745\\
NGC\,5457$^3$&2.455$\pm$0.050&1.639$\pm$0.032&2.891$\pm$0.081&7.508$\pm$0.187&8.340$\pm$0.147\\
NGC\,5907&0.760$\pm$0.015&0.519$\pm$0.010&0.834$\pm$0.021&1.882$\pm$0.046&1.713$\pm$0.030\\
NGC\,6118&0.165$\pm$0.003&0.114$\pm$0.002&--&--&--\\
NGC\,6822&1.832$\pm$0.037&1.255$\pm$0.025&0.985$\pm$0.027&1.179$\pm$0.032&1.448$\pm$0.026\\
NGC\,6946&3.157$\pm$0.064&2.187$\pm$0.043&5.096$\pm$0.134&13.221$\pm$0.329&20.177$\pm$0.357\\
IC\,342$^4$&7.680$\pm$0.155&5.168$\pm$0.101&7.806$\pm$0.218&22.219$\pm$0.553&39.486$\pm$0.698\\

\hline
\end{tabular}
\end{center}
\tablecomments{$^1$Aperture size, axis ratio and orientation are matched to WISE;  see Table \ref{tab:flux}
for coordinate position and aperture details.  $^2$Photometry of NGC\,5194/5 is uncertain due to blending.
$^3$For comparison, Gordon et al. (2008) measured the following {\it uncorrected} IRAC fluxes for M\,101:
2.84, 1.76, 3.69 and 7.26 Jy, IRAC-1, 2, 3 and 4, respectively.  The MIPS-24 flux is 10.5 Jy. Both the IRAC
and MIPS-24 measurements are within 5\% of the {\it uncorrected} measurements of this work.
$^4$IRAC imaging of IC342 saturated in the core, recovered using
short-exposure HDR images.
Measurements are not corrected for Galactic or internal extinction.}

}

\end{table*}

Most of the galaxies in the sample have integrated fluxes that have been previously published, notably from
Dale et al. (2007) for the SINGS sample, and Gordon et al. (2008) for M\,101.    Even though there has
been no attempt to match apertures with the Dale et al. (2007) measurements, there is relatively good agreement
between their flux densities and those of the isophotal photometry.
We find that for most sources, the agreement
is better than 10\%, with a few notable exceptions, NGC\,584, NGC\,6822 and M\,51b.   
The early-type galaxy NGC\,584 is much brighter
in the IRAC-4 and MIPS-24 measurements of the SINGS extraction, likely due to using an aperture
that is too large for relatively weak R-J emission in these bands; see Section 4.2 for discussion.
Similarly, a smaller aperture (compared to SINGS) was used to extract the W4 and MIPS-24
photometry for NGC6822, resulting in a much smaller flux by comparison.  NGC\,6822 is
the most difficult galaxy to characterize due to its low surface brightness, flat-profile morphology,
and proximity to the Galactic Plane with its associated foreground star and dust emission contamination.
Another outlier in the comparison,
NGC\,5195 (M\,51b) is much fainter
in the SINGS extraction, likely due to the complexity and uncertainty with de-blending
M\,51b from the larger M\,51a.  
As a final comparison, we note that for the giant spiral galaxy, M\,101, there is
excellent agreement with photometry (where the aperture measurements are uncorrected) reported by Gordon et al. (2008);
see Table \ref{tab:spitflux} notes for details.

\subsection{A.2  Comparing WISE with \Spitzer and IRAS}

Averaging over large scales, a comparison between the WISE and \Spitzer imaging
is provided by the mean radial profiles for each galaxy in each band,
plots are included in Appendix C.  The IRAC/MIPS-24 profile (magenta line)
typically matches closely with the WISE profile (black line), with offset differences
due to the bandpass differences.  For a more quantitative global analysis, we turn to
the integrated fluxes.

\begin{figure*}[ht!]
\begin{center}
\vspace{-5pt}
\includegraphics[width=17.5cm]{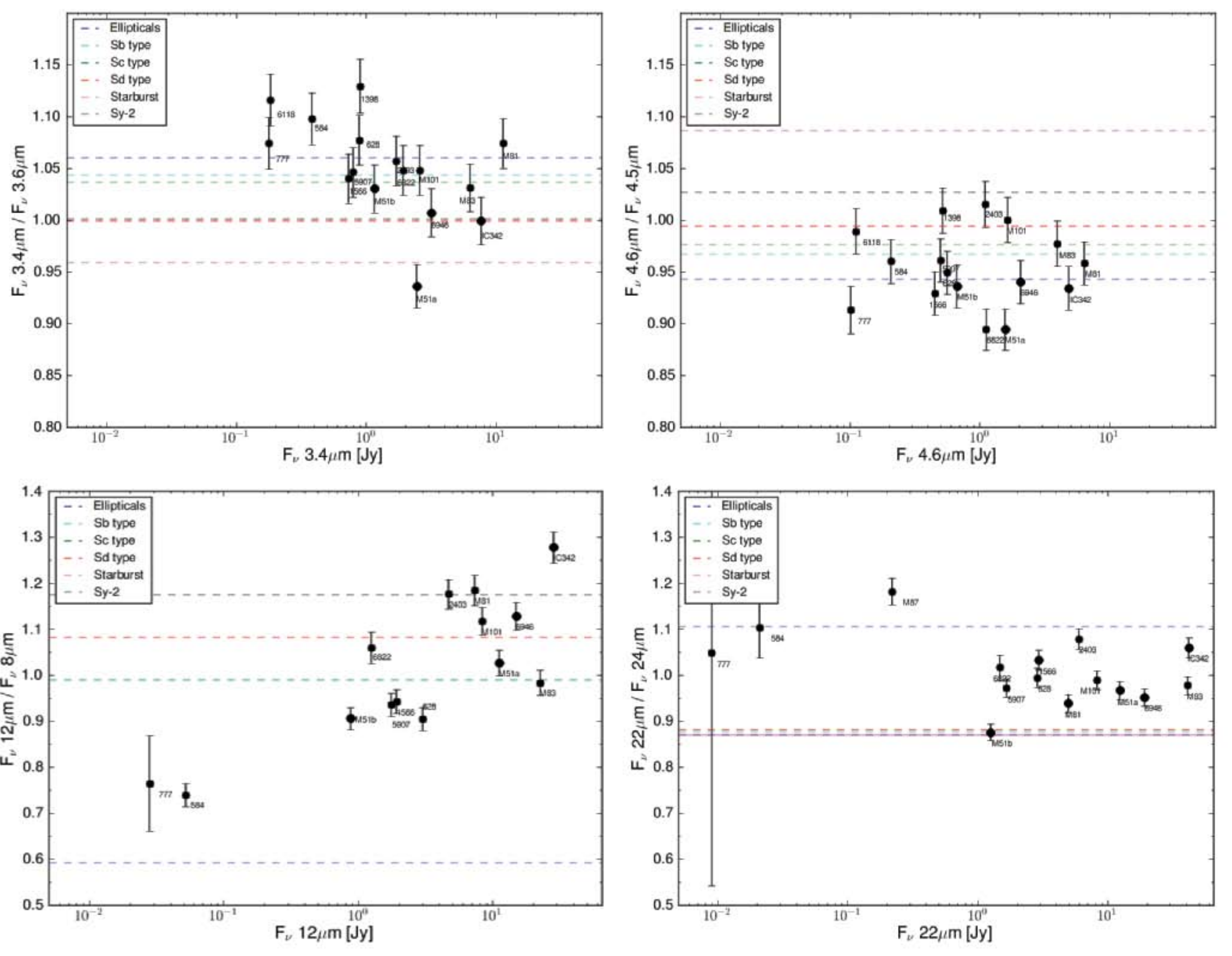}
\caption[WISE vs Spitzer]
{\small{WISE photometry compared to Spitzer IRAC and MIPS-24.
The isophotal apertures have been matched between WISE and Spitzer
in most cases (filled circles);  the open-circle symbols denotes those Spitzer measurements in which the mosaic was
too small to capture the total flux, or when the source photometry may be compromised (blending or source confusion).  The galaxy name is indicated next to the measurement, where single numbers represent the NGC \#.
The horizontal lines represent the expect flux ratios for specified Hubble types (see legend),
derived from model templates integrated over the WISE and Spitzer RSRs.
}}
\label{WISEvSpitzer}
\end{center}
\end{figure*}

Directly comparing the WISE isophotal photometry with the aperture-matched photometry of
IRAC and MIPS-24 requires taking into account the bandpass differences between the
two infrared missions.   Computing synthetic photometry of model galaxy SED templates
(GRASIL: Polletta et al. 2006 \& 2007; Silva et al. 1998)
from the WISE and \Spitzer bandpass RSRs (Jarrett et al. 2011),
we are able to predict the integrated flux ratio between
WISE and \Spitzer photometry for a range of Hubble Types.
Fig. \ref{WISEvSpitzer} presents WISE versus \Spitzer
photometry flux ratios, which are compared to predicted ratios
from early-type, late-type and starburst galaxies.
As noted in Section 3, color corrections for both WISE and
\Spitzer sources have been applied.

Comparing W1 3.4 $\mu$m to that of IRAC-1 3.6 $\mu$m,
W1 will tend to be brighter than IRAC-1
since the WISE band is relatively bluer;  i.e., it is more sensitive
to R-J light from the evolved population that is peaking in the near-IR
window.
And indeed, that is what is observed, particularly for the early-type
galaxies (NGC\,584, NGC\,777, M\,87, NGC\,1398, M\,51b) which tend to
be 5$-$10\% brighter in W1  (the expected value is $\sim$6\%).
The late-type galaxies range between 0 and 5\% brighter in W1, which
is in line with the expectation.  The notable outliers are
the S4G galaxies NGC\,6118 and NGC\,1398,
which are 10$-$15\% too bright in W1 compared to the expectation;
and M\,51a, which is too faint by 5 to 10\%, likely due to it complex blending
with M\,51b. IC342, interestingly, has a ratio that is consistent with a starburst SED
in which warm dust has inverted the ratio;  indeed, the nucleus of IC342
is undergoing a strong starburst (cf.  Laine et al. 2006; Brandl et al. 2006;
J. Turner, private communication).

Comparing W2 4.6 $\mu$m to that of IRAC-2 4.5 $\mu$m,
the WISE fluxes are expected to be slightly fainter, $\sim$5\%, than the IRAC fluxes
because of the slope in the R-J tail.  The observed scatter in the flux ratio
is on that order,  with early and late-types mixing without any obvious
segregation.  The outliers are M\,51a/b, a blended pair system,
and NGC\,6822, which is a Magellanic dwarf located behind the Milky Way.

The W3 12  $\mu$m bandpass is significantly different from the
IRAC-4 8.0 $\mu$m both in terms of the central wavelength
and the sensitivity to the ISM.    The IRAC-4 band is centered on
the 7.7 $\mu$m PAH band (and includes the 6.2 $\mu$m PAH band), which are small and charged molecules,
while the W3 band is centered on the 11.3 PAH band, which are
larger and tend toward neutral-charge molecules (cf. Tielens 2008; Draine 2011).
The broad W3 band  also includes the silicate absorption band
at 10 $\mu$m,  a typically significant feature in infrared-luminous galaxies.
The expected flux ratios are, consequently, expected to
be large depending on the galaxy type, ranging from  0.6 for early types
(R-J emission) and 1.2 for star-forming disk galaxies with AGN.  Although the range is
large, the observed ratios appear to be within 5 to 10\% of the expected values.
The highest ratio belongs to IC342, which is likely the strongest nuclear starburst
in the sample.   The other strongly star-forming galaxies, including NGC\,2403,
NGC\,6946 and M\,83, all have high ratios, indicating strong 11.3 $\mu$m PAH emission.

For the longest wavelength extractions, we compare W4 22 $\mu$m with the MIPS-24 measurements.  The expected
flux ratios range from 0.9 (late-types) to 1.1 (early-types).   The observed ratios tend
to cluster around unity, suggesting an uncertainty of $\sim$5 - 10\% between WISE and
\Spitzer measurements in these two bands.  The only outlier is M\,87,
revealing a difference in nuclear AGN emission observed by WISE and \Spitzer.
Finally, a note about the MIPS-24 observations of  M\,81, NGC\,1566 and M\,51b:
they all have \Spitzer imaging that does not fully cover the field as wide as WISE;
hence, there is a slight ($\sim$5\%) underestimate of their \Spitzer fluxes.

\begin{figure*}[ht!]
\begin{center}
\vspace{-5pt}
\includegraphics[width=17.5cm]{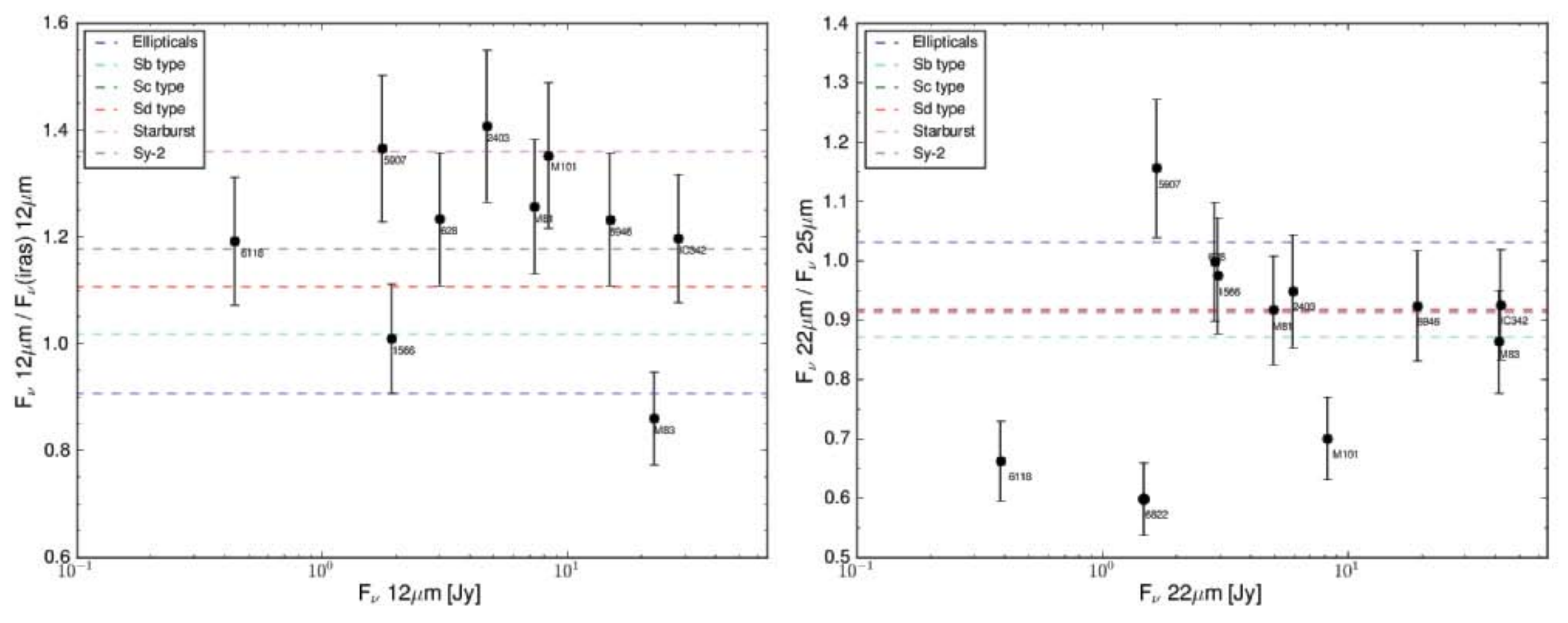}
\caption[WISE vs IRAS]
{\small{WISE photometry compared to the IRAS 12 and 25 $\mu$m bands.
The IRAS photometry comes from Rice et al. (1988),
Sanders et al. (2003) and Lisenfeld et al. (2007).
Sources NGC\,6822 and NGC\,1398 are not shown due to
confusion or poor quality IRAS measurements.
}}
\label{WISEvIRAS}
\end{center}
\end{figure*}

We now compare the WISE isophotal photometry with
published IRAS 12 and 25 $\mu$m fluxes.  As with the \Spitzer comparison, we derive
the expected ratios using model SED templates and the WISE/IRAS
bandpass RSRs.  No color corrections have been applied to the IRAS fluxes,
only the published values are used for this comparison.
The results are presented in Fig. \ref{WISEvIRAS}.
Due to the limited sensitivity of IRAS, all of the galaxies in this comparison
are gas-rich star-forming systems, roughly Sc/d types (M\,81 is the exception, it is
an early-type spiral).

For 12 $\mu$m, we expect ratios
between 1.0 (early type spirals) and 1.35 (starbursts).  The observed
ratios are 1.1 to 1.4 for most of the systems, which is $\sim$10\%
brighter than the expected range for Sc/Sd type galaxies.  M\,83 appears
to have a flux ratio that is 10\% too faint, but well within the uncertainties
(IRAS fluxes).
We would then conclude that WISE W3 photometry is systematically
$\sim$10\% brighter than IRAS estimates.

For 25 $\mu$m we expect a more narrow range in flux ratio, from 0.9 (late types)
to 1.1 (early types).   What is observed is that most sources do have ratios in that range,
but exceptions include NGC\,5907 (ratio is too high), NGC\,6118 (too low) and M\,101 (too low).
It is not clear if this is a WISE or IRAS discrepancy,
but the latter is the more likely given that WISE W4 and MIPS-24 have very good agreement.
The uncertainty between WISE W4 and IRAS 25 $\mu$m extractions is  $\sim$20\%,
consistent with the quoted IRAS flux density uncertainties.

\section{Appendix B: GALEX Photometry}

Aperture photometry extracted from the GALEX images of the sample galaxies (except NGC\,6118, which was
not observed by GALEX)
is presented in Table \ref{tab:Gflux}.  Since there is little (or weak) spatial correspondence
between the UV and the infrared properties (discussed in Section 5, M\,83 results), there is no attempt to match the UV apertures
with those of WISE;   instead, we use the optical RC3 elliptical shape values, with slight adjustments where
needed.   The reported flux densities have not been corrected for the foreground Galactic extinction, however the
table also includes the expected foreground extinction (AB mag units) in the GALEX bands
(see Treyer et al. 2007).  In Section 6
we estimated the UV star formation rate using these flux densities corrected for Galactic extinction.

\begin{table*}[ht!]
{\scriptsize
\caption{GALEX Aperture Photometry  \label{tab:Gflux}}
\begin{center}
\begin{tabular}{r r r r r r r r r r r}

\hline
\hline
\\[0.25pt]

Name  &  axis  &   p.a.  & R$ _{25}$ & F$_{FUV}$ & A$_{FUV}$ & F$_{NUV}$ & A$_{NUV}$ \\
           &  ratio &  (deg) & (arcmin) &   (mJy) & (mag) & (mJy) & (mag)   \\

\hline
\\[0.25pt]

NGC\,584   &   0.66&   69.5&    2.9&    0.4&  0.35&    1.6&  0.35 \\
NGC\,628   &   0.93&   80.0&   10.3&   75.2&  0.59&  104.7&  0.58 \\
NGC\,777   &   0.74&  145.0&    2.1&    0.3&  0.38&    0.6&  0.38 \\
NGC\,1398  &   0.71&  100.0&    5.2&    9.4&  0.11&   14.6&  0.11 \\
NGC\,1566  &   0.91&   59.0&    6.5&   51.5&  0.08&   66.1&  0.08 \\
NGC\,2403  &   0.61&  126.0&   14.9&  251.2&  0.33&  322.1&  0.33 \\
NGC\,3031  &   0.70&  157.0&   21.1&  177.0&  0.66&  248.9&  0.66 \\
NGC\,4486  &   0.94&    0.0&    7.9&    6.5&  0.19&   13.8&  0.19 \\
NGC\,5194  &   0.65&  184.0&    8.2&  131.8&  0.29&  214.8&  0.29 \\
NGC\,5195  &   0.79&   79.0&    9.3&    4.5&  0.29&   10.0&  0.29 \\
NGC\,5236  &   1.00&    0.0&   14.0&  325.1&  0.55&  544.5&  0.55 \\
NGC\,5457  &   0.95&   80.0&   17.2&  334.2&  0.07&  416.9&  0.07 \\
NGC\,5907  &   0.16&  156.0&    7.2&    8.1&  0.09&   12.5&  0.09 \\
NGC\,6822  &   1.00&    0.0&   15.1&  359.8&  1.94&  539.5&  1.90 \\
NGC\,6946  &   0.96&   80.0&    9.4&  288.4&  2.82&  461.3&  2.74 \\
IC\,342    &   0.96&   81.0&   15.0& 1037.6&  4.60& 1116.9&  4.39 \\

\hline
\end{tabular}
\end{center}
\tablecomments{The reported flux densities have not been corrected for the
foreground Galactic extinction; however, A$_{FUV}$ and A$_{NUV}$ are the
estimated extinctions (in AB mag units) for the GALEX bands and may be used to correct the flux
densities accordingly.
The formal errors are less than 1\%; but including the flat-field and
calibration errors, the actual photometric uncertainty is $\sim$5\%.  The semi-major axis,
R$_{25}$ is approximately equal to the one-half of the optical (RC3) D25 diameter.
Photometry of NGC\,5194/5 is uncertain due to blending.
}
}

\end{table*}


\section{Appendix C:  Radial Surface Brightness Profiles}

The following plots show the azimuthally-averaged, radial surface brightness profiles for
sample of galaxies, including both WISE HiRes (black solid line) and \Spitzer (magenta solid line) measurements.
The results paired in the standard way:  W1/IRAC-1, W2/IRAC-2,
W3/IRAC-4 and W4/MIPS-24, from left to right panels, respectively.
The WISE profile shapes are simply characterized using a double S\'ersic function fit:
the inner galaxy or bulge light (cyan dashed line) and the
disk (blue dashed line);  the grey-dashed line is the total contribution from
the bulge and disk.\\

For the elliptical and early-type galaxies (NGC\,584, NGC\,777, M\,87) , the WISE and \Spitzer results are solid consistent for all
radii down to the background level, but for the longer wavelengths the signal is much fainter and
the surface brightness is drops rapidly with radius.
The star forming galaxies also show good correspondence between WISE and \Spitzer profiles.
Note the for the W3 to IRAC-4 comparison, the bandpasses are sufficiently different that there
is an offset in surface brightness.  The expected offset, based on the spectral shape, ranges
from a scale factor of 0.6 (early types) to 1.1 (star-forming) to 1.2 (AGN);  see Fig. 7 for details.
One clear trend that is seen:  the \Spitzer images (notably IRAC) are sometimes too small to capture the
total flux of the galaxy, and as a consequence the local background subtraction tends to chop off the
outer most extended light. For example, the W3 profile of M\,83 (see also Fig. 13) drops to zero
at a radius of 250$\arcsec$, whereas the WISE W3 (as well as GALEX) extends beyond 500$\arcsec$.\\

\begin{figure*}[ht!]
\begin{center}
\includegraphics[width=17cm]{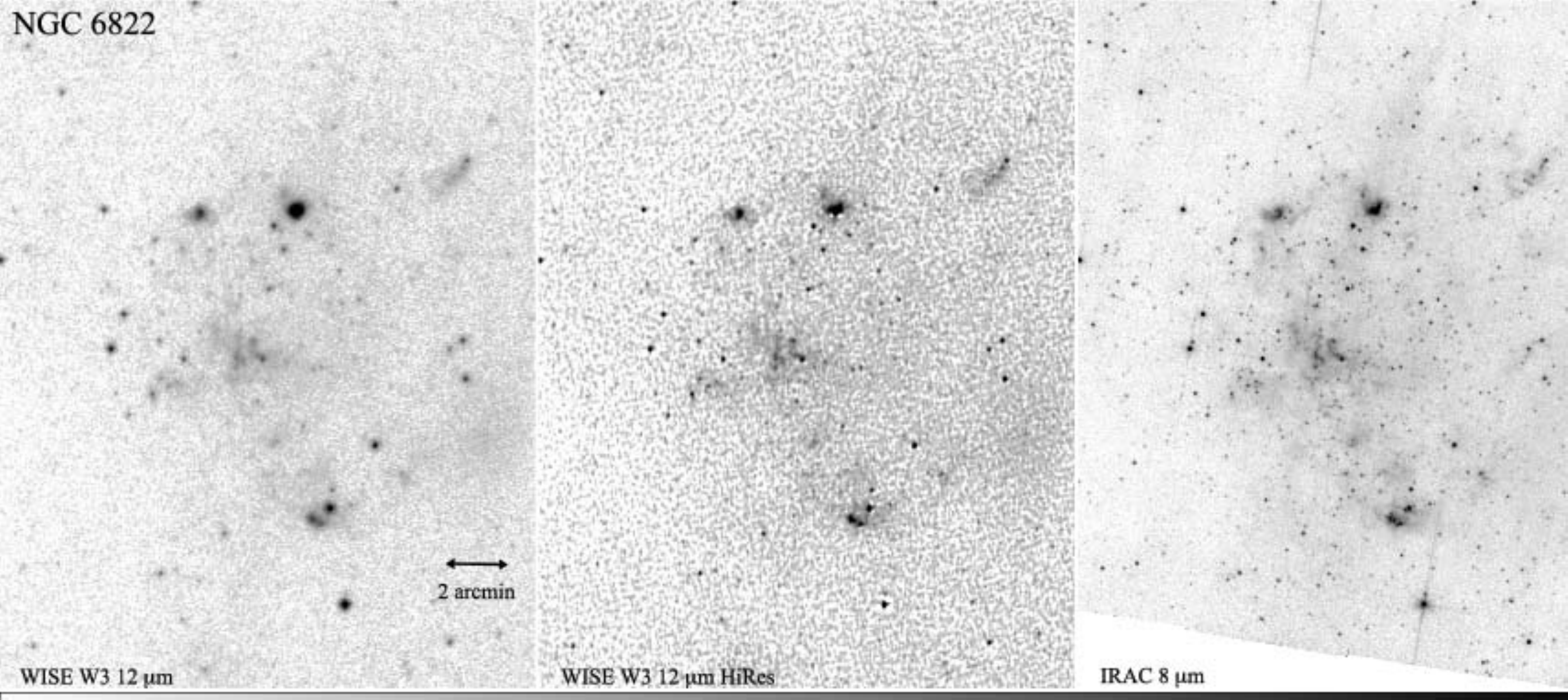}
\caption[n6822]
{\small{The dwarf galaxy NGC\,6822.
WISE W3 12$\mu$m drizzle and HiRes, and \Spitzer IRAC 8$\mu$m
imaging.
}}
\label{n6822}
\end{center}
\end{figure*}

Finally, the low surface brightness galaxy in the sample, NGC\,6822, shows the largest departure
differences between WISE HiRes and the \Spitzer imaging profiles.  In all four bands, the WISE profiles
show a saw-tooth ringing pattern, which is not seen in \Spitzer profiles.   This galaxy is characterized
by a flat diffuse component punctuated with stars and (a few) star formation regions
(see Fig. 1).  The averaged WISE profiles of NGC\,6822 are in fact dominated by noise fluctuations and
the ``ringing" artifacts that are induced by the MCM process (see Paper I for more details on the
amplified noise and ringing behavior).   However, without the radial averaging,
the differences are not so apparent when comparing images side by side; see Fig. \ref{n6822} which shows the
W3 12 $\mu$m (drizzle and HiRes versions) and IRAC 8$\mu$m images of NGC\,6822.
This case serves as a cautionary note:  for dwarf and low surface brightness galaxies,
the drizzle co-addition method is the preferred enhanced resolution process for WISE imaging.

\begin{figure*}[ht!]
\begin{center}
\includegraphics[width=17.5cm]{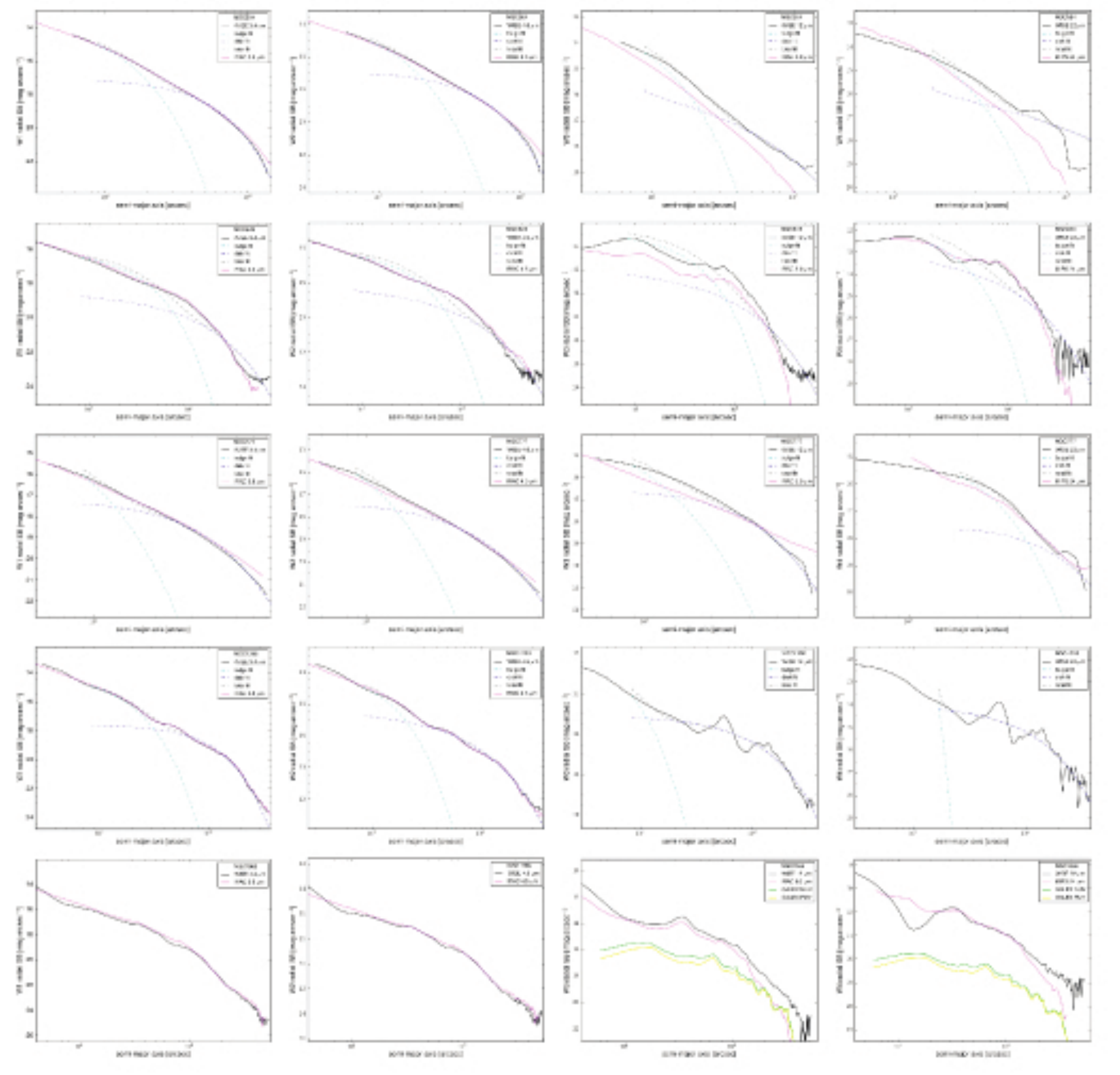}
\caption[Profiles-1]
{\small{Azimuthally-averaged elliptical-radial profiles comparing WISE (black line) with
IRAC and MIPS-24 (magenta line).  The dashed lines show the S\'ersic
fits to the bulge (cyan) and disk (blue) regions.
}}
\label{Profiles-1}
\end{center}
\end{figure*}

\begin{figure*}[ht!]
\begin{center}
\includegraphics[width=17.5cm]{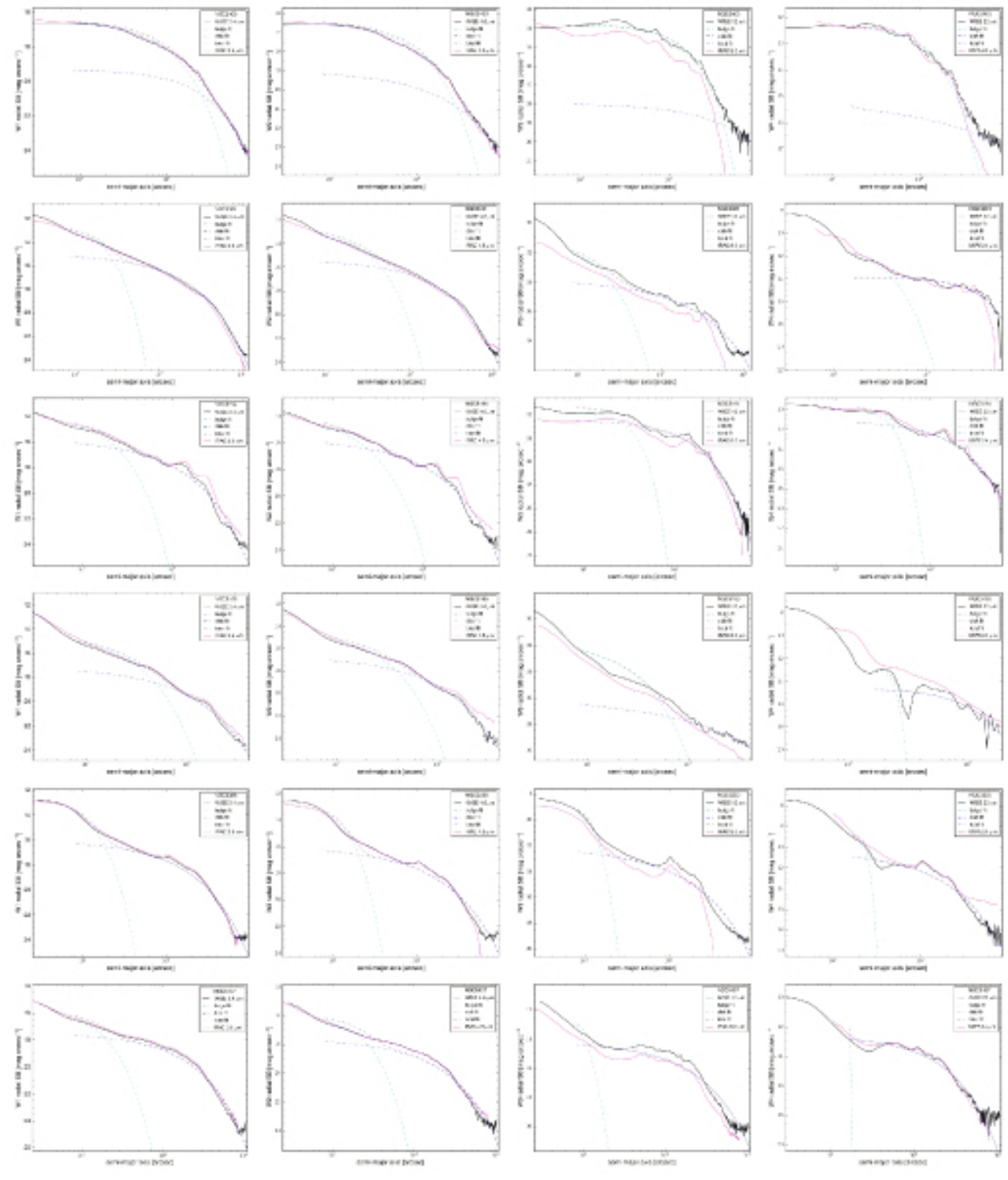}
\caption[Profiles-2]
{\small{See Fig. \ref{Profiles-1} for details.
}}
\label{Profiles-2}
\end{center}
\end{figure*}

\begin{figure*}[ht!]
\begin{center}
\includegraphics[width=17.5cm]{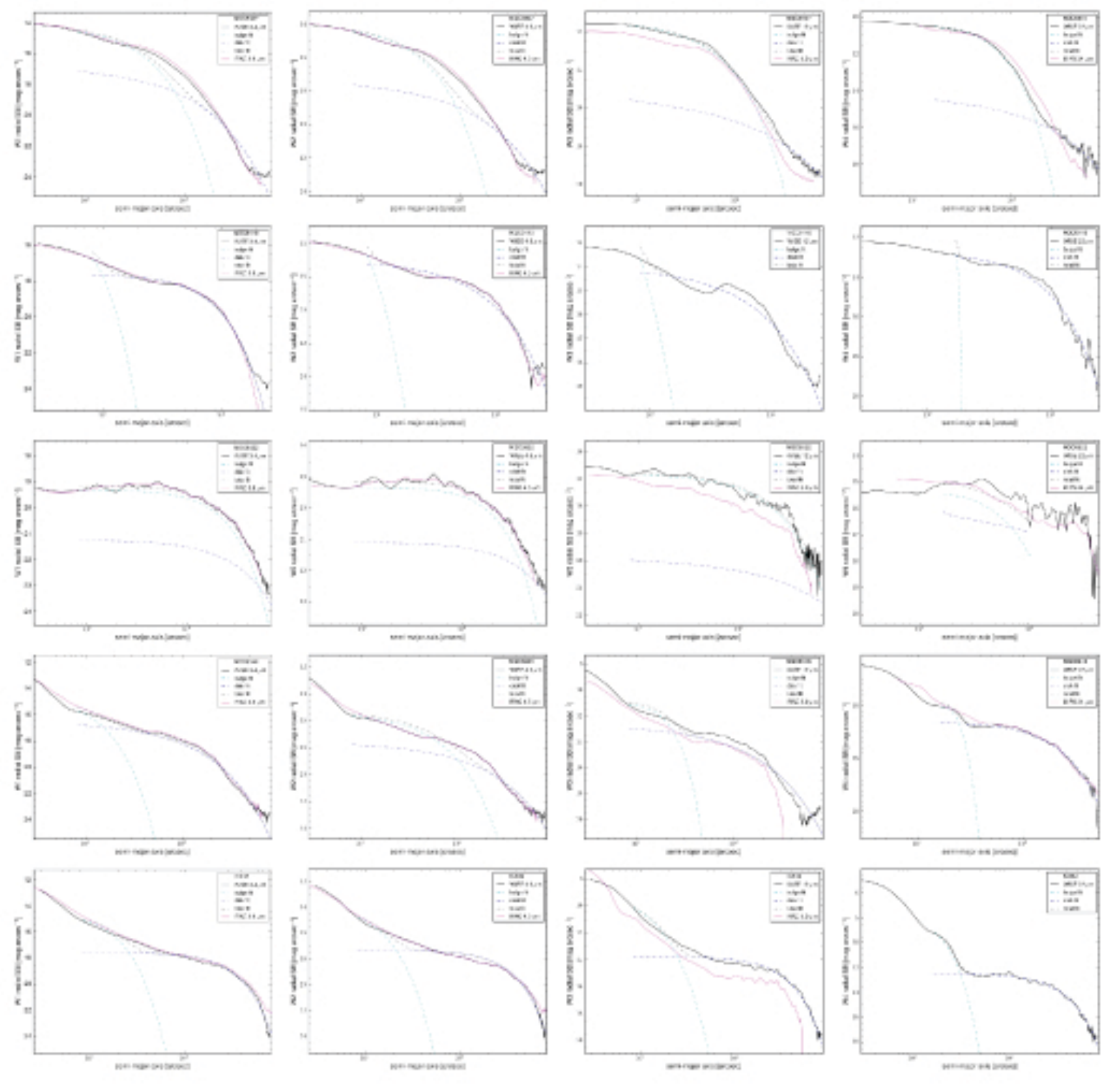}
\caption[Profiles-3]
{\small{See Fig. \ref{Profiles-1} for details.
}}
\label{Profiles-3}
\end{center}
\end{figure*}

\clearpage

\section{Appendix D:  WISE Enhanced Resolution Galaxy Atlas (WERGA)}

The results presented in this work represent the pilot study of a large program in
which
we will use the WISE image and source catalog archive to
build the
WERGA,
consisting of the largest galaxies in the sky,
that will complete the S4G IRAC-1 and IRAC-2,   D=40 Mpc
volume-limited, survey, as well as complement other large
surveys (e.g., SDSS, 2MASS, AKARI, LVIS).

A key feature of the WERGA is the construction of images with a factor of $\sim$3 to 4
improvement in spatial resolution compared to the public release mosaic imaging.
The achieved resolutions
are comparable, $\sim$25 to 30\%, to those of \Spitzer IRAC and MIPS-24.
We estimate that MCM/HiRes resolution enhancement (Paper I) is most effective for all galaxies with a near-infrared diameter greater
than $\sim$2 arcmin.    Using the 2MASS XSC (Jarrett et al 2000; 2003) to select sources by their
near-infrared angular size, we find over $\sim$10,000 galaxies with diameters greater than this limit.
For galaxies smaller than this size threshold, the most effective and practical resolution-enhancement method is
to employ Variable-Pixel Linear Reconstruction (`drizzling')
which improves spatial resolution performance compared to nominal WISE imaging by $\sim$30-40\% , but
is also much less cpu-intensive than the MCM/HiRes reconstruction.  

Both sets of images, four bands each,
will comprise the WERGA imaging that will be released to the public via NED as part of this project.
This all-sky image Atlas will have broad and enduring impact to the community, representing the only WISE imaging
that is dedicated to resolved sources, and with the HiRes reconstructions a unique and fundamental component that
transforms the all-sky survey into a powerful `observatory' not unlike that of the {\it Spitzer} Space Telescope.
The legacy value of these
high-resolution images will span decades given that
AKARI and WISE are likely to be the only mid-IR all-sky survey for many years to come;
indeed, most galaxies in the local universe have only been imaged and measured by WISE.
(e.g., compared to the \Spitzer coverage of the local universe).

The advantage of WISE to simultaneously trace the stellar mass and the star formation activity
mean the WERGA will play a crucial and complementary role in the
multi-wavelength endeavor to understand galaxy assembly and evolution.
This dual capability means that the 
the WERGA could provide a resolved 
anchor for the (typically, spatially coarse)
atomic and molecular gas
studies of the SFR to stellar mass scaling relations that are now underway (e.g., GASS/COLDGASS).
Moreover, in the coming decade we will see the
entire sky mapped with unprecedented spectral sensitivity and spatial detail by radio-wave surveys
(ASKAP, MeerKAT and Apertif SKA pathfinders).
Notably the ASKAP-WALLABY (Koribalski \& Staveley-Smith, in prep) and EMU (Norris
 et al. 2011) projects will extract neutral hydrogen content and radio continuum (e.g., synchrotron) emission from
 galaxies in the local universe,  both of which in combination with WISE provide insight to
 the present and past star formation histories.

\end{document}